\documentclass{jfm} 
\usepackage{subcaption}
\usepackage{graphicx}
\usepackage{newtxtext}
\usepackage{newtxmath}
\usepackage{natbib}
\usepackage{hyperref}
\hypersetup{
    colorlinks = true,
    urlcolor   = blue,
    citecolor  = black,
}

\newcommand{\RomanNumeralCaps}[1]

\title{Ventilated Cavitation around a sphere through surface air injection at Re = 10,000}

\author{Soundararajan R \aff{1}
  \and Anikesh Pal\aff{1} \corresp{\email{pala@iitk.ac.in}}}

\affiliation{\aff{1}Department of Mechanical Engineering, Indian Institute of Technology, Kanpur 208016, India
}

\begin{document}
\maketitle

\begin{abstract}
We perform DNS of ventilated cavitation over a sphere at subcritical $Re = 10,000$. The flow is considered a homogeneous mixture of water and air, and the interface between the two fluids is tracked using the volume-of-fluid method. Unlike the previous numerical investigations, we include the effect of surface tension in our simulations. The air is injected from a circular strip on the sphere's surface into the water domain. To investigate the effect of injection location on cavity leading-edge dynamics and formation, three injection locations are chosen based on the bulk flow dynamics around the sphere. The angular span of the injection patches is: i) front (FR025): $18^{\circ} \leq \theta \leq 63^{\circ}$, ii) mid (MD025, MD050): $75^{\circ} \leq \theta \leq 104^{\circ}$, and iii) back (BK025): $120^{\circ} \leq \theta \leq 180^{\circ}$. The air is injected in a direction perpendicular to the sphere’s axis, passing through its forward stagnation point and its center, at ventilation coefficients $C_q = 0.2$ and $0.4$. The cavity inception and its stability are determined by the location of air injection. The ventilation also alters the flow separation in the front and mid-injection cases. An unsteady bubbly cavity forms in front-injection owing to vigorous puffing, which continuously fragments the injected air, increasing drag by $\sim 56\%$ relative to the single-phase case.  In contrast, both mid- and back-injection cases form stable cavities, but with different leading-edge dynamics. In the mid-injection cases, the puffing phenomenon is dominant, and despite delayed flow separation, the cavity detaches from the leading edge of the injection patch. However, in the back-injection case, the cavity detaches upstream of the injection patch due to the adverse pressure gradient generated by cavity formation. The Kelvin-Helmholtz instabilities and the divot formation are features of the cavity formed in the back-injection case. We find that the stable cavities strongly entrain the air, and a re-entrant jet forms as the cavity closes. We achieve a $35\%$ and $25\%$ drag reduction in the mid-injection case with $C_q = 0.2$ and $0.4$, respectively, while the back-injection configuration yields a maximum drag reduction of $46\%$ with respect to the single-phase case. \\

\end{abstract}


{\bf MSC Codes }  {\it(Optional)} Please enter your MSC Codes here

\section{Introduction}
\label{sec:introduction}

Hydrodynamic drag is the resistive force that a fluid exerts on a bluff body moving through it. A significant portion of energy is spent on overcoming this drag. Drag reduction is beneficial in attaining higher speeds and improving energy efficiency. Drag force primarily originates from the shear distribution (friction drag) and the pressure distribution (form drag) around the body. The selection of a drag-reduction strategy depends on the type of drag force to be reduced and the working fluid's Reynolds number regime \citep{Gad_el_Hak_2000}. Generally, in bluff bodies, flow separation reduces the fore-aft pressure recovery, thereby increasing the form drag and making it a significant contributor to the drag force. Flow separation occurs when a low-momentum flow along the surface detaches owing to an adverse pressure gradient. Avoiding or delaying flow separation would be key to form drag reduction. Surface dimples \citep{Choi_2006,Beratlis_2019}, boundary layer tripping \citep{ Son_2011}, blowing, and suction \citep{Jeon_2004, Wilson_2013,Chen_2013, Song_2025} are methods to delay separation. On the other hand, to control skin friction, the flow must remain laminar by suppressing the flow transition to a turbulent regime. Surface roughness \citep{Achenbach_1971, Shih_1993}, surface heating \citep{Vakarelski_2011}, air-layer drag reduction \citep{Elbing_2013, Makiharju_2017}, and cavitation \citep{Wu_1972,Ceccio_2010} are some methods to reduce skin friction drag. Extensive research over several decades has focused on flow control techniques for drag reduction. Different flow control methods over bluff bodies are reviewed in \cite{Choi_2008}.\\

Among the various drag-reduction strategies, cavitation has attracted considerable attention because it replaces the liquid adjacent to the body with gas, thereby significantly reducing viscous shear stresses \citep{Ceccio_2010}. Cavitation is the process of creating a cavity near the body by injecting gas externally or by generating vapor within the flow. Replacing water (a dense, more viscous fluid) with gas (a lighter, less viscous fluid) reduces friction drag. The presence of a gaseous cavity near the body not only reduces friction but also alters its pressure distribution, thereby influencing the form drag. Maximum drag reduction is achieved in supercavitation when the gas cavity completely envelops the body \citep{Ceccio_2010}. Based on the mechanism of cavity formation, cavitation is broadly classified into natural cavitation and ventilated cavitation. According to \cite{Franc_2006}, natural cavitation refers to the breakdown of the fluid medium due to low pressure. A major limitation of natural cavitation is that cavitation inception typically requires very high flow speeds. In ventilated cavitation, this limitation is overcome by injecting a non-condensable gas near the body to form a gaseous bubble around the surface. In contrast to the requirement of high traversing speeds in natural cavitation, the characteristics of ventilated cavities can be actively controlled by regulating the gas injection rate at relatively lower speeds.\\ 

Different non-dimensional parameters govern the formation and dynamics of ventilated cavitation. The cavitation number ($\sigma_c$) is the primary parameter that quantifies the pressure drop in the cavity region \citep{Logvinovich_1972}. It is expressed as $\sigma_c = (p_{\infty}-p_c)/(0.5\rho_{\infty} U_{\infty}^2)$ where $p_c$ is the pressure inside the cavity, $p_{\infty}$, $\rho_{\infty}$, and $U_{\infty}$ are the freestream pressure, density, and velocity, respectively. As the freestream pressure is close to the cavity pressure, the extent of cavitation increases, and lower values of $\sigma_c$ generally favor the supercavity formation. This non-dimensional number is not known a priori for ventilated cavitation. Another important parameter is the ventilation coefficient ($C_q$), also known as the air entrainment coefficient, which characterizes the volume of gas injected into the bulk flow \citep{Ceccio_2010}, and it is defined as $C_q = Q_{in}/(U_{\infty}d_c^{2})$ where $Q_{in} = V_{in}A_{in}$ is the volumetric injection rate of gas with $V_{in}$ as the velocity of injection and $A_{in}$ as the surface area of injection. The density ratio between the injected gas and the surrounding bulk liquid is large, typically $\sim O(10^3)$. Owing to this significant density contrast, gravitational effects become important in cavitating flows. The effect of gravity with respect to the inertial effects is characterized using the Froude number($Fr$), $Fr = U_{\infty}/(\sqrt{gd_c})$, where $g$ is the acceleration due to gravity and $d_c$ is the characteristic diameter \citep{Wu_1972}. Since the ventilated cavities are formed at low traversing speeds, gravitational forces strongly influence the cavity shape \citep{Franc_2006}. The Reynolds number($\Rey$) is defined as $\Rey = \rho_{\infty}U_{\infty}d_c/\mu_{\infty}$ where $\mu_{\infty}$ is the bulk fluid dynamic viscosity at freestream. The formation of a cavity around the body also modifies the hydrodynamic forces acting on it. In particular, the drag force($D$) which is commonly expressed in terms of drag coefficient, $C_{d}$ computed as $8D/(\pi\rho_{\infty} U_{\infty}^2 d_c^2)$ based on the frontal area. These non-dimensional parameters are strongly coupled, and their combined influence on cavity dynamics is complex \citep{Chung_2018}.\\ 

At low ventilation rates, the injected gas is insufficient to form a single cavity around the body. This occurs primarily due to interaction between the gas and wake, which convects the gas downstream as bubbles within the shed vortices \citep{Gawandalkar_2025}. At times, the presence of dispersed bubbles produces a foamy appearance, and such cavities are referred to as foamy cavities \citep{Karn_2016b, Chung_2018, Pham_2024, Gawandalkar_2025}. The drag force does not decrease in foamy cavities compared to the corresponding single-phase flow \citep{Jiang_2019}. An important drawback of the bubbly cavity is that the highly unsteady and and the bubbly structure makes experimental measurements difficult \citep{Wu_2021}. Hence, a stable cavity attached to the bluff body is preferred over a bubbly cavity for drag reduction. As ventilation increases in a bubbly cavity, more bubbles are generated, which subsequently coalesce to larger bubbles \citep{Wu_1972}, and with sufficient ventilation, a single attached cavity forms. \\ 

Once a stable cavity is formed, its subsequent dynamics, including its ability to remain attached to the body, are important. The drag reduction achieved during cavity formation is sustained while the body remains enveloped by the lighter fluid. The rear portion of an attached cavity, called the cavity closure, is a core mechanism that affects its shape and dynamics. The mode of cavity closure regulates the gas evacuation from the cavity, making it important for the sustenance and growth of a ventilated cavity \citep{Gawandalkar_2025}. In natural cavitation, the water vapor can exit the cavity through the cavity wall by condensation. On the contrary, in ventilated cavities, the air exits the cavity mainly through the closure region \citep{Spurk_2002}. In the earlier literature, the most commonly observed closure modes are: re-entrant jet (RJ), twin vortex (TV), quad vortex (QV) and pulsating cavities (PC). Different closure modes were observed at different flow parameters. \cite{Campbell_1958} conducted ventilated cavitation experiments at $5\leq Fr\leq25$ and they obtained closure mechanism transition along the curve $\sigma \times Fr = 1$. In particular, RJ closure occurs when $\sigma \times Fr < 1$ and TV closure occurs when $\sigma \times Fr > 1$. At higher Froude numbers, where the gravitational effects are weak, a cavity forms a re-entrant jet at closure with gas exiting the cavity through toroidal vortices \citep{Semenenko_2001}. \cite{Brennen_2014} describes a re-entrant jet as "frothy turbulent mass tumbling back into the cavity." \cite{Spurk_2002} quantified the gas loss from ventilated cavities at high Froude numbers. They postulated that the injected air is entrained along the interface boundary layer and reaches the rear of the cavity. At the closure, the gas, on mixing with the fluid, is shed through periodic toroidal vortices. \cite{Kinzel_2009} extended this theory to twin vortex cavities. According to \cite{Kinzel_2009}, the pressure at the closure of the twin vortex cavity is high, leading to an adverse pressure gradient at the closure. This adverse pressure gradient is responsible for shear layer thinning, leading to gas exiting the cavity surface. A portion of this exiting air reenters the cavity from the closure region, and the remaining air enters the twin vortices. At lower Froude numbers, gravity effects on the cavity become significant, lifting the rear portion. The lifting up of cavity forms two counter-rotating vortices through which the injected gas exits the cavity \citep{Wu_2019}. \cite{Kawakami_2011} observed two types of RJ closure at low ventilation coefficients, and a third closure type at the transition from RJ to TV. At low $C_q$, the cavity was short and opaque, and periodically shedding toroidal vortices; at the transition, the major portion of the cavity was clear, with a recirculation region at the rear. Additionally, at low Fr and small $C_q$, they observed the splitting of the twin vortices to QV closure.
\cite{Karn_2016b} proposed that the pressure difference across the cavity interface is key in determining the mode of cavity closure. The cavity's internal and external pressure difference depends on the non-dimensional parameters, leading to different closure modes. \cite{Cao_2017} observed that as the cavity elongates, the adverse pressure gradient at the closure decreases. \cite{Chung_2018} observed double-layer cavities with different closures for the outer and inner cavities. However, the mechanism responsible for this structure was not explained. \\

\cite{Wu_2019} visualized the internal flow of a ventilated supercavity generated on a backward-facing cavitator under re-entrant jet and twin vortex closure modes using particle image velocimetry. The internal flow is divided into three regions: the ventilation influence region, where the injected gas exerts a major influence; the internal boundary layer along the gas-water interface, which entrains air downstream of the cavity closure; and the reverse flow region, which is directed upstream. In the RJ closure mode, the cavity elongates to accommodate the increased ventilation as gas leakage through the toroidal vortices decreases. On the other hand, the vortex tubes widened during TV closure, removing excess air and causing no elongation of the cavity length. \cite{Lv_2021} experimentally investigated a ventilated cavity over an axis-symmetric body, and their results were consistent with \cite{Wu_2019}.\\

Pulsating cavity closure was first observed by \cite{Silberman_1959} at high ventilation rates while performing experiments with axis-symmetric bodies in a water tunnel with a vertical test section. The resonance between the surface wave generated at the cavity wall and the gas/liquid system frequency triggers cavity pulsation that further causes periodic variation in cavity volume due to the separation of air pockets, leading to oscillation in cavity pressure \citep{Michel_1984}. Oscillation in cavity pressure generates noise. \cite{Skidmore_2015} conducted experiments to measure noise in ventilated cavities. The noise generated by pulsating cavities was $40$ dB higher than the RJ and TV closure modes. To control the noise generated in pulsating cavities, \cite{Skidmore_2016} shifted the gas/liquid system frequency away from the surface wave frequency. This led to the formation of TV cavities by suppressing the pulsating cavity.\\ 

In ventilated cavitation, a hysteresis effect where the ventilation required to create a stable supercavity exceeds that required to sustain the cavity. \cite{Kawakami_2011} reported ventilation hysteresis in the experiments of artificial supercavitation behind a sharp-edged disk. Though many studies have also observed ventilation hysteresis around different bodies \citep{Karn_2016a, Karn_2016b, Jiang_2019, Hao_2022, Gawandalkar_2025}, only a few have justified its occurrence. \cite{Karn_2016b} postulated the mechanism of ventilation hysteresis based on the bubble coalescence and pressure difference at the cavity closure. In a recent work, \cite{Gawandalkar_2025} investigated the ventilated cavities in the wake of a two-dimensional bluff body experimentally and explained the ventilation hysteresis in terms of the difference in gas leakage from a foamy cavity and a stable cavity. The hysteresis effect introduces path selectivity to the cavitation problem, with ventilation demand depending on the path used to achieve ventilation. \cite{Karn_2016b} could form an RJ closure only by reducing ventilation from TV closure. This is in contrast to \cite{Gawandalkar_2025}, who obtained RJ closure by increasing ventilation from a foamy cavity. Moreover, \cite{Karn_2016b} observed that after the formation of a supercavity, a reduction in ventilation was accompanied by a change in closure mode, whereas \cite{Gawandalkar_2025} did not find any change in the cavity closure with $C_q$ reduction.\\

Although drag reduction is a primary motivation for ventilation, relatively few studies have directly examined drag reduction in cavitating flows. 
Generally, a ventilated cavity is formed by injecting gas in a low-pressure region, created by attaching a cavitator to the front of an object \citep{Karn_2016a, Karn_2016b, Chung_2018, Jiang_2019, Wu_2019, Li_2025}. \cite{Chung_2018} conducted a ventilated cavitation experiment on a moving body in still water within a free-surface bounded flow. They found that the presence of a cavitator increased the drag force in the single-phase flow. However, for their ventilated cases, the formation of a supercavity results in $\sim 73\%$ drag reduction, and the drag force becomes lower than that of the body without the cavitator in single-phase flow. \cite{Jiang_2019} experimentally studied drag characteristics of an axis-symmetric supercavitating bodies(blunt and slender) by varying $C_q$ and $Fr$. They reported maximum drag reduction for higher Fr cases because the supercavity completely covers the body. Similar to \cite{Chung_2018}, \cite{Jiang_2019} initially observed an increase in the drag coefficient on attaching a cavitator in their experiments. However, upon air injection, they also achieved a significantly lower drag coefficient than single-phase flow.\\

The drag force associated with a cavitating flow over a bluff body is primarily dependent on the location of the cavity detachment. \cite{Brennen_1969b} calculated the position of cavity separation based on potential flow analysis and compared it with experiments. The experimentally observed cavity separation occurred downstream of the theoretical prediction. The cavity did not separate smoothly, possibly due to viscous effects. The delay in separation is reflected in the experimental drag coefficient, which was higher than the theoretical prediction. In subsequent works \citep{Brennen_1970a, Brennen_1970b}, they studied the appearance and stability of natural cavity surfaces. \cite{Brennen_1970a} investigated the wave patterns generated by the growth of instabilities in the cavity interface of spheres and hemispheres, whereas \cite{Brennen_1970b} found that the addition of drag-reducing polymers distorted the cavity separation line and introduced additional instabilities on the cavity wall. More recently, \cite{Brandner_2010} conducted experiments on natural cavitation over a sphere at $Re = 1.5\times10^6$ from inception of cavity to supercavitation. At higher cavitation numbers ($0.95 < \sigma < 1$), the cavity breaks down due to Kelvin-Helmholtz (KH) interfacial instabilities. At moderate cavitation numbers ($0.4<\sigma <0.9$), the cavity is broken by the re-entrant jet and shed downstream with a foamy appearance, and at low cavitation numbers ($\sigma <0.4$), interfacial instabilities are damped and a stable supercavity forms. Extending the previous work \cite{Brandner_2010}, \cite{De_graaf_2017} analyzed the spectral content of cloud cavitation. They observed a unimodal spectral regime at high cavitation numbers due to the breakup of the cavity by the surface KH instabilities. At lower cavitation numbers, they saw two bimodal regimes due to RJ instability and propagation of a condensation shockwave at cavity break-off.\\

Several experimental studies focused on the location of cavity separation and flow dynamics around the cavity detachment region. \cite{Arakeri_1975} examined the influence of viscosity on the cavity separation over elongated bluff bodies. They showed that the laminar flow separation occurs upstream of the cavity flow separation. They also formulated the location of cavity detachment, and their formulation correlated well with the experiments. Later \cite{Franc_1985} used a variety of test objects, such as a hydrofoil, square, and elliptical cylinders, to experimentally demonstrate that stable cavities cannot exist in the absence of upstream flow separation. They also noted that an upstream adverse pressure gradient promotes flow separation. \cite{Tassin_Leger_1998a, Tassin_Leger_1998b} experimentally studied axisymmetric ventilated cavitation over hydrophilic and hydrophobic spheres and a two-dimensional natural cavity over a hydrofoil. \cite{Tassin_Leger_1998a} reported the presence of a recirculation region just upstream of the cavity in hydrophilic surfaces. This region was absent in the hydrophobic bodies due to the proximity of laminar separation and cavity detachment. \cite{Tassin_Leger_1998b} observed the formation of divots in the hydrophilic test objects. Triangular peaks were observed at the cavity leading edge for the hydrophilic counterparts.\\

Despite the challenges, cavitation has been widely studied experimentally. Cavitating flows contain bubbles spanning a wide range of spatial and temporal scales. As a result, numerical studies of cavitation around bluff bodies remain relatively limited. \cite{Gnanaskandan_2016} performed a compressible large eddy simulation (LES) on the effect of natural cavitation in the near wake of a circular cylinder. As $\sigma_c$ decreases, the periodic cavity shedding transitions to an attached cavity with periodic leakage. The cylinder's force coefficients exhibit a primary peak associated with cavity shedding. However, secondary peaks resulting from the impingement of pressure waves generated by the cavity collapse in the high-pressure wake are also reported. \cite{Wang_2021} simulated ventilated cavitating flow around a cylinder by introducing a spherical ventilation source downstream of the cylinder. They found that increasing ventilation suppresses turbulence kinetic energy at the closure, leading to fewer bubbles. The shear layers interact at the cavity closure, delaying vortex shedding. Vortex formation is mainly affected by instability at the cavity surface and re-entrant jet formation. With a similar ventilation setup, \cite{Liu_2023} simulated ventilation behind a circular disk to study the air leakage and turbulence dynamics at the cavity closure. They concluded that violent, turbulent mixing of the two phases and the generation of vortex structures occur at the stagnant, high-pressure closure region. The entrained air is transported downstream by vortices shed from the cavity closure. \cite{Gnanaskandan_2016, Wang_2021, Liu_2023} have ignored the influence of surface tension in their simulations. \\

The previous ventilated cavitation studies around a bluff body generate a cavity either by gas injection at the low-pressure region created by a nose-attached cavitator \citep{Karn_2016b, Karn_2016a, Chung_2018, Jiang_2019, Wu_2019, Li_2025} or by gas injection at the wake of the bluff body \citep{Tassin_Leger_1998a, Tassin_Leger_1998b, Wang_2021, Liu_2023}. To the best of our knowledge, this is the first computational study in which we generate a stable cavity around a bluff body by injecting gas from its surface. Unlike the previous computational studies, we include the influence of surface tension in our simulations. The inclusion of the surface tension significantly constrains the size of the time-step, owing to the requirement of fulfilling the capillary wave criterion for numerical stability and accurately capturing the air-water interface evolution, thus making these simulations challenging and expensive. Additionally, we report the cavity leading-edge dynamics, cavity internal dynamics, cavity closure mode, and quantify the loads on the bluff body. The present work enhances the understanding of a ventilated cavity formation over a bluff body and its detachment dynamics. \\

Single phase flow around a sphere at high Reynolds numbers has extensively been investigated using experimental methods (\cite{Achenbach_1972}, \cite{Achenbach_1974}, \cite{Taneda_1978}, \cite{Kim_1988} and \cite{Sakamoto_1990}) and numerical methods (\cite{Constantinescu_2003}, \cite{Yun_2006}, \cite{Rodriguez_2011}, and \cite{Rodriguez_2013}). In the present work, we choose $Re = 10,000$ in the subcritical regime, where the boundary layer is laminar, and the wake is turbulent \citep{Rodriguez_2011}. 
We perform direct numerical simulation (DNS) of flow past a sphere with air injected from the sphere’s surface at different positions and injection velocities to generate a stable ventilated cavity. We study the effect of flow separation on the cavity detachment dynamics and stable cavity formation. Further, we study the effect of ventilation on the surface quantities such as pressure, friction, and forces acting on the sphere.
We conclude by studying the internal dynamics of the ventilated cavities and the re-entrant jet dynamics. 
The problem formulation, the governing equations, and the problem setup with both numerical and grid discretization are discussed in $\S$ \ref{sec:problemformulation}. The validation of the solver for both single-phase and multiphase flows is presented in §\ref{sec:val}. The results and discussion are provided in §\ref{sec:result}. Finally, the summary is given in §\ref{sec:conc}.\\

\section{Problem Formulation}\label{sec:problemformulation}
 \subsection {Governing Equations}
The unsteady, incompressible, three-dimensional two-phase Navier-Stokes equations govern the flow. The Volume of Fluid (VoF) method is used to capture the interface by solving an additional conservation equation for the volume fraction. The governing equations are as follows:
 \begin{equation}
   \bnabla\bcdot\boldsymbol{u} = 0
\end{equation}
\begin{equation}\label{eq:Mom}
    \frac{\partial (\rho \boldsymbol{u})}{\partial t}+ (\boldsymbol{u} \bcdot \rho\bnabla) \boldsymbol{u} = -\bnabla p_{d} - (\boldsymbol{g} \bcdot \boldsymbol{x})\bnabla \rho + \bnabla \bcdot \mathsfbi{T} + \boldsymbol{f_{\sigma}} 
\end{equation}

\begin{equation}\label{eq:Alp_adv}
    \frac{\partial \alpha_w }{\partial t}+ \bnabla \bcdot \boldsymbol{u}\alpha_w  + \bnabla \bcdot \boldsymbol{u}_r\alpha_w(1-\alpha_w) = 0 
\end{equation}

where $\rho$, $\boldsymbol{u}$, $p_d$, and $\mathsfbi{T}$ are density, velocity, dynamic pressure, and the deviatoric viscous stress tensor of the mixture, respectively. The dynamic pressure is related to the total pressure ($p$) by, $p_d = p - \rho\boldsymbol{g}\bcdot\boldsymbol{x}$, where $\boldsymbol{g}$ is the acceleration vector due to gravity (\cite{Krishna_2025}). In the present study, the effect of gravity is neglected; hence, the dynamic and total pressures are equal, and the Froude number in the present study is infinite. Henceforth, $p_d$ is referred as $p$. The volume fraction of water is $0\leq \alpha_w \leq 1$. The volume fraction of air ($\alpha_a$) is the complement of the water volume fraction, and it is calculated as $\alpha_a=1-\alpha_w$. The solution of $\alpha_w$ on solving the advection equation\eqref{eq:Alp_adv} needs to be non-diffusive and bounded. To counter the smearing of the interface, i. e., diffusive interface, an artificial compression term, $\bnabla \bcdot \boldsymbol{u}_r\alpha_w(1-\alpha_w)$, is added to the LHS of the advection equation \citep{Klostermann_2013}, where $\boldsymbol{u}_r$ is the compression velocity between the phases. This term, being physically meaningless, is active only at the interface and sharpens the interface. Multi-dimensional Universal Limiter for Explicit Solution (MULES) \citep{Rusche_2002, Deshpande_2012} algorithm ensures the boundedness of $\alpha_w$ by limiting fluxes. MULES is an explicit algorithm based on the flux-corrected transport method \citep{Zalesak_1979}, and a detailed explanation is given in \cite{Rusche_2002} and \cite{Berberovic_2009}. The shear stress tensor ($\mathsfbi{T}$) in the momentum equation is given by  \\ 
 \begin{equation}
    \mathsfbi{T} = \mu ( \nabla \boldsymbol{u} + (\nabla \boldsymbol{u})^T ) = 2 \mu \mathsfbi{S}  
\end{equation}
where, $\mathsfbi{S}$ is the strain rate tensor.\\
The mixture density ($\rho$) and dynamic viscosity ($\mu$) are given as
\begin{equation}
   \rho = \alpha_w \rho_{w} + \alpha_a \rho_{a} 
\end{equation}
\begin{equation}
   \mu = \alpha_w \mu_{w} + \alpha_a \mu_{a} 
\end{equation}
where,  the subscripts $_a$ and $_w$ refer to air and water. The densities and viscosities are taken as $\rho_{w} = 10^3 kg/m^3$, $\rho_{a} = 1.2 kg/m^3$, $\mu_{w} = 10^{-3} Ns/m^2$ and $\mu_{a} = 1.776\times10^{-5} Ns/m^2$. The surface tension force between the fluids is modeled using the Continuum Surface Force (CSF) model \citep{Brackbill_1992}. The CSF model assumes the surface tension force as a continuous, three-dimensional force across the interface rather than a surface value. This assumption simplifies the application of the force with interfacial tracking methods. The surface tension force $\boldsymbol{f_{\sigma}}$ acting at the interface is given by \\
\begin{equation}
   \boldsymbol{f_{\sigma}} = \gamma \kappa \bnabla \alpha_w, 
\end{equation}
where $\gamma$ and $\kappa$ are the surface tension between air and water, and the radius of curvature of the interface. The surface tension at the interface is taken as $\gamma = 0.072 N/m$. The radius of curvature $\gamma$ is given by
\begin{equation}
   \kappa = \bnabla \bcdot \Bigl( \frac{\bnabla \alpha_w}{|\bnabla \alpha_w|}  \Bigr), 
\end{equation}
The volume fraction gradient term $\bnabla \alpha_w$ vanishes in the bulk fluid region, making the $\boldsymbol{f_{\sigma}}$ act only in the interfacial region.\\

\subsection {Numerical Schemes}

OpenFOAM \citep{Weller_1998, Jasak_1996, Rusche_2002} solves the Navier-Stokes equations using the finite volume method (FVM) on unstructured meshes with a collocated grid, where field values are stored at cell centers. In FVM, the governing equation is integrated over each cell volume to obtain a discretized set of equations. This set of equations is arranged as a system of algebraic equations and subsequently solved iteratively using matrix solver algorithms. The temporal derivative terms in the governing equations are discretized using the blended Crank-Nicolson method and the second-order implicit Euler method. The convective schemes in the governing equations are integrated over control volumes and converted into surface integrals using Gauss Theorem \\

\begin{equation}
    \int_V \bnabla \bcdot (\rho \boldsymbol{u} \Phi) dV= \sum_f \boldsymbol{S}_{f}\bcdot(\rho \boldsymbol{u}\Phi)_{f} = \sum_f  \boldsymbol{S}_f \bcdot(\rho \boldsymbol{u} )_{f} \Phi_f
    = \sum_f \rho_f F \Phi_f,
\end{equation}
where, $\Phi$ is the variable, $\boldsymbol{S}_f$ is the face area vector, with the magnitude of face, $F= \boldsymbol{S}_f\bcdot\boldsymbol{u}_f$ is the volumetric flux and in the direction perpendicular to the face and the subscript $f$ denotes the values at face centers. The discretization of convective terms requires field values interpolated to the face. The convective terms in the momentum equation use a linear upwind scheme, which is bounded and partially second-order. The flux for $\bnabla \bcdot(\boldsymbol{u}\alpha_w)$ term is interpolated using the van Leer scheme \citep{van_leer_1974}. The compression velocity ($\boldsymbol{u}_r$) cannot be directly obtained as we solve for the fluid velocity ($\boldsymbol{u}$). The compression velocity ($\boldsymbol{u}_r = min [c_\alpha |\boldsymbol{u}|, max(|\boldsymbol{u}|)]\boldsymbol{n}$) is calculated as \\
\begin{equation}
    (\boldsymbol{u}_r)_f = \boldsymbol{n}_f min \left(c_{\alpha}\left| \frac{F}{|\boldsymbol{S}_f|}\right|, \left| \frac{F}{|\boldsymbol{S}_f|}\right|_{max} \right),
\end{equation}
where, $\boldsymbol{n}_f$ is the normal vector to the interface \citep{Berberovic_2009}. $\boldsymbol{u}_r$ is bounded by the maximum velocity in the domain, and it acts normal to the interface. $c_{\alpha}$ is the compressive coefficient, which controls the effect of compression and hence the smearing of the interface, and it is set to unity for a sharp interface to avoid parasitic currents (\cite{Klostermann_2013}; and \cite{Krishna_2025}). The normal vector($\boldsymbol{n}_f$) is computed as \\
\begin{equation}
    \boldsymbol{n}_f = \frac{(\bnabla \alpha_w)_f}{|(\bnabla \alpha_w)_f+\gamma_n|}\quad\text{with} \quad
    \gamma_n = \frac{10^{-8}}{\sqrt[3]{\frac{\sum_{N}V_P}{N}}},
\end{equation}
where $\gamma_n$ is the stabilization factor for unstructured meshes, $V_P$ is the cell volume and $N$ is the total number of cells. The flux for the compressive term is calculated using the linear interpolation scheme.\\ 

The Laplacian terms in the governing equation are discretized as 
\begin{equation}
    \int_V \bnabla \bcdot (\rho \Gamma_{\Phi} \bnabla\Phi)dV = \sum_{f} \boldsymbol{S}_f \bcdot (\rho \Gamma_{\Phi} \bnabla \Phi)_{f} = \sum_f (\rho \Gamma_{\Phi})_{f} \boldsymbol{S}_{f} \bcdot (\bnabla \Phi)_{f}
\end{equation}
where $\Gamma_{\Phi}$ is the diffusion coefficient of the field $\Phi$. The calculation of the Laplacian terms requires gradients at the face centers. These are linearly interpolated from the adjacent cell centers. The gradients are calculated using the Green Gauss Cell-based method \citep{Jasak_1996}. The gradient at a cell center P, calculated from the  face values given by \\ 
\begin{equation}
   (\nabla \Phi)_{P}= \frac{1}{V_{P}} \sum_{f} [\boldsymbol{S}_{f}\Phi_{f}]
\end{equation}
where $V_{P}$,$\boldsymbol{S}_f$, and $\Phi_{f}$ are the volume of the cell, area vector of the face. The face center values of $\Phi$ are interpolated from the neighboring cell centers. All interpolations are done using linear interpolation. \\

The time resolution of the simulation is important to resolve the interface in a multiphase problem. The cases were run with the timestep adjusted according to the Courant–Friedrichs–Lewy (CFL) criteria \citep{Rusche_2002}, and the Courant number is calculated as \\
\begin{equation}
   Co = \frac{|\boldsymbol{u}_f\bcdot\boldsymbol{S}_f|}{\boldsymbol{d}\bcdot\boldsymbol{S}_f}\Delta t
\end{equation}
 where $\boldsymbol{d}$ is the vector joining the cell center and the neighboring cell center, and $\Delta t$ is the time step size. The $\Delta t$ for all the simulations is of the order of $10^{-7}$. The time steps were lower than Brackbill's criteria \citep{Brackbill_1992} to resolve the capillary wave, $\Delta t < \sqrt[3]{h^3(\rho_1+\rho_2)/(4\pi\sigma)}$. \\
 
\subsection{Computational Setup}\label{Comp_Setup}

\begin{figure}
\begin{center}
    \begin{subfigure}[c]{\linewidth}
        (a) \includegraphics[width=1\linewidth]{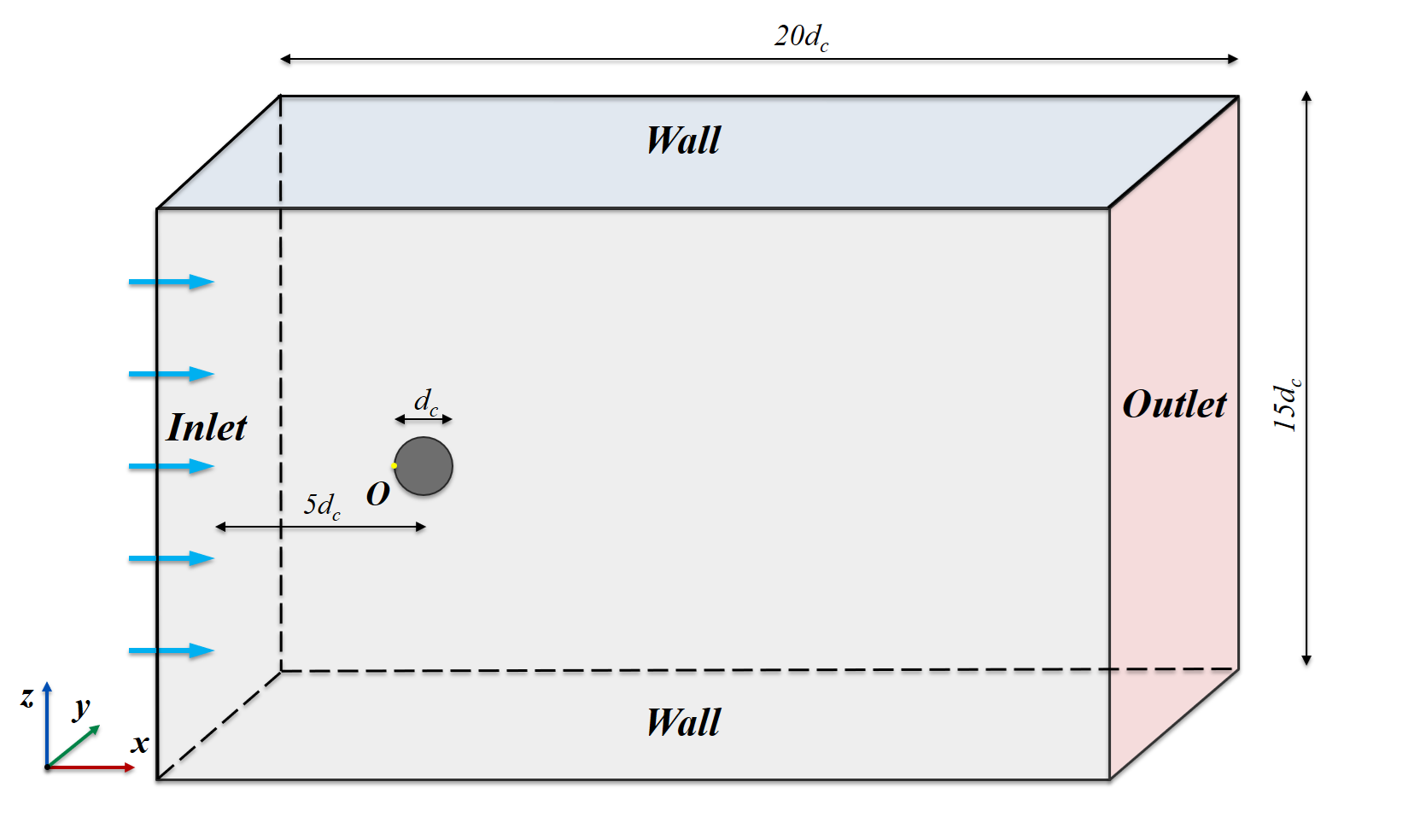}\\
    \end{subfigure}
\end{center}    
    \begin{subfigure}[c]{0.32\textwidth}
        (b)\includegraphics[width=\textwidth,trim=0 0.2cm 0 0, clip]{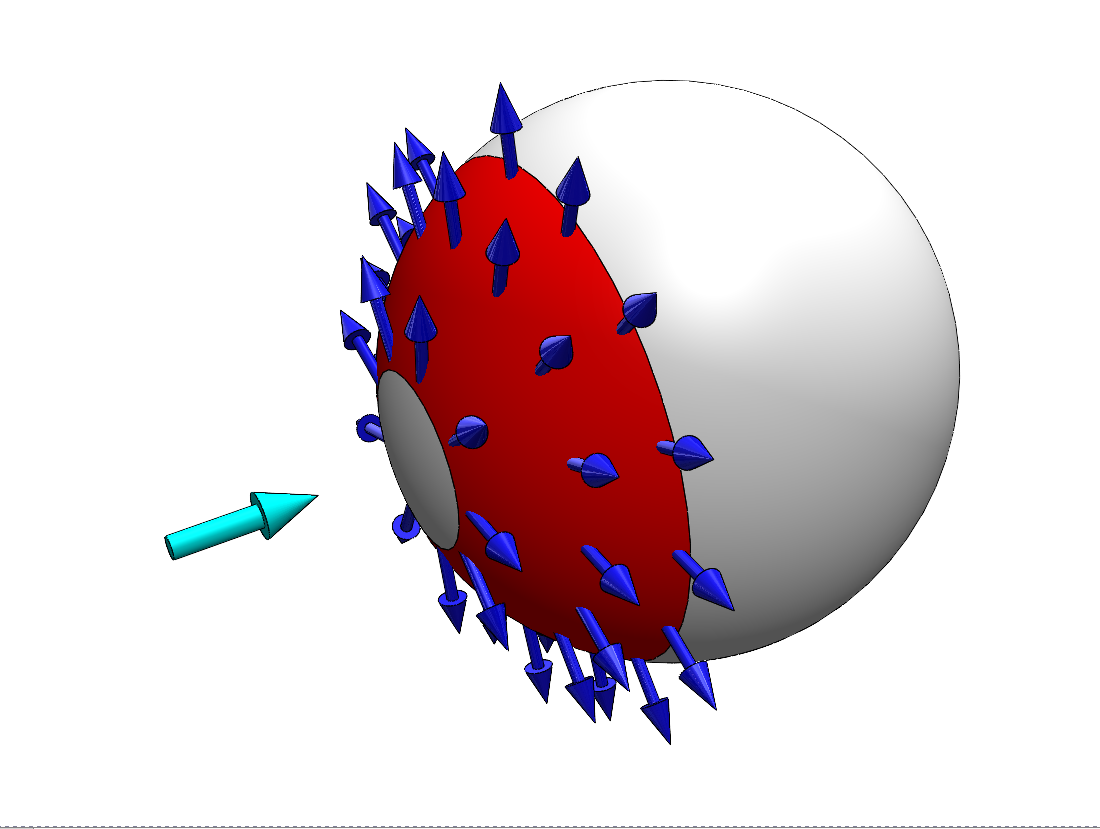}%
    \end{subfigure}  
    \begin{subfigure}[c]{0.32\textwidth}
        (c)\includegraphics[width=\textwidth,trim=0 0.2cm 0 0, clip]{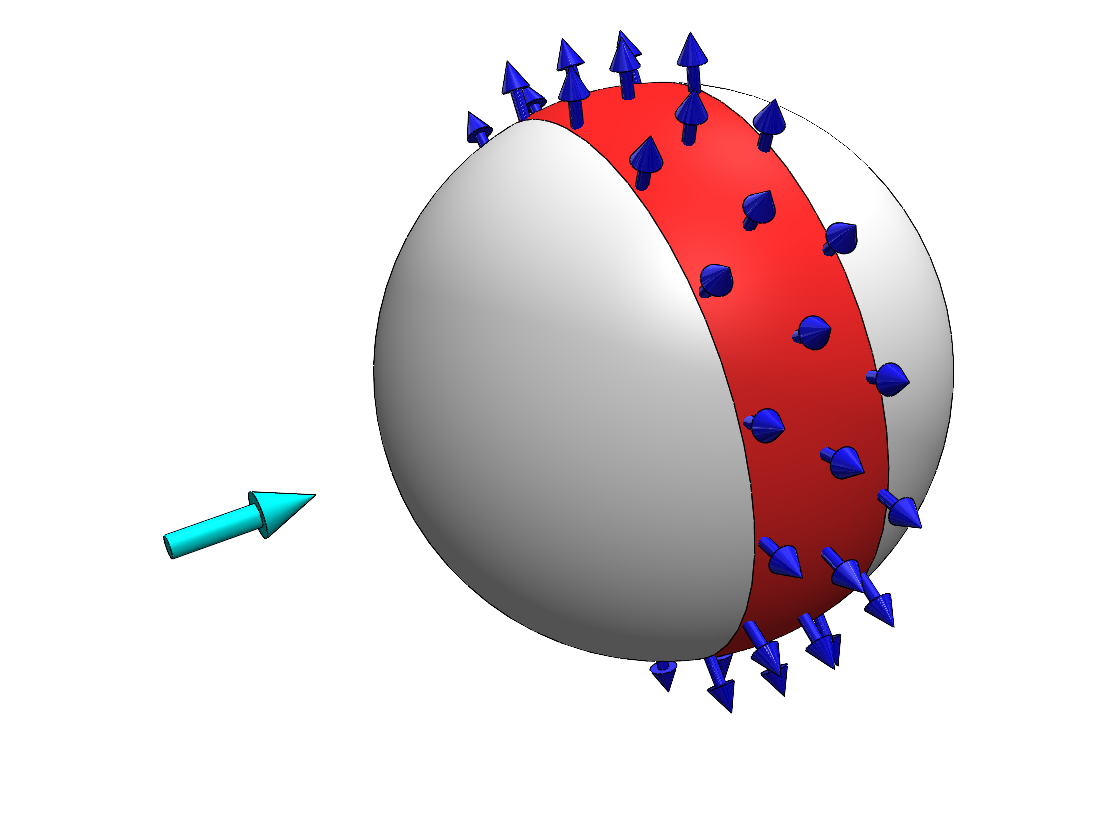}
    \end{subfigure}
    \begin{subfigure}[c]{0.32\textwidth}
        (d)\includegraphics[width=\textwidth,trim=0 0.2cm 0 0, clip]{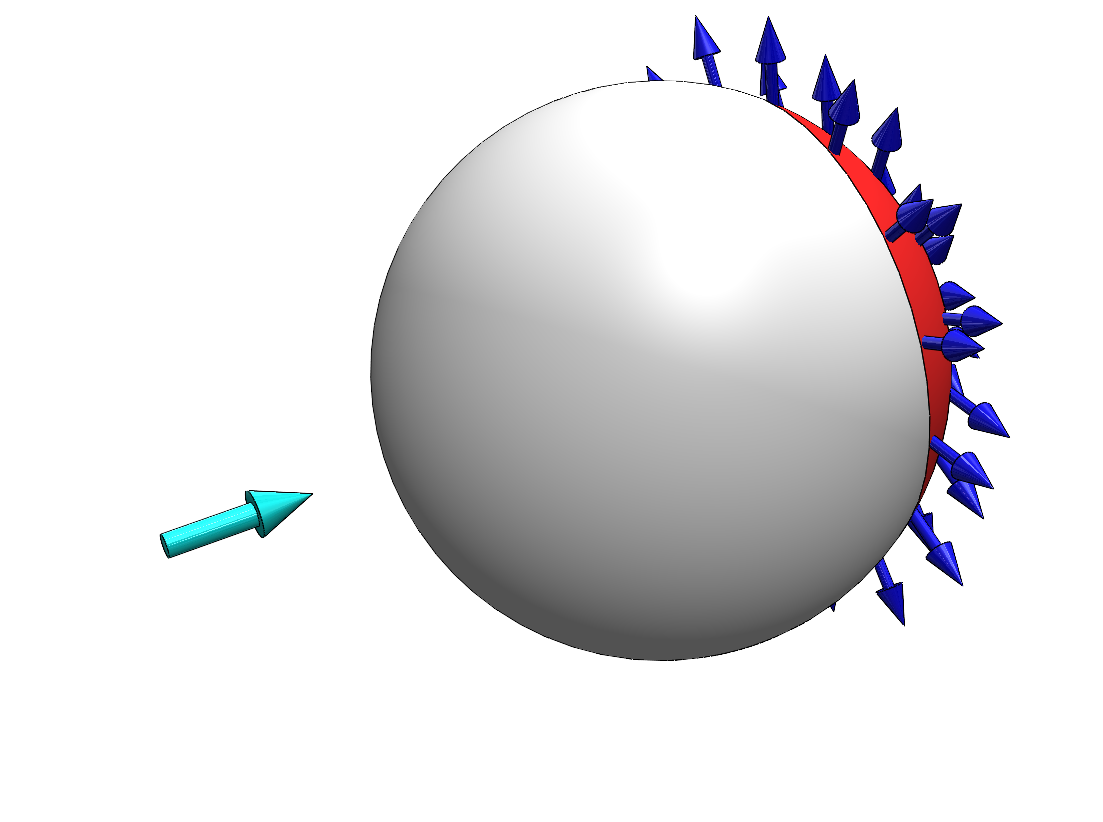}
    \end{subfigure}
  \captionsetup{justification=justified, singlelinecheck=false,width=\textwidth}
  \caption{{\textit{(a)}} 3D Schematic of the flow domain. The direction of flow is shown by the light blue arrows. The origin (O) of the domain is at the stagnation point of the sphere. (b)-(d) 3D Schematic of air injection from the sphere surface. The direction of fluid flow is shown by the cyan arrow. The red surface is the injection patch, the gray surface is the solid sphere surface, and the dark blue arrow shows the direction of air injection. {\textit(b)} Front-Injection, {\textit(c)} mid-injection and {\textit(d)} back-injection.}
\label{fig:3d_schem}
\end{figure}

The computational domain is a cuboid with dimensions $25d_c \times 15d_c \times 15d_c$, $d_c$ is the diameter of the sphere, where $d_c = 0.01 m$, as shown in figure \ref{fig:3d_schem} (a). The sphere is positioned in such a way that the forward stagnation point of the sphere coincides with the origin (O in figure \ref{fig:3d_schem} (a)). The flow direction is along the positive $x$ direction as shown by the light blue arrows in the figure \ref{fig:3d_schem} (a). The domain walls in the lateral directions are $7.5d_c$ away from the sphere. This distance is sufficiently large to neglect blockage effects. The inlet and outlet are $5d_c$ upstream and $15d_c$ downstream of the sphere, respectively. The outlet is sufficiently far from the sphere so that it doesn't affect the cavity formation. \\

The flow domain is meshed using ICEM CFD (Release 18.1). An O-grid blocking is done with the sphere at its center. The O-grid ensures a body-conforming mesh near the sphere and allows grid refinement in its vicinity, with gradual coarsening away from it. The grid spacing in the O-grid block increases according to the geometric progression of cell length. This sizing method requires specifying the first grid spacing, and the remaining nodes are spaced with a constant specified growth rate. The distance $s_{i}$ of the $i^{th}$ grid from the starting side is given as
\begin{equation}
   s_{i} = \frac{R-1}{R^{N-1}-1} \sum_{j=2}^{i} R^{i-2},
\end{equation}
where, $R$ is the total number of nodes and $R$ is the growth rate . The grid spacing near the sphere is $ 0.0005d_c$. Further downstream of the wake, for  $x/d_c > 4.2d_c, \Delta x/d_c \sim 0.03$ and $ \Delta z/d_c = \Delta y/d_c \sim 0.03$. Injection of air from the sphere’s surface causes large pressure and velocity gradients around the sphere that need to be resolved; hence, the thickness of the first layer around the sphere is kept small. 
\cite{Liu_2023} demonstrated mesh resolution studies for 3.69 (M1), 29.5 (M2), and 236 (M3) million cells and concluded that the statistics and cavity dynamics are similar for M2 and M3. 
We have performed the grid-independent study, and it is discussed in Appendix \ref{appA}. We choose $\sim 179$M hexahedral cells for our domain, consistent with the mesh-resolution study of \cite{Liu_2023}, and therefore, it is sufficient to capture all relevant scales of motion. \\

To generate a cavity, air is injected from the sphere's surface at different azimuthal positions and injection velocities. The injection direction is axially outward to the sphere and into the domain. Figure \ref{fig:3d_schem} (b), (c), and (d) shows a 3D schematic of air injection. Three positions of air injection are: front-injection ($18^{\circ} \leq \theta \leq 63^{\circ}$), mid-injection ($75^{\circ} \leq \theta \leq 104^{\circ}$) and back-injection ($120^{\circ} \leq \theta \leq 180^{\circ}$). Different injection positions are chosen based on the bluff body flow dynamics around the sphere. The pressure around the sphere is maximum at the stagnation point. Downstream of the stagnation point, the flow accelerates, causing a pressure drop. Injecting air from the stagnation point would be ideal for obtaining a cavity that engulfs the entire sphere. But upon injection from the stagnation point, the air would be completely swept downstream by the higher local ambient pressure exerted by the bulk fluid. The air swept will be replenished and accumulated due to its constant injection into the domain. These accumulation and sweeping processes will occur cyclically, leading to high-pressure oscillations around the sphere, which, in turn, affect the drag coefficient. These drag fluctuations are undesirable. \\

Additionally, in one of the cases of \cite{Chung_2018}, when air was injected from the nose of the bluff body without a cavitator, only a bubbly cavity was obtained despite increasing the ventilation rate. Hence, the front-injection, as shown in figure \ref{fig:3d_schem} (b), is shifted slightly downstream from the vicinity of the stagnation point. Further downstream to the stagnation point, the pressure reaches a minimum, causing flow deceleration and separation. This position is chosen for the mid-injection case (figure \ref{fig:3d_schem} (c)). Downstream of the flow separation, the pressure becomes nearly constant owing to the wake influence, and this location is chosen for back-injection (figure \ref{fig:3d_schem} (d)). The direction of air injection is perpendicular to the sphere’s axis, passing through its forward stagnation point and its center (as shown in figure \ref{fig:3d_schem} (b)-(d)). The magnitude of this axially outward velocity is calculated based on the respective $C_q$. The area of injection in all the cases is $A_{in} = A_{s}/4$, where $A_{s}$ is the surface area of the sphere. Table \ref{tab:cs} summarizes different cases. The difference in the local ambient pressure when air is injected at different locations enables us to study the cavity formation systematically. \\

Dirichlet boundary condition is implemented for velocity at the inlet boundary, and its value is set to $U_{\infty} = 1$ $m/s$. This value is chosen so that the diameter-based Reynolds  $\Rey = 10^4$. A homogeneous Neumann boundary condition is applied to the inlet pressure. Slip boundary conditions are implemented at the walls. At the outlet, the Dirichlet boundary condition is used for pressure. At the sphere surface, the no-slip condition is used for velocity, and zero gradient is used for pressure. For the injection cases, the injection velocity, on the injection patch, is set in the axial outward direction, and its magnitude is computed based on the corresponding $C_q$ values. A mixed boundary condition is used for the water volume fraction on the sphere surface. It is set to a homogeneous Dirichlet condition in the ventilation patch to ensure air is injected into the domain, and it is set to a homogeneous Neumann boundary condition on the rest of the sphere.\\ 

\begin{table}
  \begin{center}
\def~{\hphantom{0}}
  \begin{tabular}{lcccc}
       Cases & Injection Angular Span &   Injection Velocity ($V_{in}/U_{\infty}$)  & Ventilation Coefficient & \\[3pt]
       & ($\theta$ in $^\circ$)  & ($V_{in}/U_{\infty}$) & ($C_{q}$) & \\

      MD025 & 75 - 104& 0.26 & 0.2  \\
      MD050 & 75 - 104 & 0.52 & 0.4  \\
      FR025 & 18 - 63  & 0.37 & 0.2  \\
      BK025 & 120 - 180 & 0.41 & 0.2  \\
      SP & - & 0 & 0& \\
  \end{tabular}
  \caption{Details of the simulation cases}
  \label{tab:cs}
  \end{center}
\end{table}

\section{Validation of the numerical approach}\label{notstyle}\label{sec:val}

\subsection{Flow Past a Sphere at $Re = 10,000$}
\begin{figure}
       (a) \includegraphics[width=0.5\linewidth]{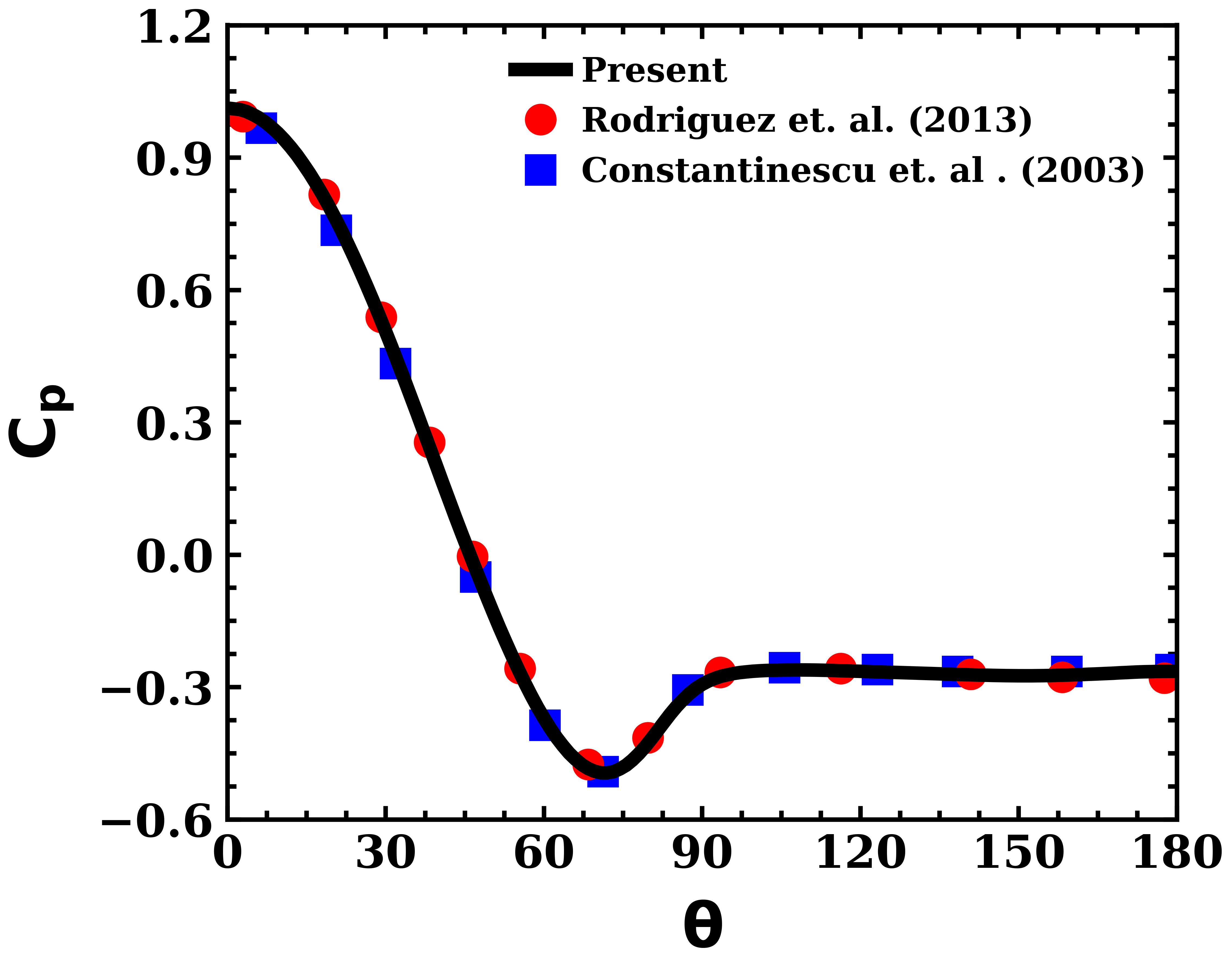}
       (b)\includegraphics[width=0.5\linewidth]{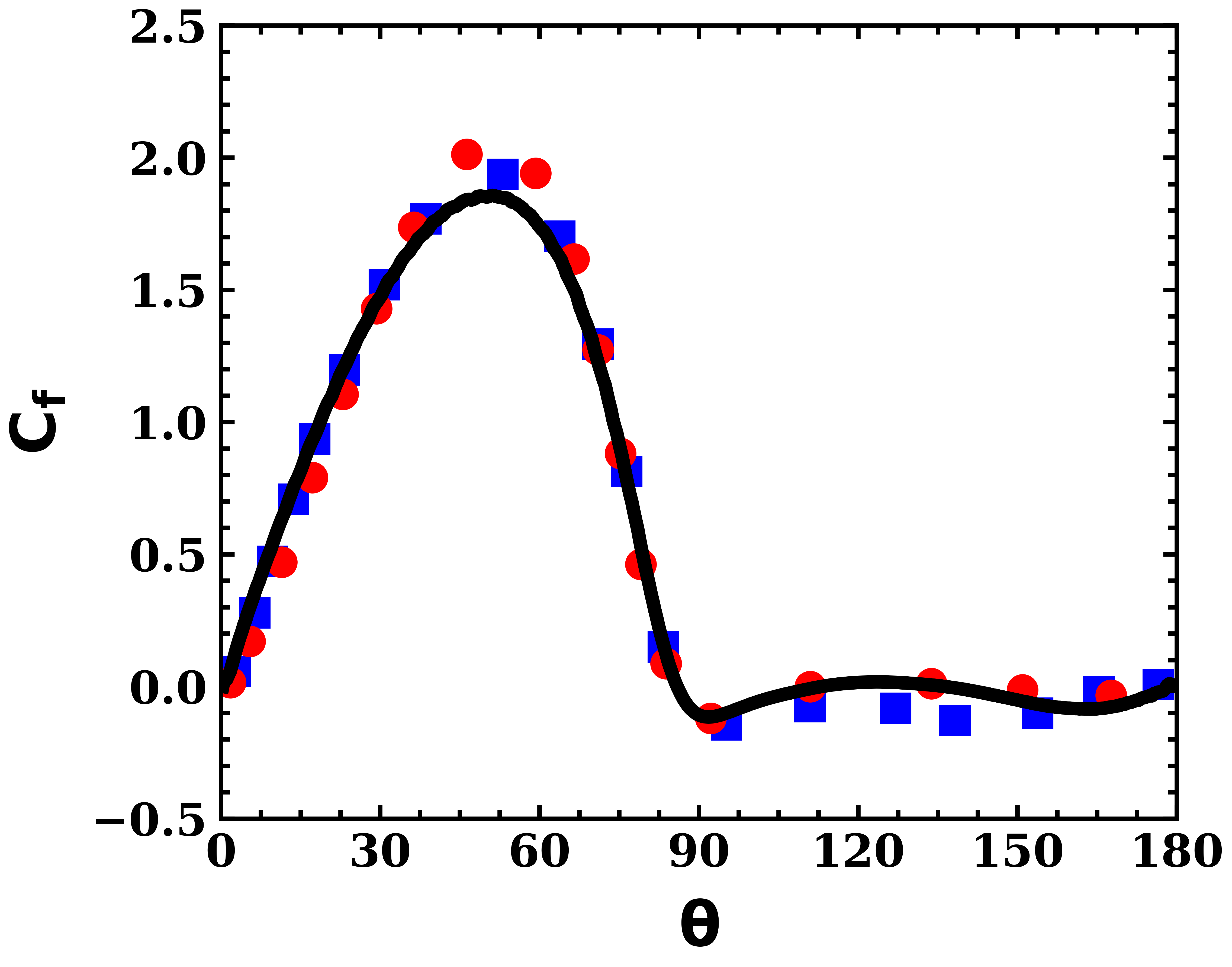}
       (c)\includegraphics[width=0.5\linewidth]{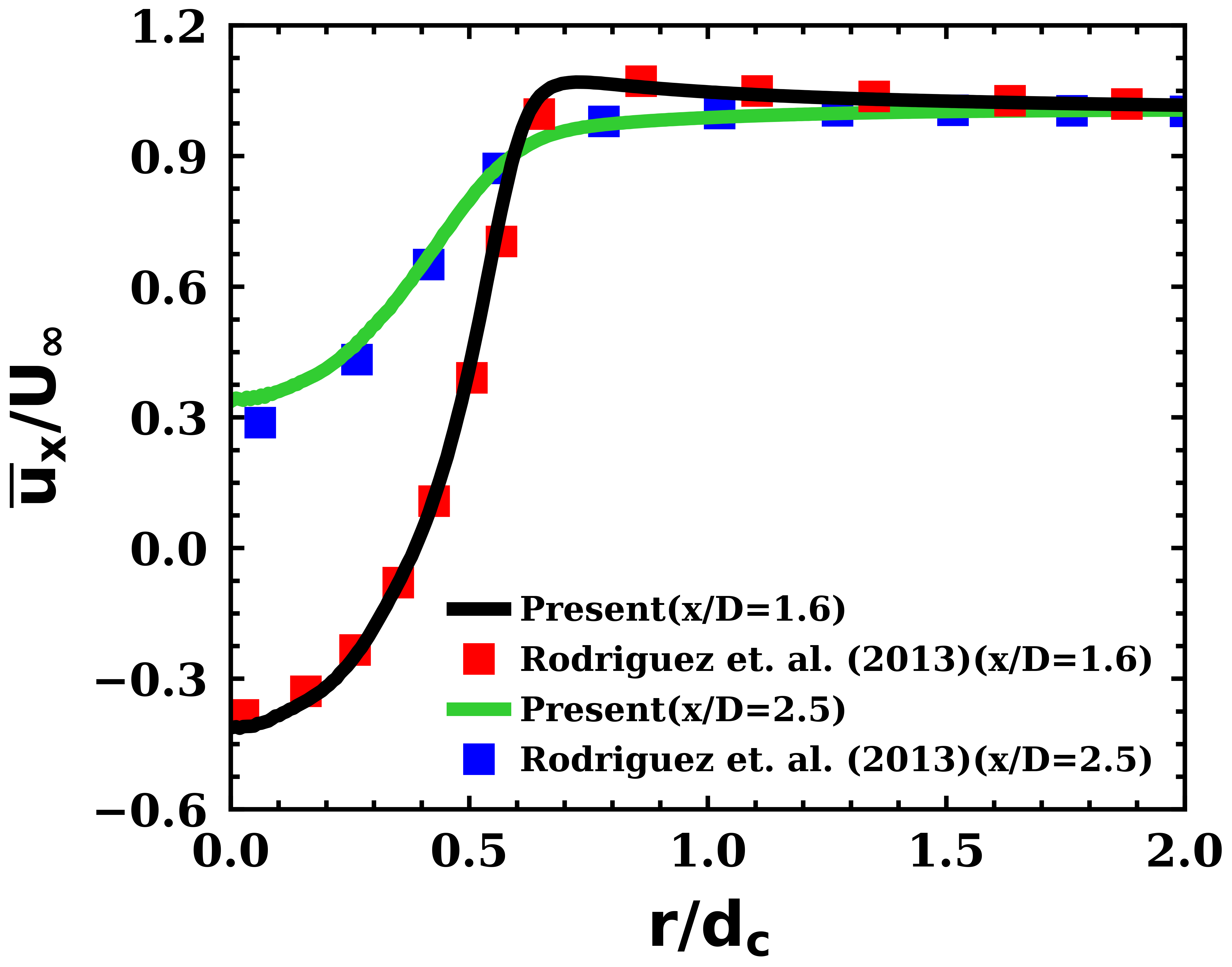}
       (d)\includegraphics[width=0.5\linewidth]{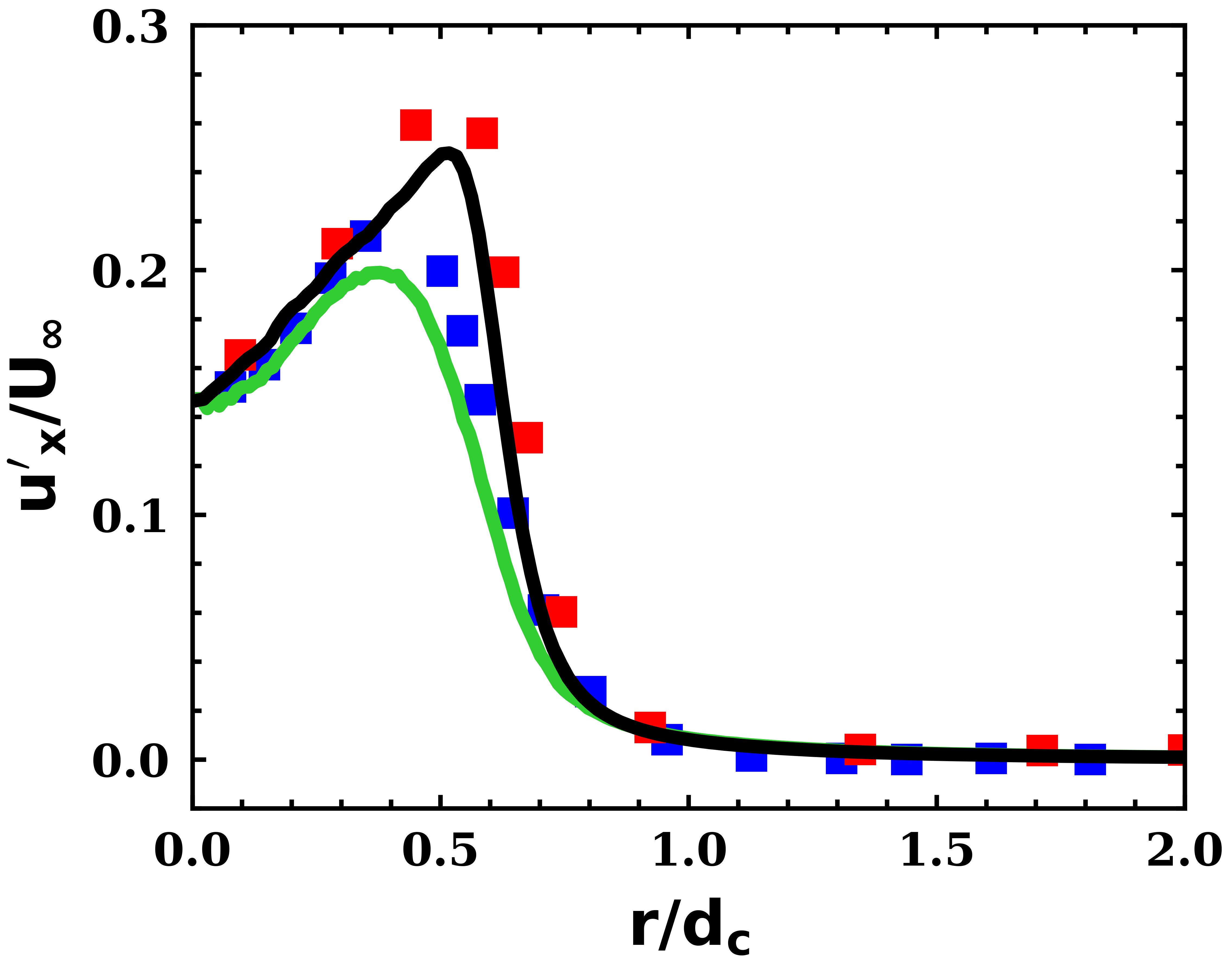}
    \captionsetup{justification=justified, singlelinecheck=false,width=\linewidth}
    \caption{Validation of single phase flow past a sphere. Azimuthal variation of (a) Pressure Coefficient $ (Cp = (p-p_{\infty})/(0.5 \rho_{\infty} U_{\infty}^2))$ and (b) Friction Coefficient $ (C_f= \tau_w/(\rho_{\infty} U_{\infty}^2\Rey^{-0.5}))$. Variation of (c) mean streamwise velocity and (d) root mean square streamwise velocity fluctuations along the radial coordinate ($r=\sqrt{y^2+z^2}$).}
    \label{fig:SP_Val_1}
\end{figure}

We validate the numerical solver by simulating the flow past a sphere at $Re = 10,000$, and compare our results with those of \cite{Constantinescu_2003} (LES) and \cite{Rodriguez_2013} (DNS). Figure \ref{fig:SP_Val_1} (a,b) shows the variation of time-averaged pressure coefficient ($C_p$) and friction coefficient ($C_f$) with azimuthal angle ($\theta$), with $\theta = 0$ corresponding to the leading stagnation point. We achieve an excellent match with the DNS results, implying that the surface gradients are accurately resolved. The time-averaged mean velocity and rms velocity fluctuations at the wake are plotted in figure \ref{fig:SP_Val_1} (c,d). The mean and rms velocity profiles near the wake also match well with the DNS of \cite{Rodriguez_2013}. \\
 
\subsection{Single Bubble rise Benchmark}
We simulate a rising bubble case from \cite{Gamet_2020} to validate the Volume of Fluid model and interface tracking technique. The case is set up in a cuboidal domain of size $1 m \times 2 m \times 1 m$. The domain contains a denser fluid (liquid) of density $1000 kg/m^{3}$ and dynamic viscosity $10 kg/m/s$.  A spherical bubble of diameter $0.5m$ is initialized at the center of the horizontal plane, which is at a distance of $0.5m$ from the bottom wall. The bubble is made up of a lighter fluid (gas) of density $1 kg/m^{3}$ and dynamic viscosity $0.1 kg/m/s$. The surface tension between the fluids is $1.96 N/m$. The walls in the vertical direction are no-slip boundaries, and the walls in the horizontal direction are slip boundaries. The simulations are performed with a grid resolution of $\Delta x = 1/160m$ and time step size of $\Delta t = 2.5 \times 10^{-4} s$ \citep{Gamet_2020}.\\

\begin{figure}
(a)\includegraphics[width=0.5\linewidth]{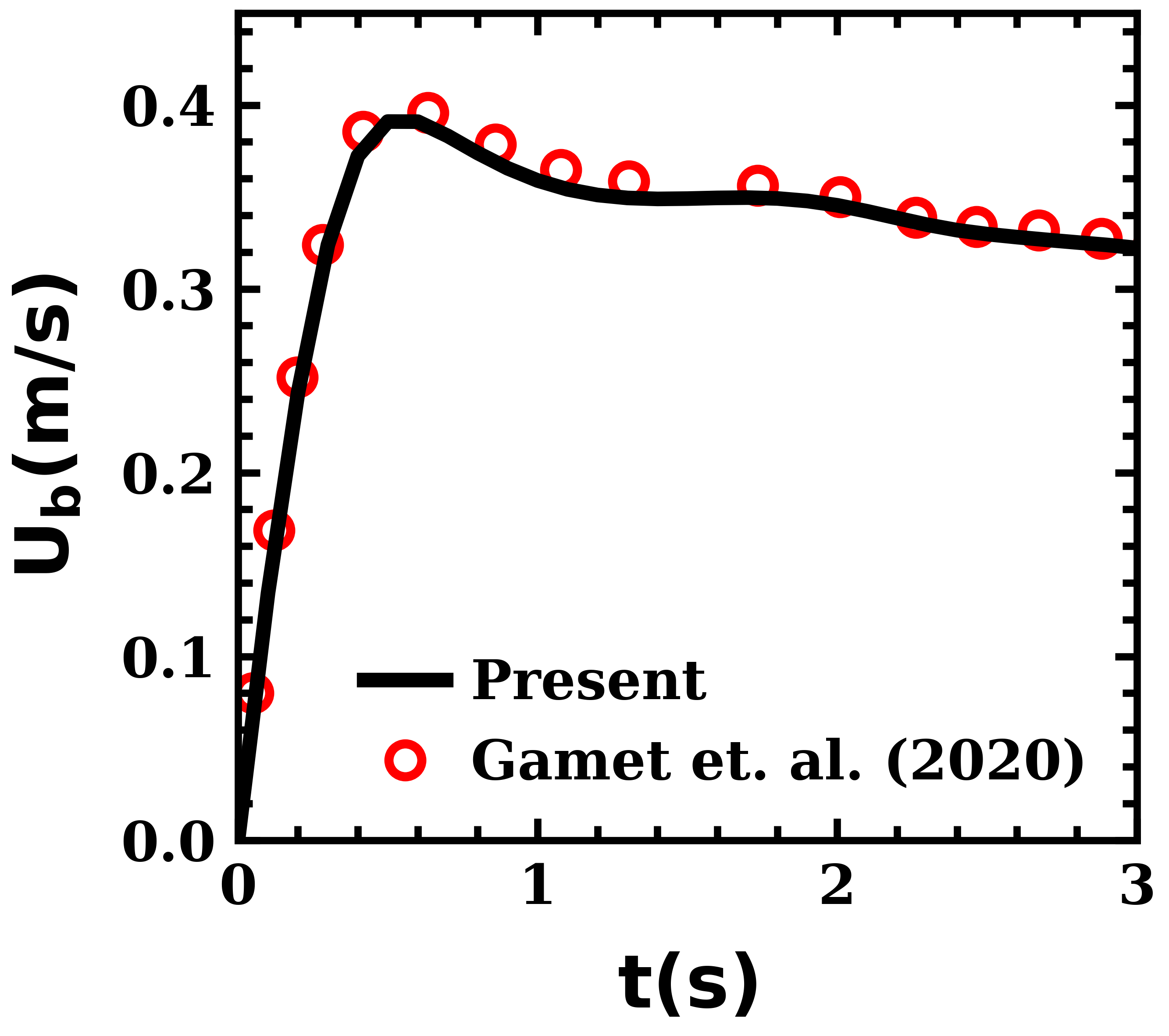}
(b)\includegraphics[width=0.5\linewidth]{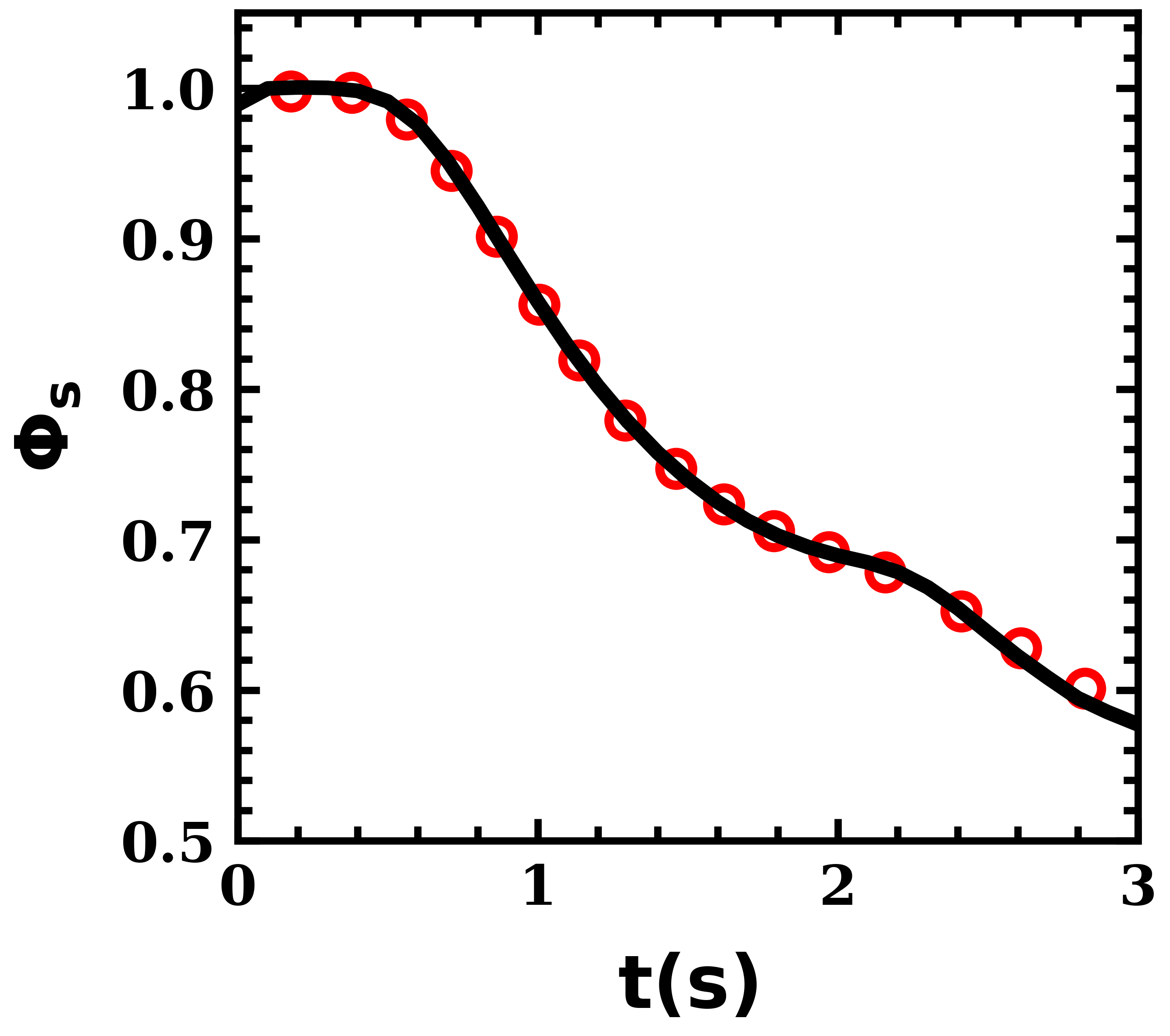}
\begin{center}
(c)\includegraphics[width=0.5\linewidth]{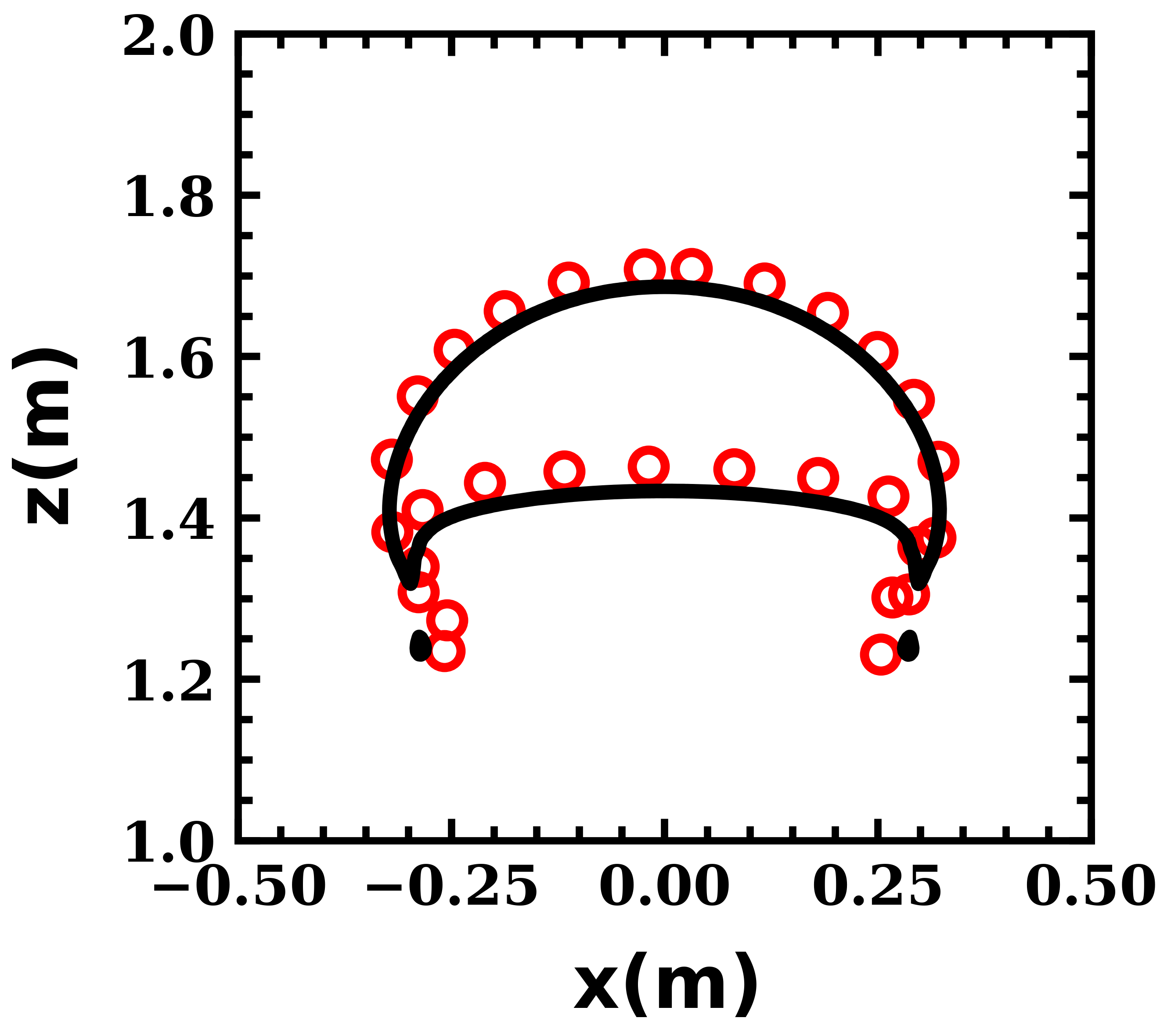}
\end{center}
\captionsetup{justification=justified, singlelinecheck=false,width=\textwidth}
\caption{Validation of bubble rise benchmark. Time evolution of (a) bubble rise velocity, (b) sphericity ($\Phi_s$), and (c) volume fraction isocontour ($\alpha_a = 0.5$) at $t = 3s$.}
\label{fig:MP_Val_2}
\end{figure}

Figure \ref{fig:MP_Val_2} (a) shows the time evolution of bubble rise velocity. Bubble rise velocity is calculated as $U_b = \frac{1}{V_b} \int_{\Omega} \alpha_a u_z dV$ where $V_b$ is the volume of bubble ($V_b = \int_{\Omega} \alpha_a dV$), $\Omega$ is the 3D Domain and $\boldsymbol{u}$ is the instantaneous velocity. Figure \ref{fig:MP_Val_2} (b) shows the time evolution of bubble sphericity ($\Phi_s$). Sphericity is calculated as $\Phi_s = \frac{(3V_b/(4\pi))^{2/3}}{A_b/(4\pi)}$. $A_b$ is the area of isosurface of  $\alpha_w = 0.5$. Initially, sphericity is unity for a spherical bubble, and decreases as the bubble deforms over time. Figure \ref{fig:MP_Val_2} (c) shows the bubble shape at time $t = 3s$. Figure \ref{fig:MP_Val_2} shows a good agreement with the literature for the bubble rise benchmark case. \\

\section{Results and Discussion}\label{sec:result}

We discuss the ability of the injected air to form a cavity by plotting the volume fraction distribution (figure \ref{fig:MP_Alp}) on the sphere and the location of cavity detachment in stable cavities (MD025, MD050, and BK025). Then we study the cavity leading-edge dynamics, which are affected by the injection location along the sphere's surface. We demonstrate the puffing phenomenon (figure \ref{fig:MP_Puffing}) in front and mid-injection cases, and divot formation (figure \ref{fig:MP_Divots}) in the back-injection cases. Next, we examine the surface statistics on the sphere post-injection, followed by a discussion on the surface pressure (figure \ref{fig:MP_Cp}), skin friction distribution (figure \ref{fig:MP_Cf}), and the location of flow separation around the sphere in different cases. We further report the dependence of the location of flow separation and the local pressure imbalance on cavity formation.  Subsequently, we analyze the effect of injection on the loads of the sphere (figure \ref{fig:Load_History}). Finally, we plot the wake axial pressure and velocity profiles to study the internal dynamics of cavities. \\

\subsection{Distribution of air on the sphere surface}
\begin{figure}
    \centering
    \begin{subfigure}[b]{0.75\textwidth}
        \includegraphics[width=\textwidth]{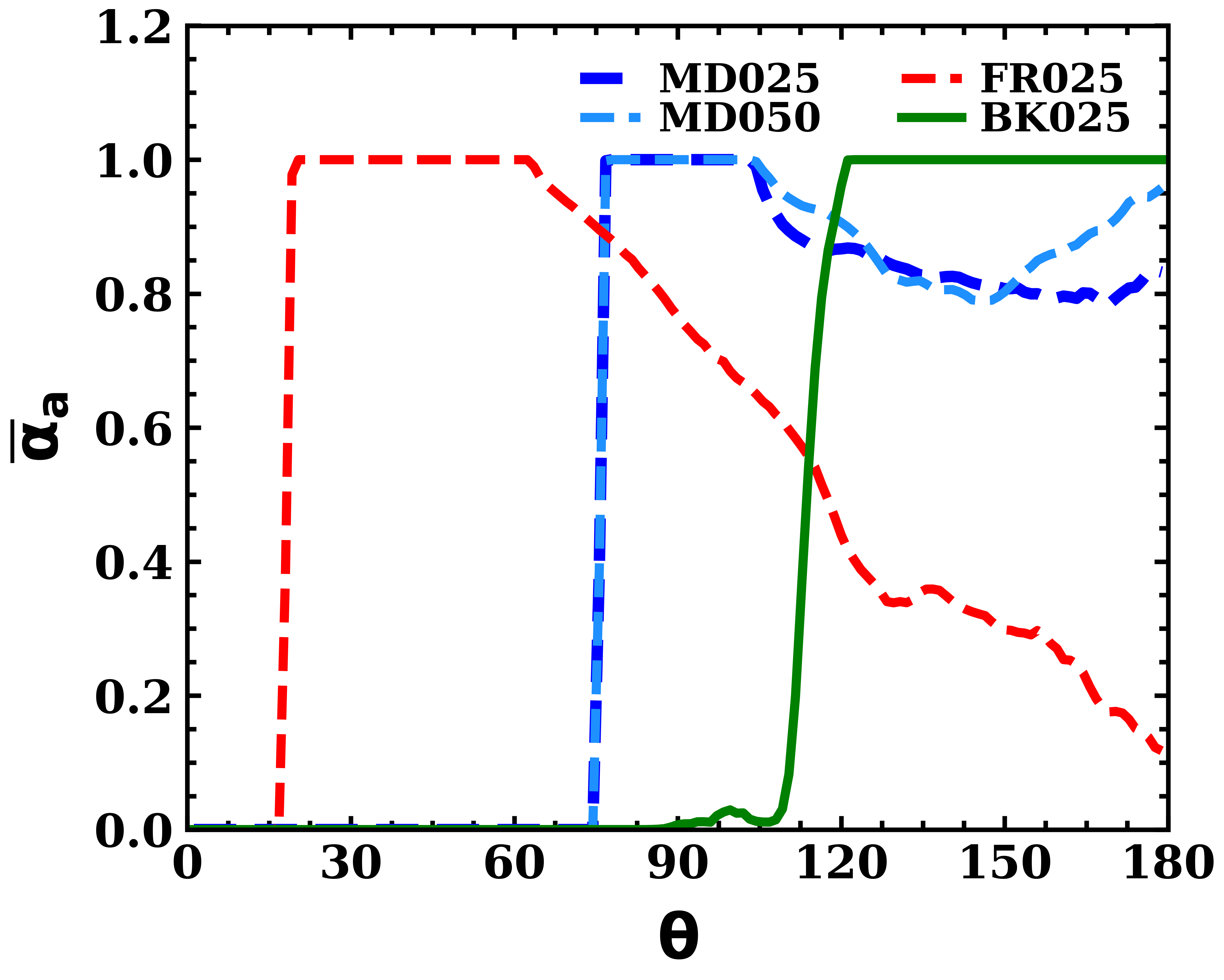}
    \end{subfigure}
    \captionsetup{justification=justified, singlelinecheck=false,width=\textwidth}
    \caption{Time and azimuthal averaged air volume fraction ($\overline{\alpha}_a$) distribution over the sphere surface for different injection cases}
    \label{fig:MP_Alp}
\end{figure}
Figure \ref{fig:MP_Alp} plots the time-averaged and azimuthal averaged air volume fraction ($\overline{\alpha}_a$) with respect to the azimuthal angle ($\theta$) on the sphere surface for various injection cases to illustrate the distribution of air. The air volume fraction is unity in the respective injection patches in all the cases. In mid-injection cases (MD025, MD050), beyond the injection patch, the air volume fraction remains well above $0.5$, indicating that the surface is majorly covered by air due to a stable cavity formation.  Despite a cavity formation, there is a slight drop in $\overline{\alpha}_a$ from unity inside the cavity. This is due to the impingement of water entrained by the re-entrant jet from cavity closure onto the sphere's base (discussed later). In FR025, $\overline{\alpha}_a$ decreases gradually from the trailing edge of the injection patch to the base of the sphere. This shows the surface is partially covered by air, and a stable cavity does not form in this case. In BK025, $\overline{\alpha}_a$ reaches unity before the leading edge of the injection patch. This shows a stable cavity formation with an upstream shift in the cavity detachment location. There is a small peak (at $\theta = 98^{\circ} $) adjacent to the cavity detachment location owing to the bubbles suspended upstream of the cavity.  All injection cases except FR025 form a stably attached cavity. \\

\begin{table}
  \begin{center}
\def~{\hphantom{0}}
  \begin{tabular}{cccccccccc}
       Case & $\varphi_s$ & $\varphi_c(in  ^\circ)$& $l_c/d_c$ & $(2 \times r_c)/d_c$& ${C_d}$ & Drag Reduction& ${C_l}$& ${C_s}$& $\sigma_c$   \\
             & $(in  ^\circ)$ &$(in  ^\circ)$ &   &   &  &  ($\%$) & ($\times 10^{-3}$) & ($\times 10^{-3}$) &  \\[3pt]
      MD025  & 101 &75   & 4.055 & 1.485 & 0.263 &  35 & 4.54   & -36.53    & 0.115 \\
       MD050  & 104 &75   & 7.625 & 1.730 & 0.302 &  25 & 5.74  & 3.06   & 0.060 \\
      FR025  & 21  & -    & -   &  -    & 0.632 & -57 & 9.83  & 17.18  & 0.280 \\
      BK025  & 87  & 114  & 5.75 & 1.362 & 0.217 &  46 & 5.51  & 2.93   & 0.066 \\
      SP     & 85  & -   & -   &  -    & 0.402 &  -  & -2.66 &1.12   &   -\\
  \end{tabular}
  \caption{Flow and cavity parameters. Separation angle ($\varphi_s$), cavity detachment angle ($\varphi_c$), cavity length ($l_c/d_c$), cavity diameter ($2*r_c/d_c$), maximum cavity diameter ($(2\times r_c)/d_c$), drag coefficient ($C_d$), drag reduction, lift ($C_l$) and side force ($C_s$) coefficient, and cavitation number ($\sigma_c$)} %
  \label{tab:fl_param} 
  \end{center}
\end{table}

Once a stable cavity forms, the position of cavity detachment from the stagnation point becomes important. The closer the cavity detachment point is to the stagnation point, the larger fraction of the sphere surface is covered by air, which in turn affects the skin friction coefficient. Table \ref{tab:fl_param} shows the angular position of cavity detachment ($\varphi_{c}$) for the injection cases. The position of cavity detachment relative to the stagnation point will be discussed in the upcoming stable cavity cases. We find that in the mid-injection cases (MD025, MD050), the cavity detachment position coincides with the leading edge of the injection patch, and this location does not vary with injection velocities. The cavity detachment occurs well ahead of the flow separation for the mid-injection case. The absence of the laminar flow separation upstream of the cavity results in immediate detachment of the cavity from the leading edge for the mid-injection cases. In BK025, the air is injected after the flow is separated, and the location of cavity detachment remains downstream of the laminar flow separation. This case follows the "viscous laminar separation" described by \cite{Arakeri_1975}. \\

\cite{Arakeri_1975} formulated the location of cavity detachment based on their experiments for cylinders and spheres by considering the viscous and surface tension effects. They gave the angular position of cavity detachment ($\gamma_C$) from the stagnation point as \\
\begin{multline}  
    \varphi_C = (\gamma_B)_{NC}-2.37\bigg[1+\frac{\sigma_c}{(C_s)_{NC}}\bigg]\times[(\gamma_B)_{NC}-(\gamma_{Cmin})_{NC}] \\
                    +\bigg(\frac{360}{\pi}\bigg)\bigg[\bigg(\frac{\theta_s}{d_c}\bigg)_{NC}\bigg]\bigg\{130-312\frac{\mu_{\infty} U_{\infty}}{T}[1-(C_s)_{NC}]\bigg\}.
\end{multline}
Here, $\gamma_B$,$C_s$, $\gamma_{Cmin}$, and $\theta_s$ are the location of boundary layer separation, pressure coefficient at the location of flow separation, location of minimum pressure coefficient, and $\theta_s$ is the momentum thickness of the boundary layer at flow separation, respectively. The subscript $()_{NC}$ denotes the non-cavitating condition. The cavity in BK025, detaches at an angular position of $114^\circ$, consistent with the formulation and with the experiments of \cite{Arakeri_1975} with an error of $\sim 7\%$. The cavity detachment location is in good agreement with the case where air was injected from the object's wake \cite{Arakeri_1975} and the natural cavitation experiment around the sphere by \cite{Brennen_1970b} ($\varphi_s=115^{\circ}$). This shows that for gas injection beyond the flow separation, the cavity detachment location is independent of the air injection location patch and solely dependent on the flow parameters. \cite{Arakeri_1975}'s formulation is only applicable to cavities undergoing a viscous laminar separation, such as BK025. Hence, their formula does not apply to the stable cavities in mid-injection cases.\\  

\subsection{Cavity Leading Edge Dynamics} \label{sec:Cav_Lead_Dyn}

\begin{figure}
    
        (a) \includegraphics[width=0.49\textwidth]{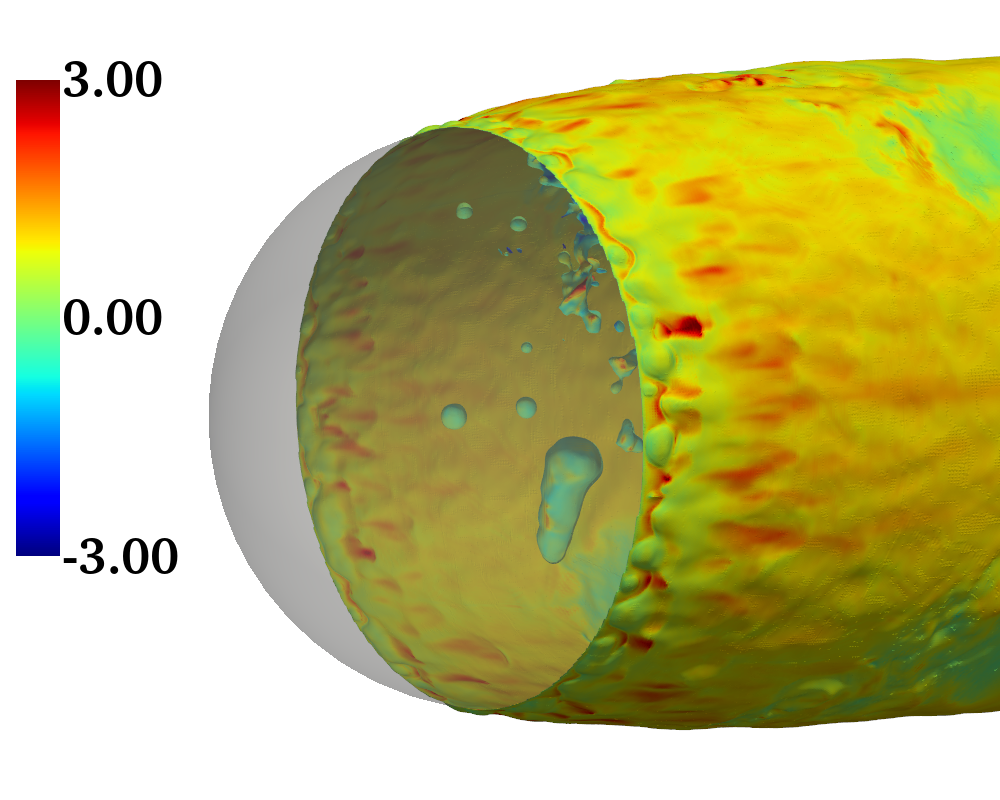}
        (b) \includegraphics[width=0.49\textwidth]{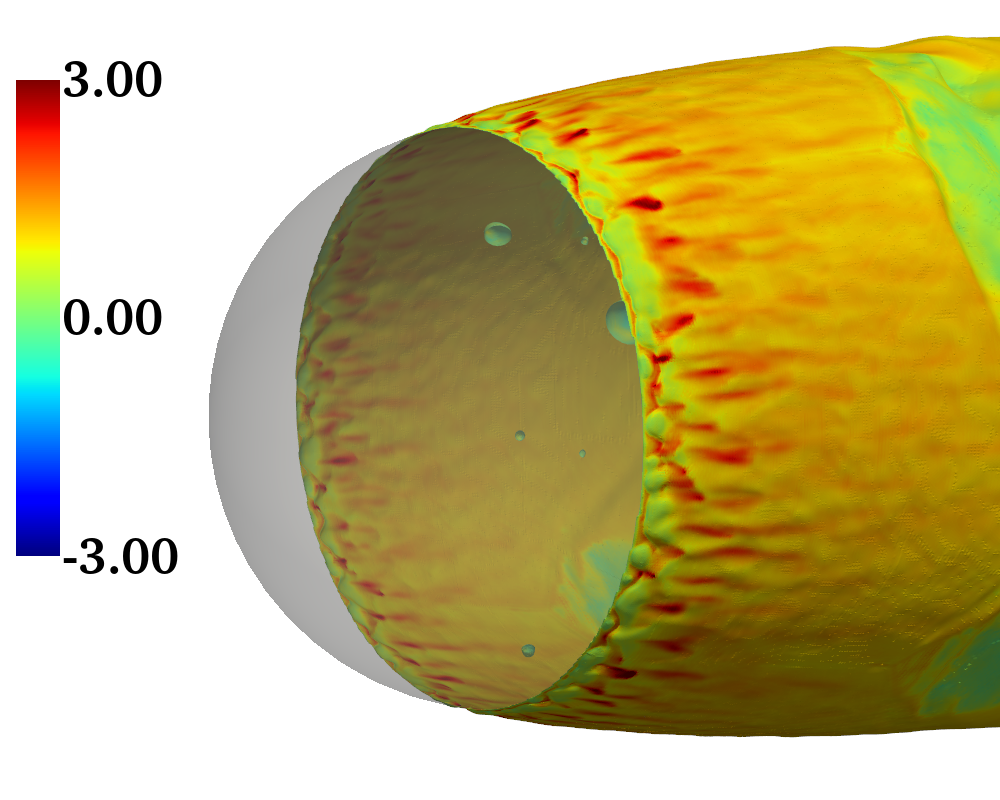}
        (c) \includegraphics[width=0.49\textwidth]{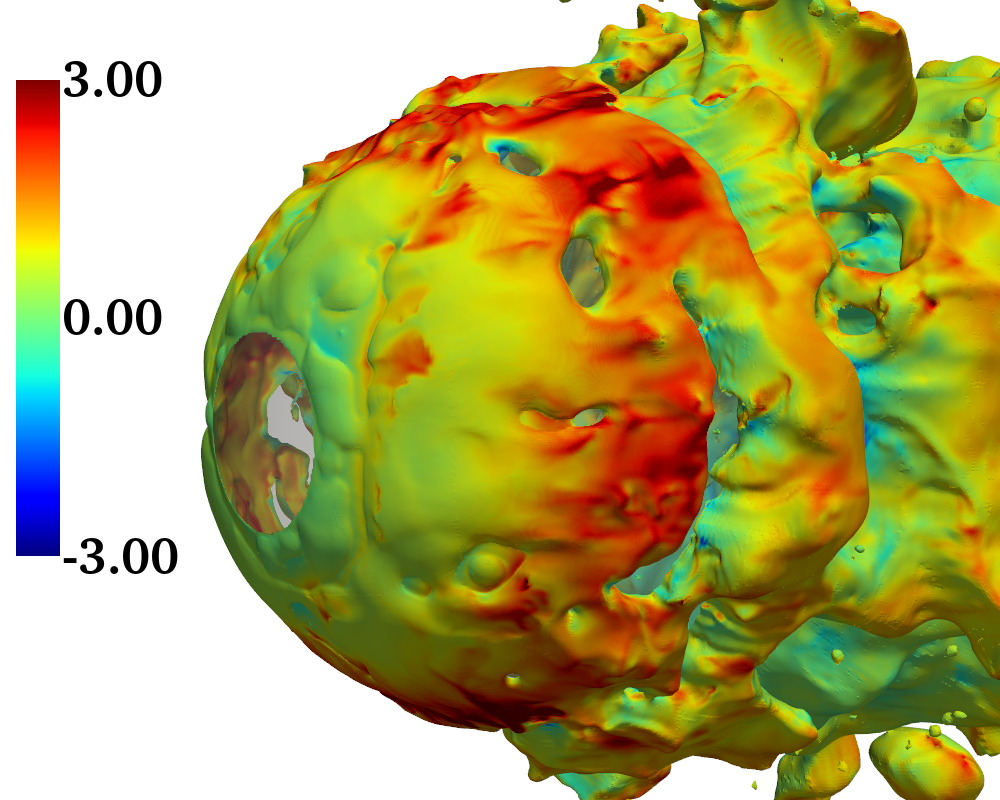}
        (d) \includegraphics[width=0.49\textwidth]{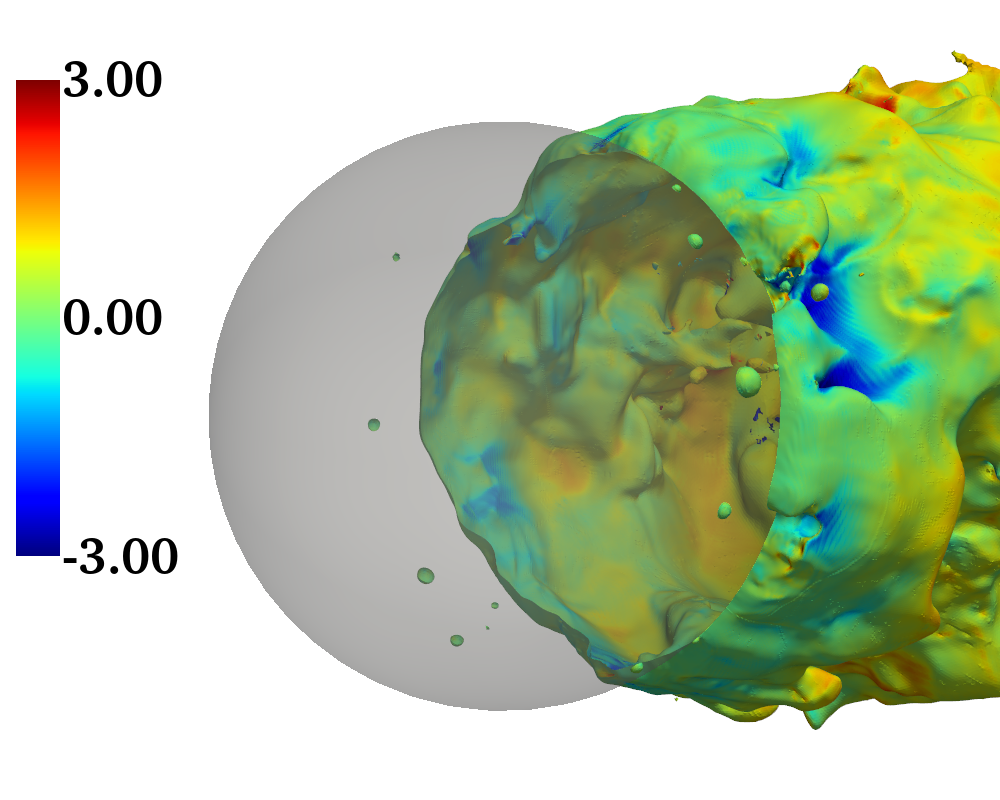}
  \captionsetup{justification=justified, singlelinecheck=false,width=\textwidth}
  \caption{Instantaneous 3D isocontour of air volume fraction ($\alpha_a$=0.5) colored by axial velocity component ($u_x/U_\infty$). (a) MD025, (b) MD050, (c) FR025 and (d) BK025.}
\label{fig:MP_Cav_Detach}
\end{figure}
A stable cavity does not form for all injection cases. The reason lies in the dynamics of the cavity detachment. Figure \ref{fig:MP_Cav_Detach} shows a magnified view of the cavity interface (volume fraction isocontour ($\alpha_a = 0.5$)) colored by the axial velocity component ($u_x$) to show the interfacial velocity of the cavity. Stable cavities were formed behind the sphere during mid-injection and during back-injection. In both MD025 and MD050, axisymmetric elliptic cavities are formed downstream to the injection patch as shown in figure \ref{fig:MP_Cav_Detach} (a,b). Apart from a large cavity formation, the leading edge contains small air pockets followed by longitudinal velocity streaks (prominent in MD050, figure \ref{fig:MP_Cav_Detach} (b)). The small air pockets span the entire circumference of the leading edge. These small air pockets form owing to the interaction of the injected gas and the water crossflow. This phenomenon is known as puffing \citep{Makiharju_2017, Xiong_2024}, and is defined as the cyclic oscillation of volumetric flux around a mean due to the blocking and blow-out of the injected gas by the bulk cross flow. The small gas pockets in mid-injection cases are present only near the leading edge, demonstrating that puffing is limited to the cavity leading edge. In FR025 (figure \ref{fig:MP_Cav_Detach} (c)), there is no attached cavity, and the injected air quickly breaks into bubbles as it moves downstream from the injection patch. Bigger air pockets form in FR025 in comparison to the mid-injection case. These air pockets span the injection patch, followed by a high-velocity region. Here, the extent of puffing behavior spans across the injection patch, and the injected air is entirely swept downstream, preventing a stable cavity formation. The blocking of the injected gas by the water flow results in the formation of air pockets in the front and the mid-injection cases. The high interfacial velocity downstream of these air pockets is due to the sweeping of the air pockets from the previous puffing cycle (see movies S1, S2, and S3 for reference). \\

\subsubsection{Pressure imbalance in puffing phenomenon}\label{sec:puff}

\begin{figure} 
        \begin{subfigure}{0.5\linewidth}
            \centering
            \textbf{MD025}
        \end{subfigure}
        \begin{subfigure}{0.5\linewidth}
            \centering
            \textbf{FR025}
        \end{subfigure} \\
  
        (a) \includegraphics[width=0.48\textwidth]{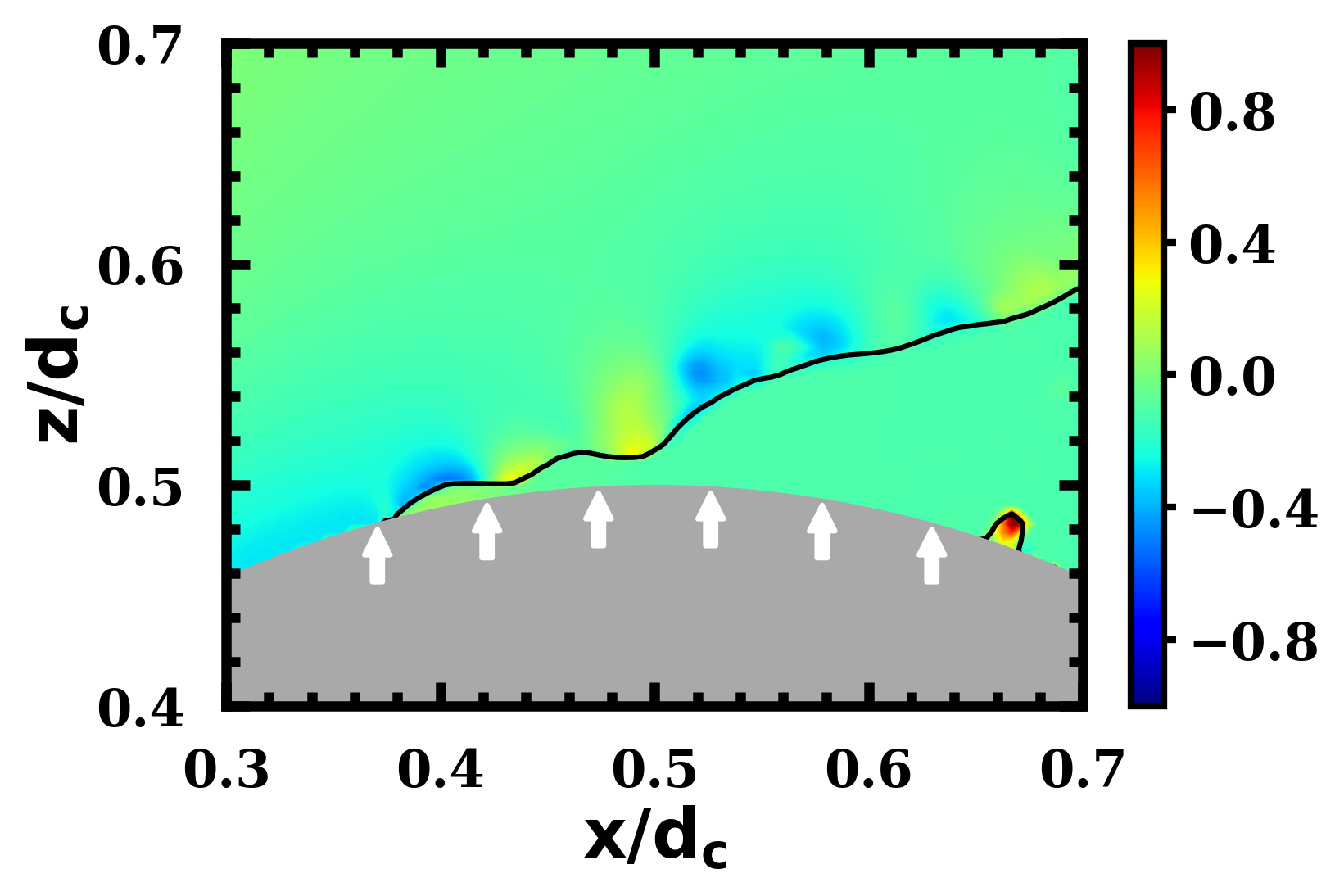}
        (e) \includegraphics[width=0.48\textwidth]{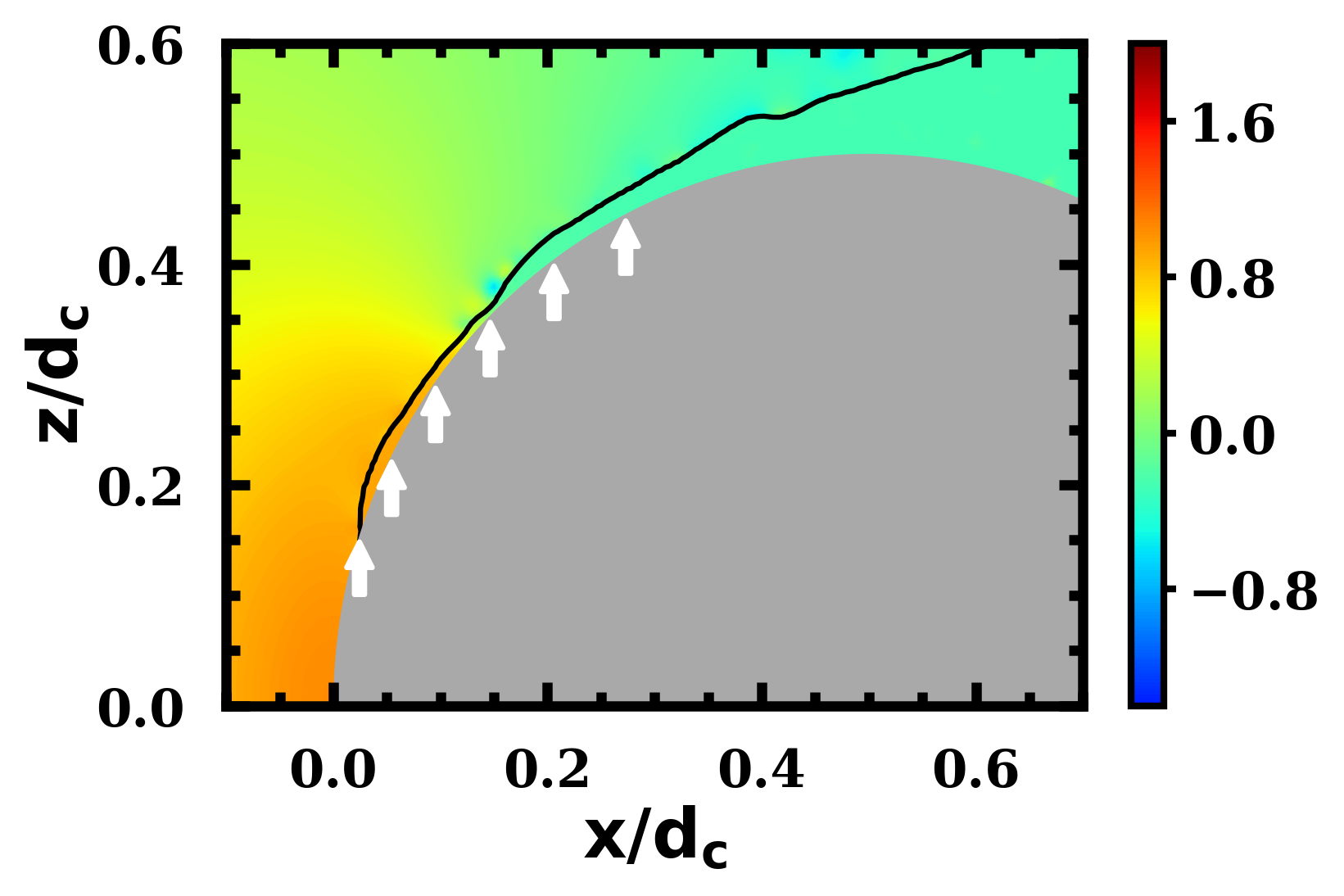}\\
        (b) \includegraphics[width=0.48\textwidth]{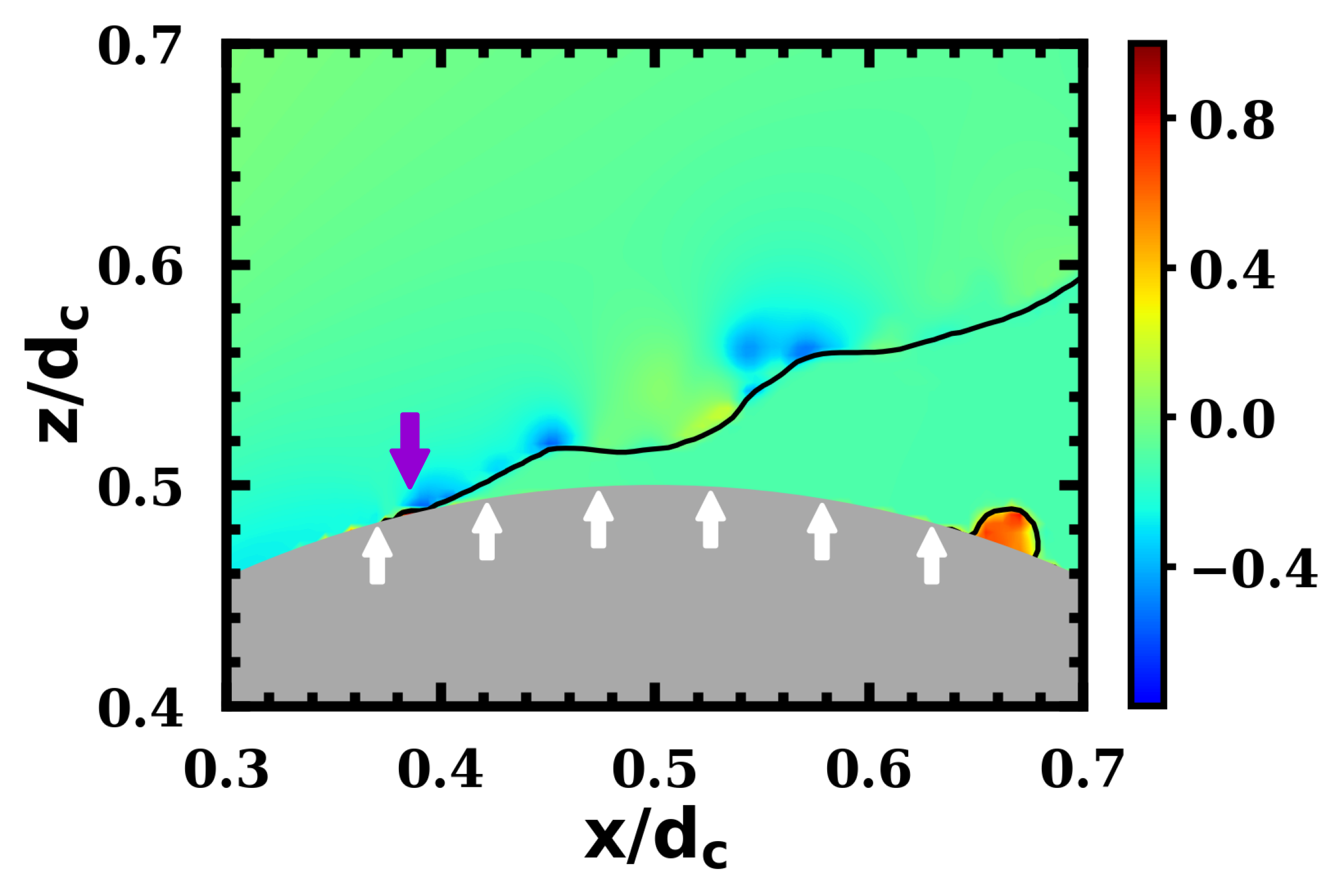}
        (f) \includegraphics[width=0.48\textwidth]{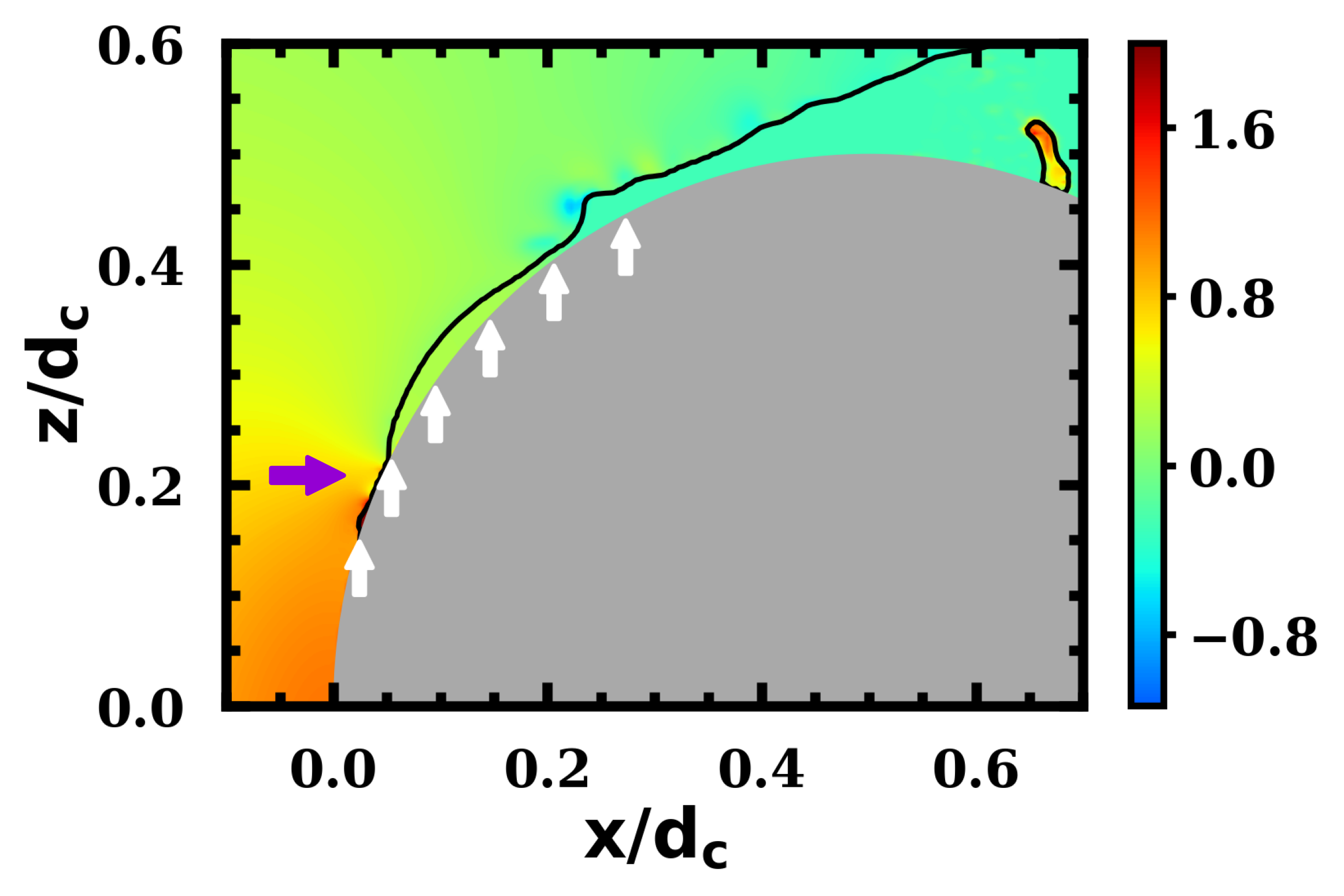}\\
        (c) \includegraphics[width=0.48\textwidth]{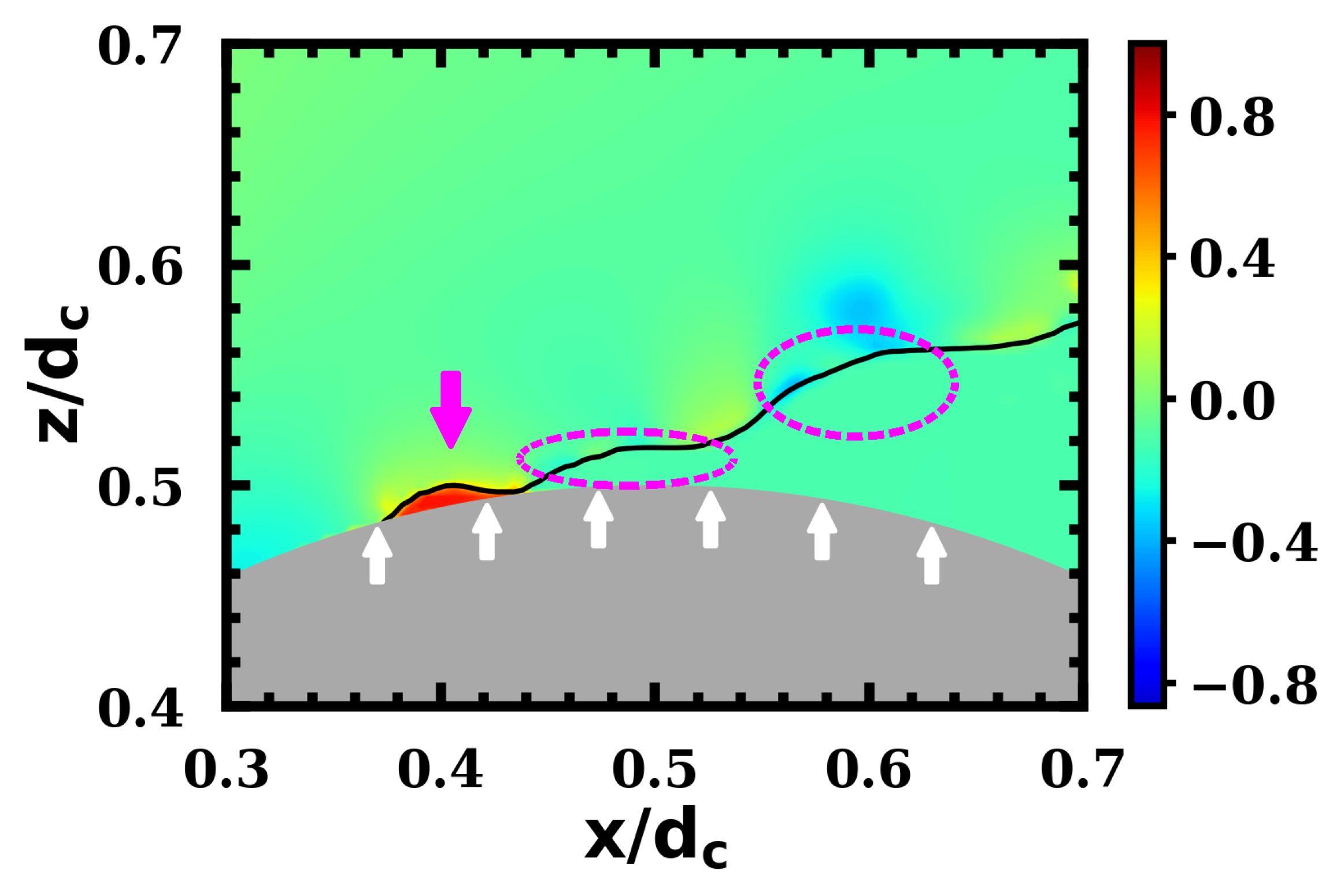}
        (g) \includegraphics[width=0.48\textwidth]{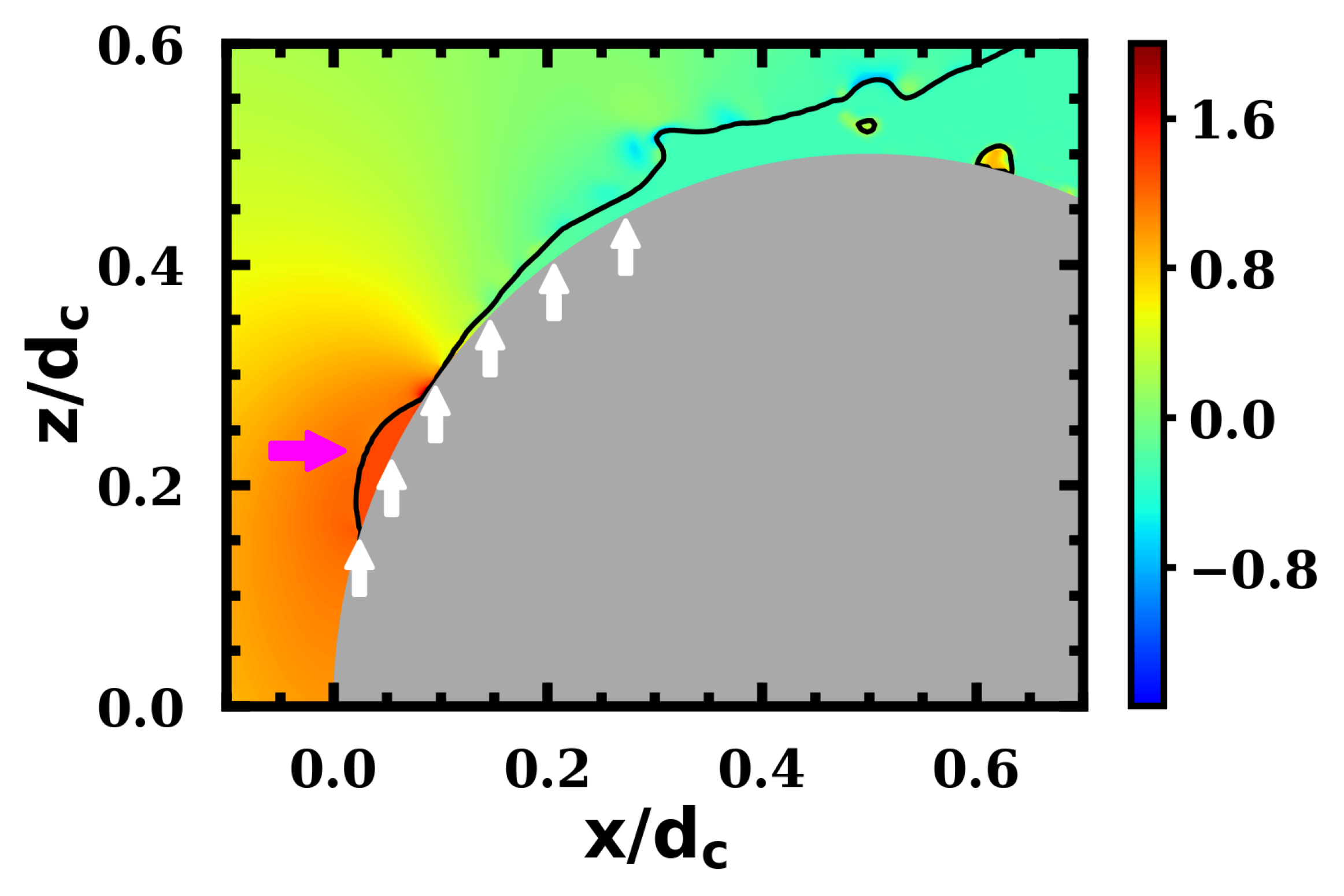}\\
        (d)\includegraphics[width=0.48\textwidth]{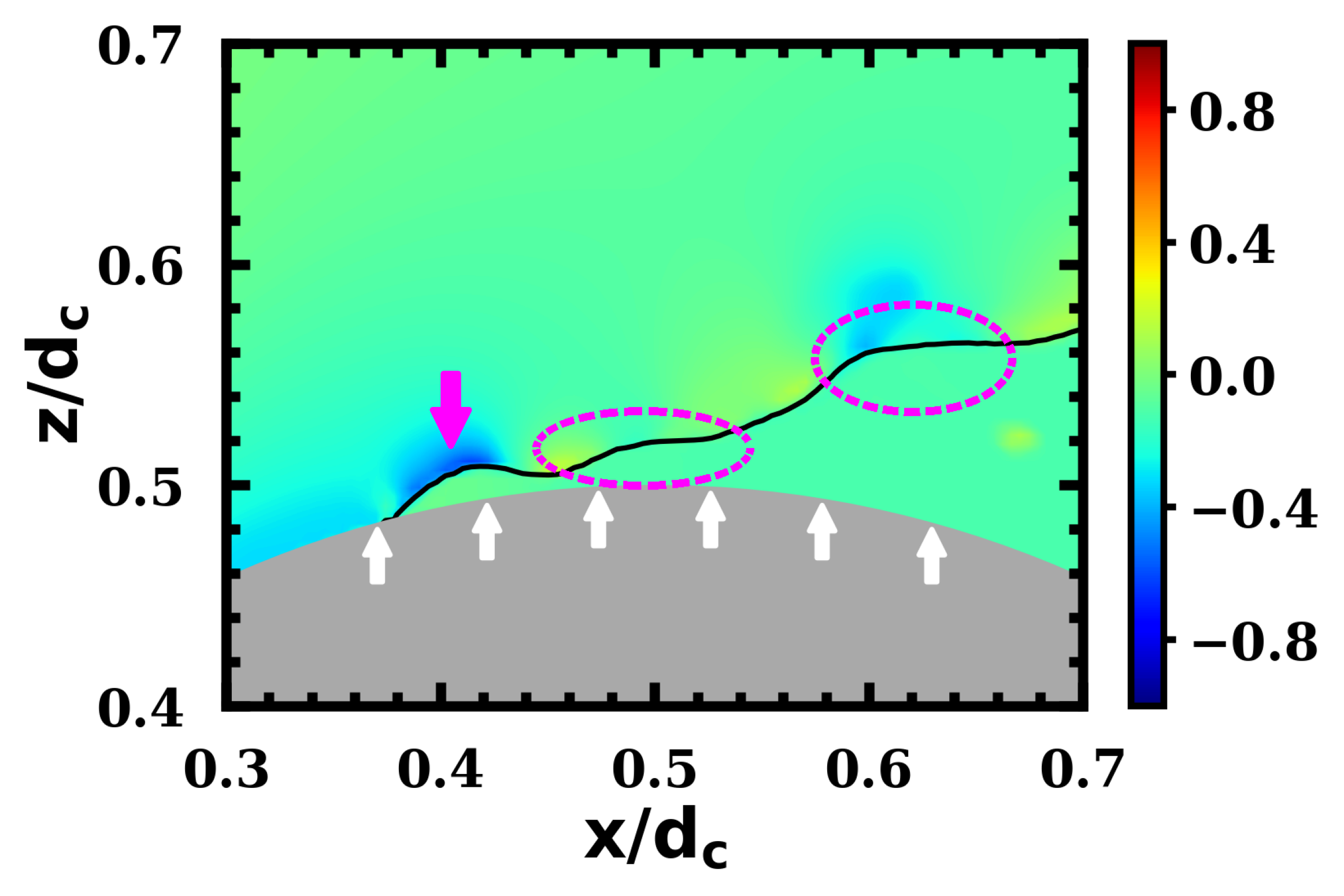}
        (h) \includegraphics[width=0.48\textwidth]{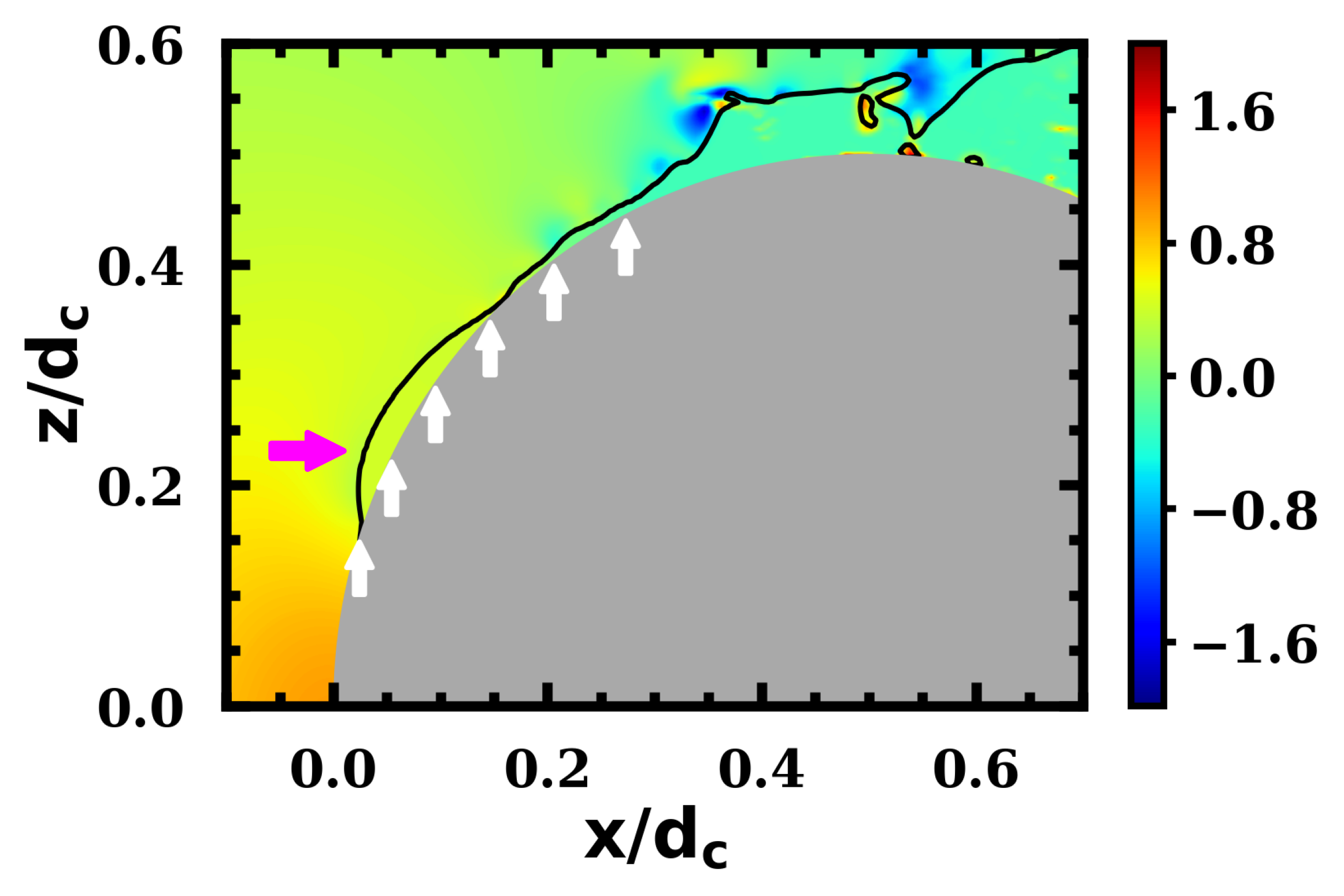}\\
    \captionsetup{justification=justified, singlelinecheck=false,width=\textwidth}
    \caption{Pressure imbalance near the leading edge of the injection patch due to puffing in FR025 and MD025. Pressure coefficient ($C_p$) contour in the $x-z$ plane plotted with air volume fraction isocontour ($\alpha_a =0.5$) in black. The white arrows show the direction of air injection from the surface. MD025: (a-d) and FR025: (e-h). The magenta arrow shows the air pockets. Magenta circles show the puffing instability in MD025.  
    }
    \label{fig:MP_Puffing}
\end{figure}
To better understand the puffing phenomenon in the front-injection and mid-injection cases, the air volume fraction isocurves are superimposed on the pressure-coefficient contour in the $x-z$ plane near the cavity detachment location in figure \ref{fig:MP_Puffing} (a)-(d) for MD025 and \ref{fig:MP_Puffing} (e)-(h) for FR025 at different time instants. Since puffing is limited to the leading edge of the cavity, and is similar in MD025 and MD050, we show the pressure contours only for MD025. Figure \ref{fig:MP_Puffing} (a,e) shows no constriction of the injected air due to the water crossflow. The constriction of the air begins in figure \ref{fig:MP_Puffing} (b,f) (the constricted interface as shown by a violet arrow) with water crossflow occurring right as the interface becomes attached to the sphere surface. A local low-pressure region is visible in figure \ref{fig:MP_Puffing} (b) above the restricted air owing to the acceleration of the water crossflow. The constriction of air causes an increase in local pressure air (figure \ref{fig:MP_Puffing} (c,g)) as air is continuously being injected. This leads to the formation of air pockets shown by magenta arrows in figure \ref{fig:MP_Puffing} (c,g). Once the air pressure becomes sufficiently high, the water crossflow is blocked, and the air is pushed downstream (figure \ref{fig:MP_Puffing} (d,h)), thus reducing the pressure near the leading edge. This blocking and sweeping near the leading edge occur in repeated cycles. The puffing spans the entire injection patch for FR025 as demonstrated by the pressure contours, and causes high pressure, $\sim 60 \%$ higher than the stagnation pressure, near the injection patch, thus affecting the loads on the sphere. In MD025, puffing is limited to the leading edge. Instabilities develop along the cavity interface for MD025 as marked by magenta circles in figure \ref{fig:MP_Puffing} (c,d) owing to the downstream sweeping of small air pockets. Thus, in mid-injection cases, the puffing phenomenon drives the growth of puffing instabilities along the interface. \\ 

\begin{figure}
        (a) \includegraphics[width=0.5\textwidth]{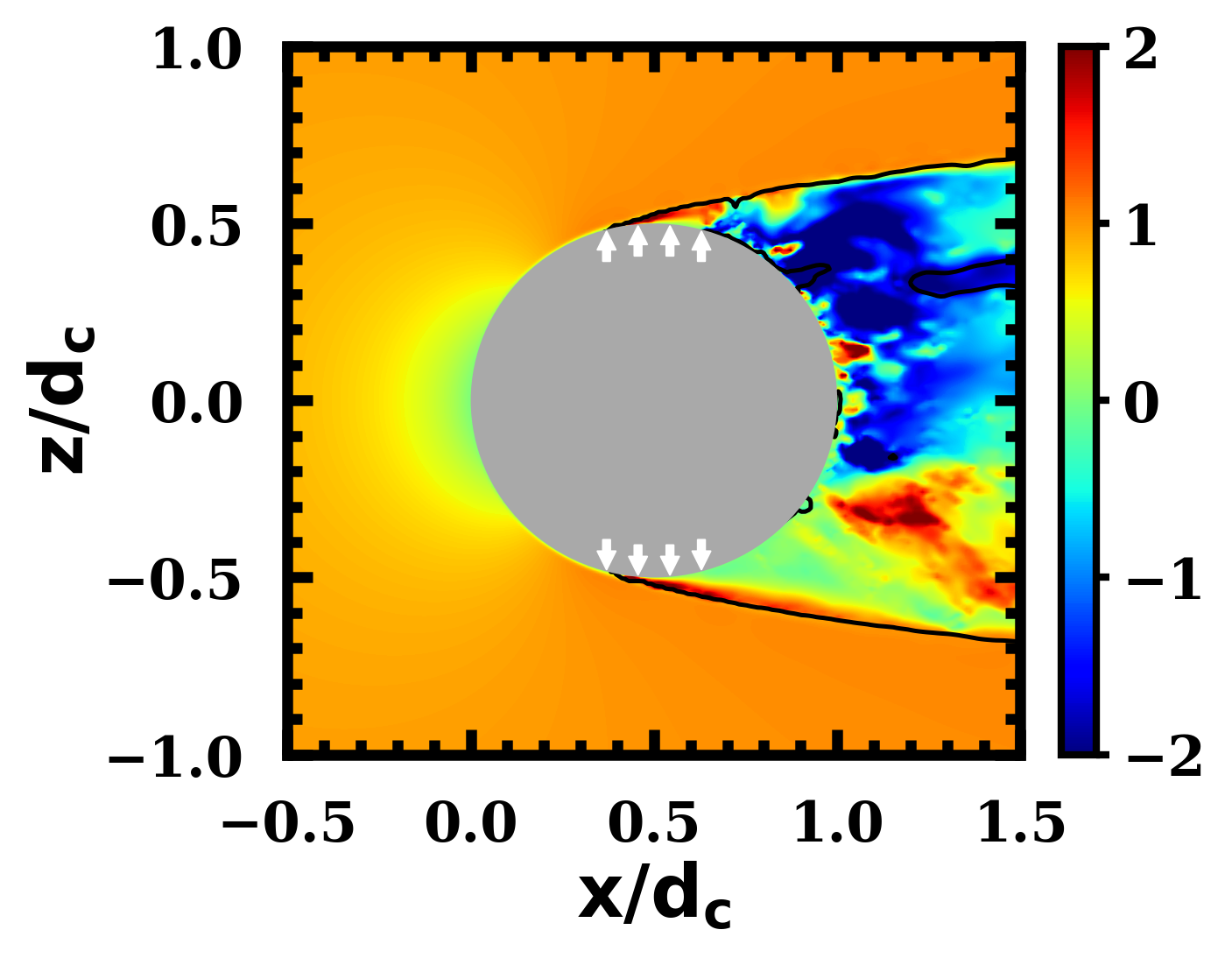}
        (b) \includegraphics[width=0.5\textwidth]{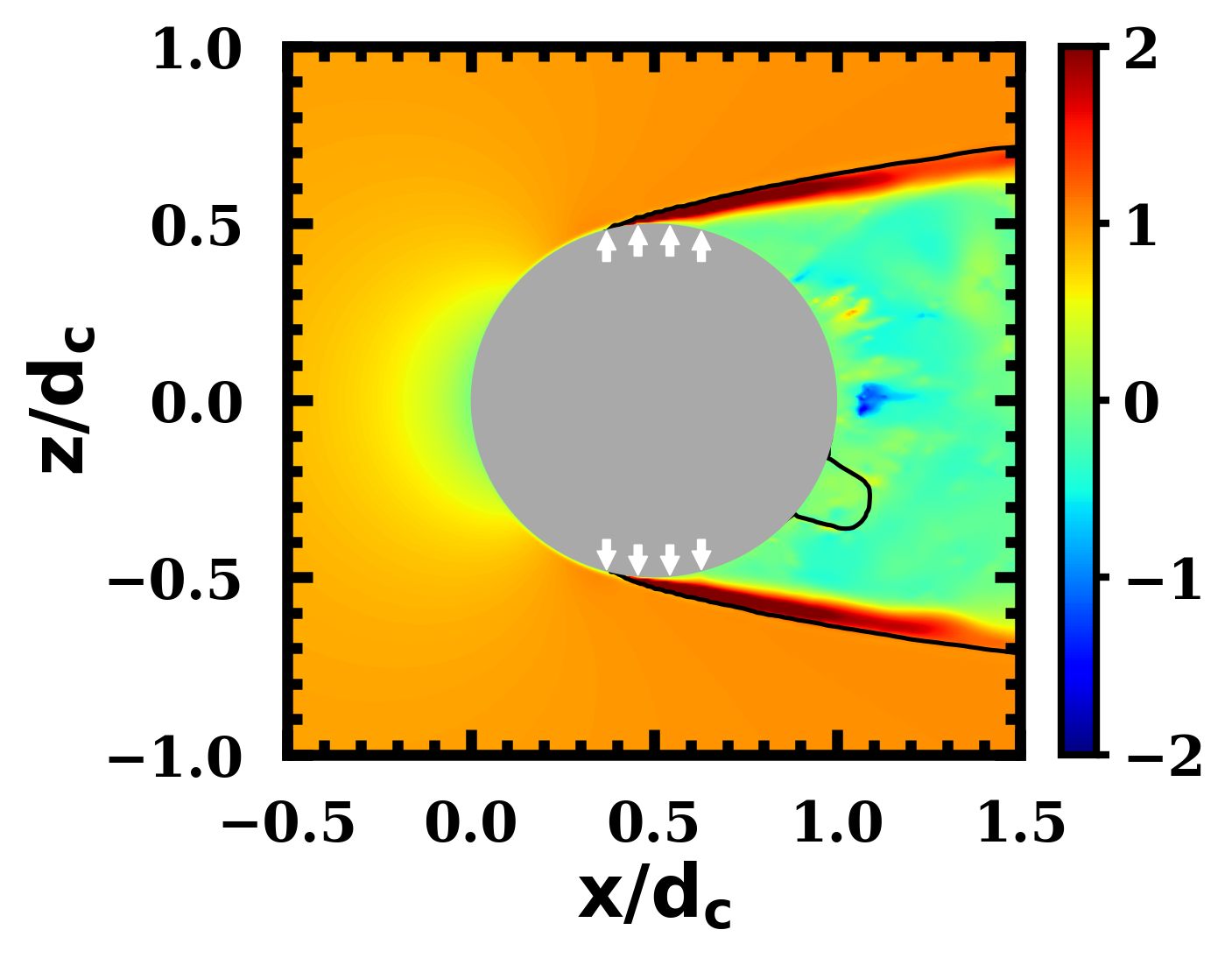}\\
        (c) \includegraphics[width=0.5\textwidth]{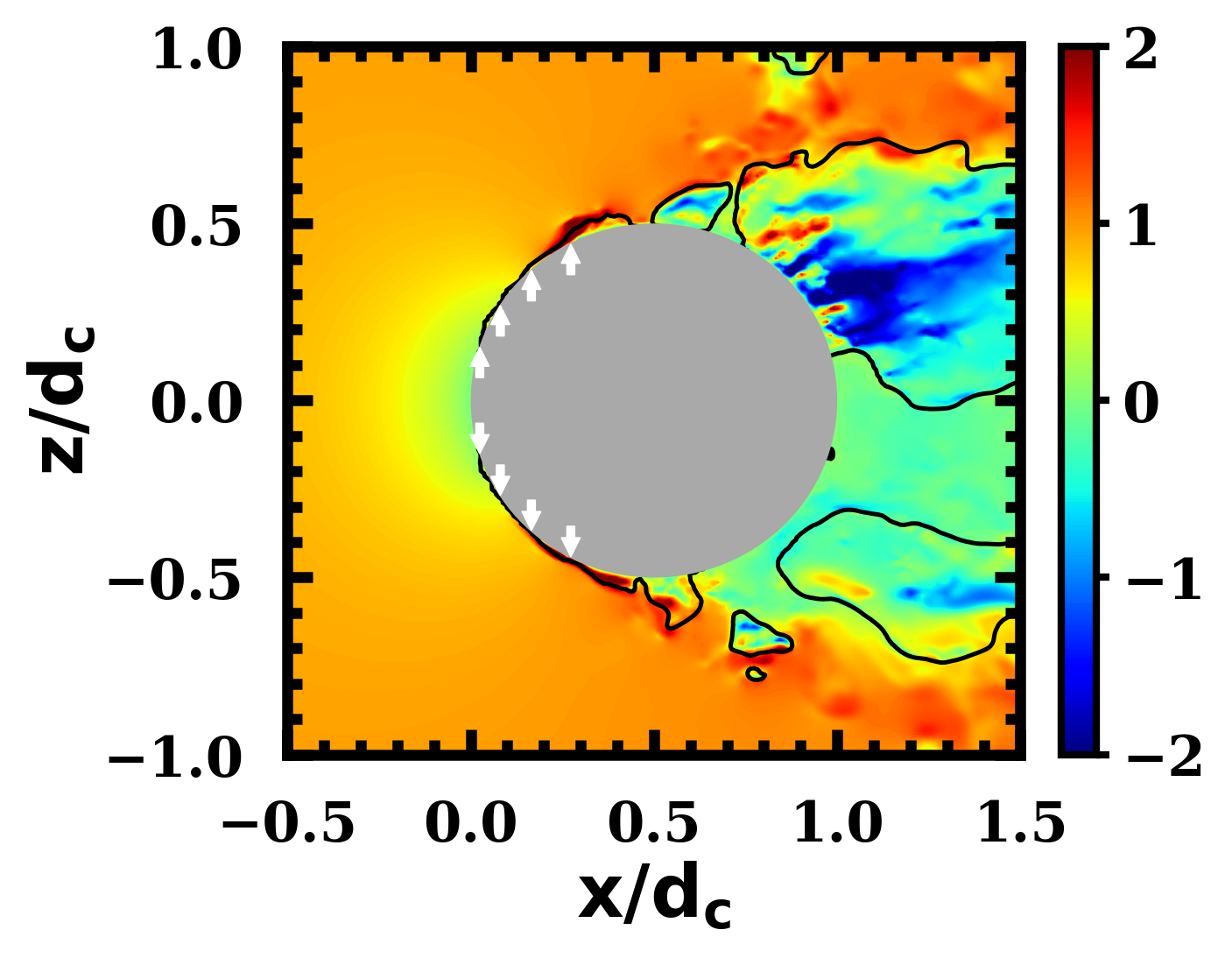}
        (d) \includegraphics[width=0.5\textwidth]{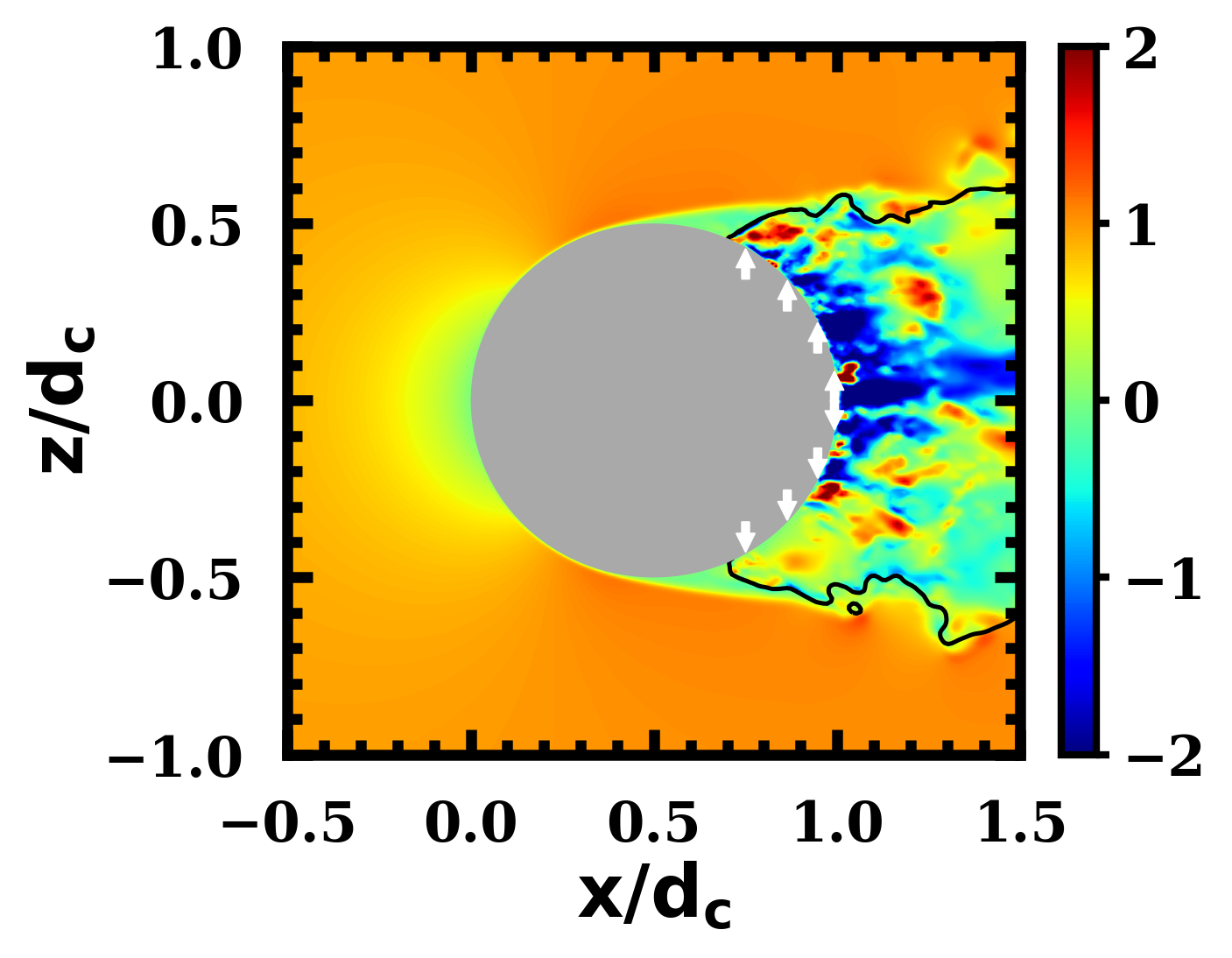}
  \captionsetup{justification=justified, singlelinecheck=false,width=\textwidth}
  \caption{Instantaneous isocontour of air volume fraction ($\alpha_a$=0.5) over axial velocity component ($u_x$) on $x-z$ plane. (a) MD025, (b) MD050, (c) FR025 and (d) BK025. The white arrows show the direction of air injection.}
\label{fig:MP_Cav_Lead_Edge}
\end{figure}

\cite{Xiong_2024} conceptualized the generation of the aforementioned instabilities due to the imbalance of the stagnation pressure of the crossflow and the static pressure of the injected gas. It is evident from the pressure coefficient distribution of SP (figure \ref{fig:SP_Val_1} (a)) that the pressure exerted on the sphere's surface owing to the bulk flow across the injection patch is maximum in the front-injection case and minimum in the mid-injection cases. The reduced pressure imbalance observed in the mid-injection cases can be attributed to two factors. First, the stagnation pressure due to crossflow across the injection patch is lowest in the mid-injection case, because of its chosen location. Second, with a stable cavity formation (figure \ref{fig:MP_Cav_Lead_Edge} (a,b)), the static air pressure also increases. With a decrease in the stagnation pressure of the crossflow and an increase in the static pressure of the injected gas due to cavity formation, the pressure imbalance is reduced, leading to reduced puffing behavior in mid-injection cases. The location of the injection patch near the high-pressure region in FR025 creates a greater pressure imbalance, resulting in the highest puffing in FR025. Puffing causes periodic buildup and evacuation of air downstream \citep{Makiharju_2017}. This air supplied intermittently during puffing is convected in the mean flow direction, preventing the formation of a cavity in FR025. Downstream to the injection patch, as seen in figure \ref{fig:MP_Cav_Lead_Edge} (c), the air was swept as bubbles, forming a bubbly cavity. \\

\subsubsection{Divots in BK025}\label{sec:divot}

\begin{figure}
     (a) \includegraphics[width=0.49\textwidth]{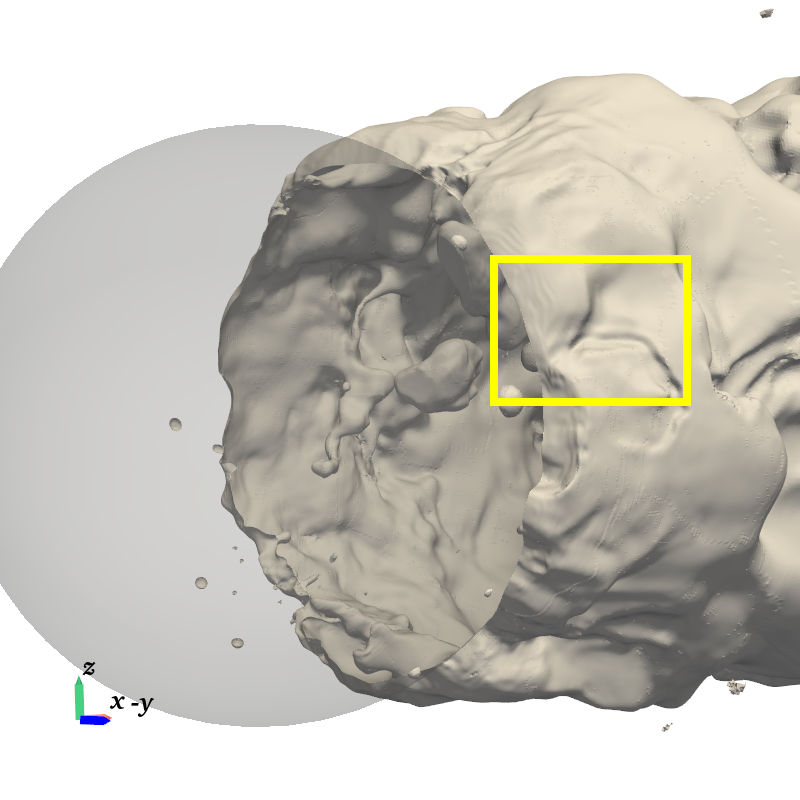}
      (b) \includegraphics[width=0.49\textwidth]{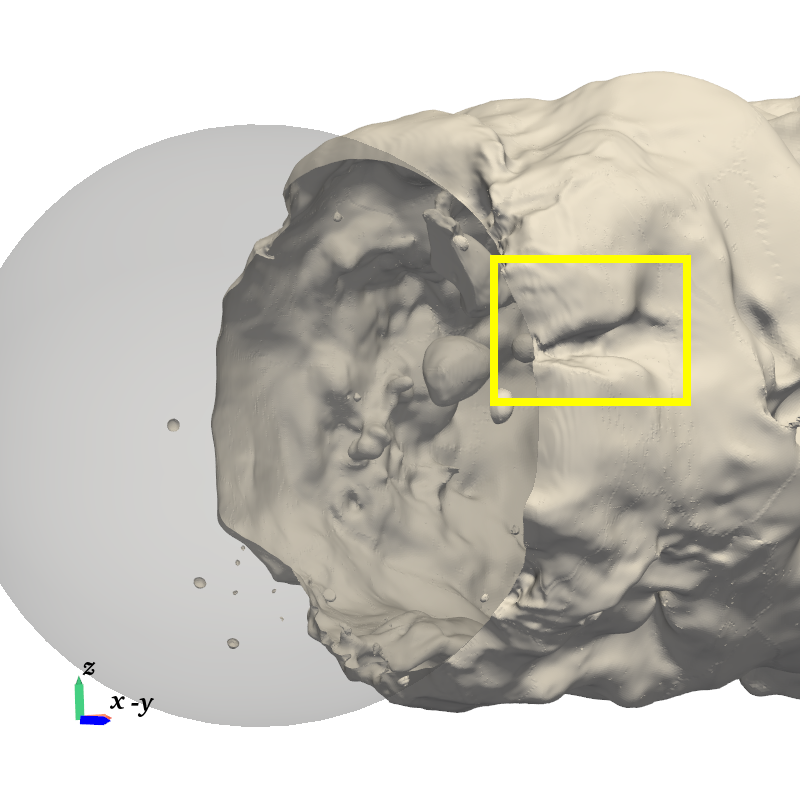}\\
     (c) \includegraphics[width=0.49\textwidth]{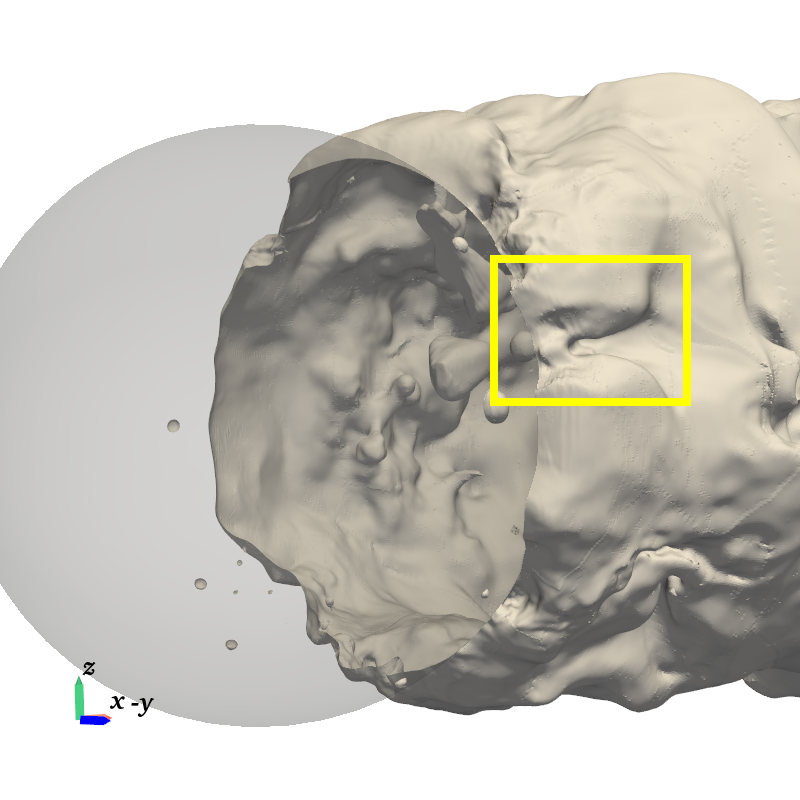}
     (d) \includegraphics[width=0.49\textwidth]{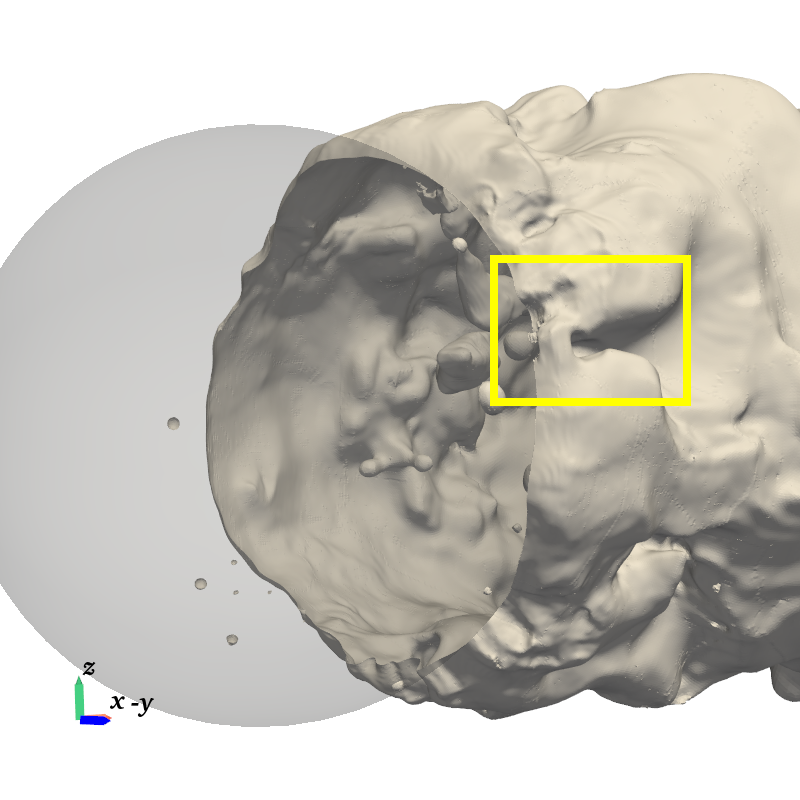}\\
     (e) \includegraphics[width=0.49\textwidth]{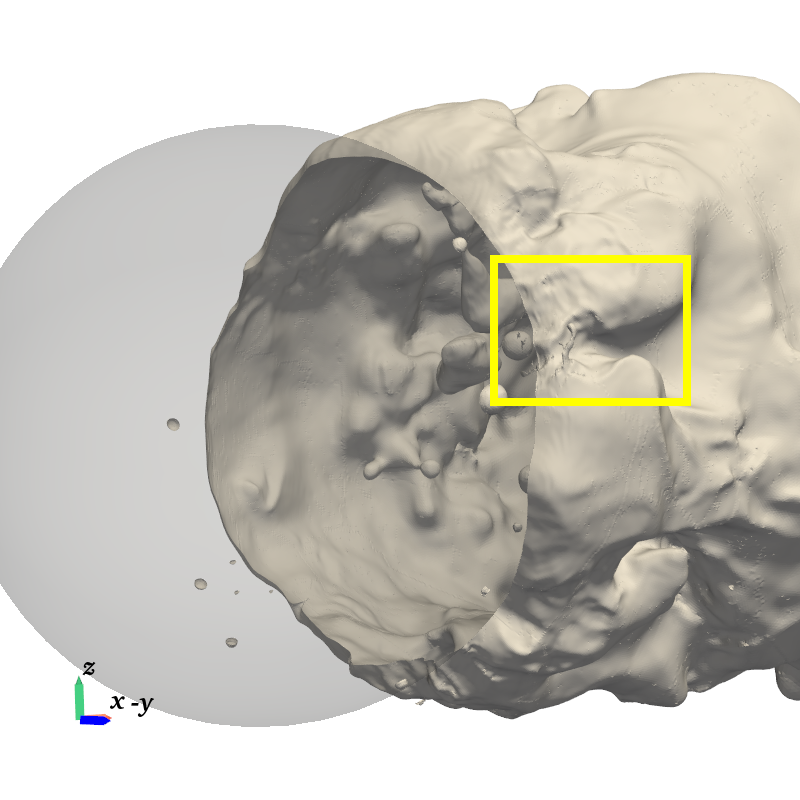}
     (f) \includegraphics[width=0.49\textwidth]{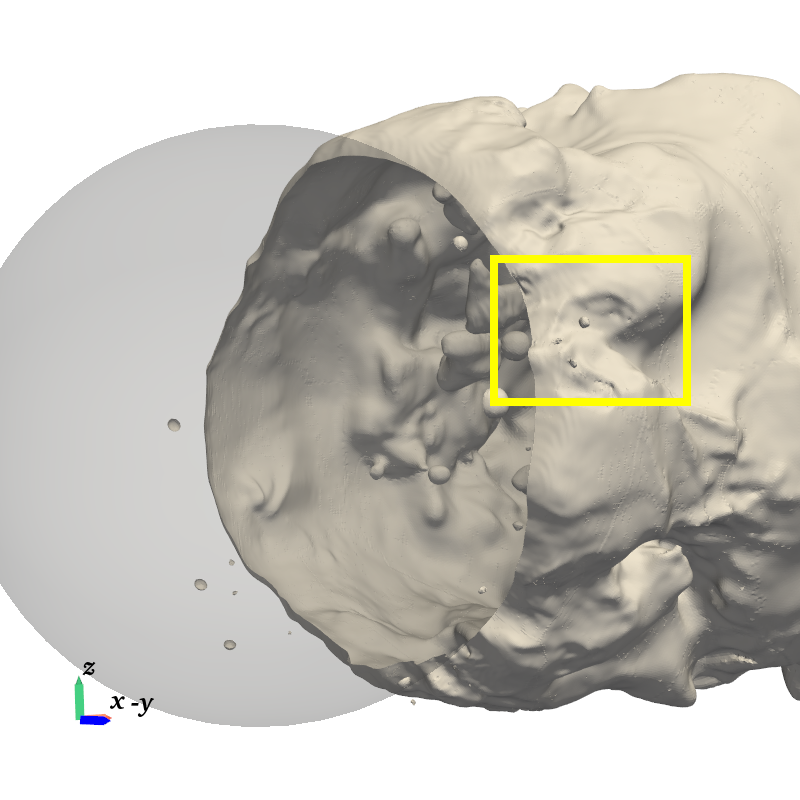} \\
    \captionsetup{justification=justified, singlelinecheck=false,width=\textwidth}
    \caption{Divot formation and Collapse in BK025. The yellow box shows the interested divot. Volume fraction isocontours are plotted at different time instants. 
    }
    \label{fig:MP_Divots}
\end{figure}

\begin{figure}
    \centering
    \begin{subfigure}[b]{1.0\textwidth}
        \includegraphics[width=\textwidth]{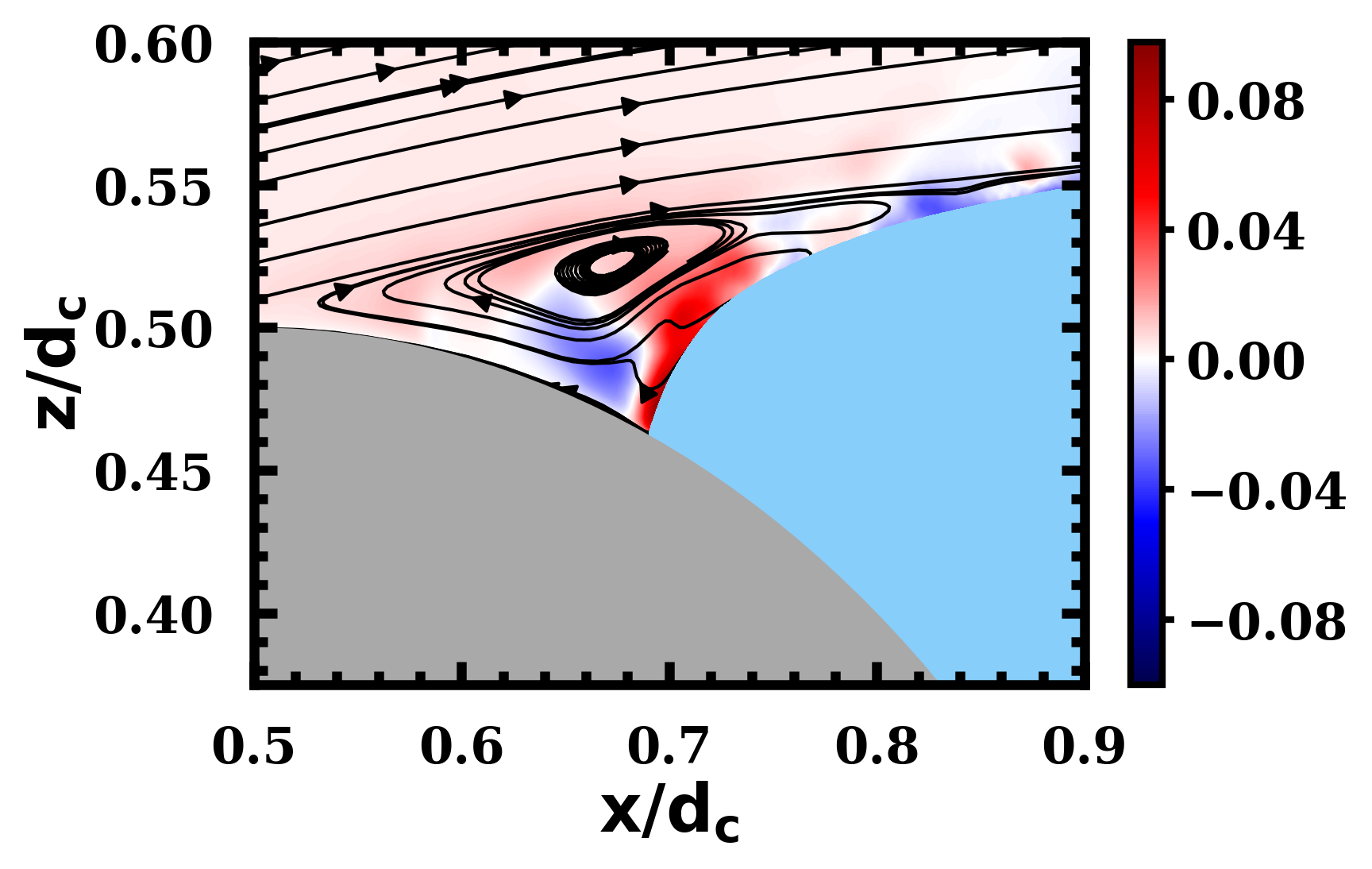}
    \end{subfigure}
 
    \captionsetup{justification=justified, singlelinecheck=false,width=\textwidth}
    \caption{Recirculation zone upstream to the cavity in BK025. Flow streamlines (black) are plotted over the $y$ component of mean velocity ($\overline{u}_{y}$) contour for the water phase in the $x-z$ plane.}
    \label{fig:MP_Recirc}
\end{figure}

Puffing is absent near the cavity detachment in BK025; instead, large-scale indentations were observed at the leading edge of the cavity in BK025 (figure \ref{fig:MP_Cav_Detach} (d)). Since the flow around the injection patch in BK025 is separated, the interaction between the injected air and the separated crossflow is not strong enough to cause puffing. The indentations were similar to the "divots" observed in the experiments of \cite{Tassin_Leger_1998b} and, more recently, in \cite{Brandner_2010}. Multiple divots occurred at random points around the circumference of the sphere. The primary reason for the divot formation is the disturbance caused by the water droplets from the inside of the cavity hitting the cavity interface \citep{Tassin_Leger_1998b}. The formation and collapse of a divot is shown in figure \ref{fig:MP_Divots} and is also evident in the movie S4. These water drops were created due to the splashing of the re-entrant jet onto the sphere's surface. Additionally, divots can form due to perturbations generated upstream in the separated shear layer incident on the cavity surface. Since the Reynolds number of the freestream flow is lower than the experiments of \cite{Tassin_Leger_1998a} and \cite{Tassin_Leger_1998b}, the divot formation due to upstream flow perturbation is seldom observed. Once the interface is perturbed by either disturbance (figure \ref{fig:MP_Divots} (a)), a jet of water impinges on the divot location \citep{Tassin_Leger_1998b} (figure \ref{fig:MP_Divots} (b)). The surface disturbance results in the formation of divots that tend to grow as shown in figures \ref{fig:MP_Divots} (c,d). The localized indentations begin to collapse quickly (figure \ref{fig:MP_Divots} (e)), and their circumferential motion is not visible. The strength of the liquid jet at the divot depends solely on the Reynolds number, which is relatively low in the present case, leading to the rapid disappearance of the divot. The collapse of divots forms small air bubbles as seen in figure \ref{fig:MP_Divots} (f).  The bubbles shed from the leading edge either travel downstream or travel upstream and attach to the sphere's surface (as seen in figure \ref{fig:MP_Divots}). \\

The upstream motion is aided by the presence of the recirculation zone formed between the separated shear layer and the cavity near the sphere surface. The recirculation zone upstream of the cavity detachment is shown in the figure \ref{fig:MP_Recirc} with the help of the streamlines of the liquid phase over the $y$ component of mean velocity ($\overline{u}_{y}$) contour for the water phase in the $x-z$ plane. The cavity is masked with sky blue color. There is a strong $\overline{u}_y$ (or the plane normal velocity component for the particular $x-z$ plane) just near the cavity wall which shows the three-dimensionality of the flow upstream to the cavity, similar to smearing of dye in the transverse direction in \cite{Tassin_Leger_1998b}'s experiments. The presence of a cavity in the flow creates a stagnation region near its leading edge, thereby increasing the circumferential flow velocity and making the flow three-dimensional. In BK025, the cavity detaches locally perpendicular (figure \ref{fig:MP_Recirc}) to the surface, due to the given boundary condition for $\alpha_w$ on the sphere surface. Hence, there is no prominent "separation meniscus" (curving of the cavity interface near cavity detachment \citep{Brennen_1970b}) in BK025. \\

\subsubsection{Cavity surface appearance}
Owing to distinct mechanisms at the cavity leading edge under different injection conditions, we study the emergence of the cavity interface. The cavity interfaces in MD025 and MD050 are relatively smooth compared to the cavity in BK025 (figure \ref{fig:MP_Cav_Detach} (a,b,d)). There are two reasons associated with this. First, the decelerated flow region in the laminar separated flow upstream of the cavity, which generates instability \citep{Brennen_1970b}, and the downstream propagation of this instability to the cavity interface. The second reason is that the interfacial velocity in the cavity is high during mid-injection, making the interface less susceptible to instabilities \citep{Brandner_2010}. Higher velocity of the interface in mid-injection cases, as seen in figure \ref{fig:MP_Cav_Detach} (a,b), reduces the velocity difference between the cavity interface and the liquid shear layer overlying the cavity leading edge. Hence, the shear is reduced along the liquid side of the interface, thereby suppressing the growth of Kelvin-Helmholtz ($K-H$) instabilities. Despite the suppression of the $K-H$ instabilities in mid-injection cases, the puffing instability generated by the leading-edge puffing is dominant on the cavity surface. In BK025, the lower interface velocity promotes the growth and propagation of the $K-H$ instability, resulting in a rougher cavity surface (figure \ref{fig:MP_Cav_Detach} (d)). Few suspended bubbles are visible in all the stable cavity cases near the sphere's base inside the cavity (figure \ref{fig:MP_Cav_Detach} (a,b,d)) due to interaction of the re-entrant jet with the sphere's base. The details on the dynamics of the re-entrant jet are discussed in \S \ref{sec:splash}. \\

\subsection{Surface pressure and skin friction}\label{sec:cp_cf}

\begin{figure}
    \centering
    \begin{subfigure}[b]{1.0\textwidth}
        \includegraphics[width=\textwidth]{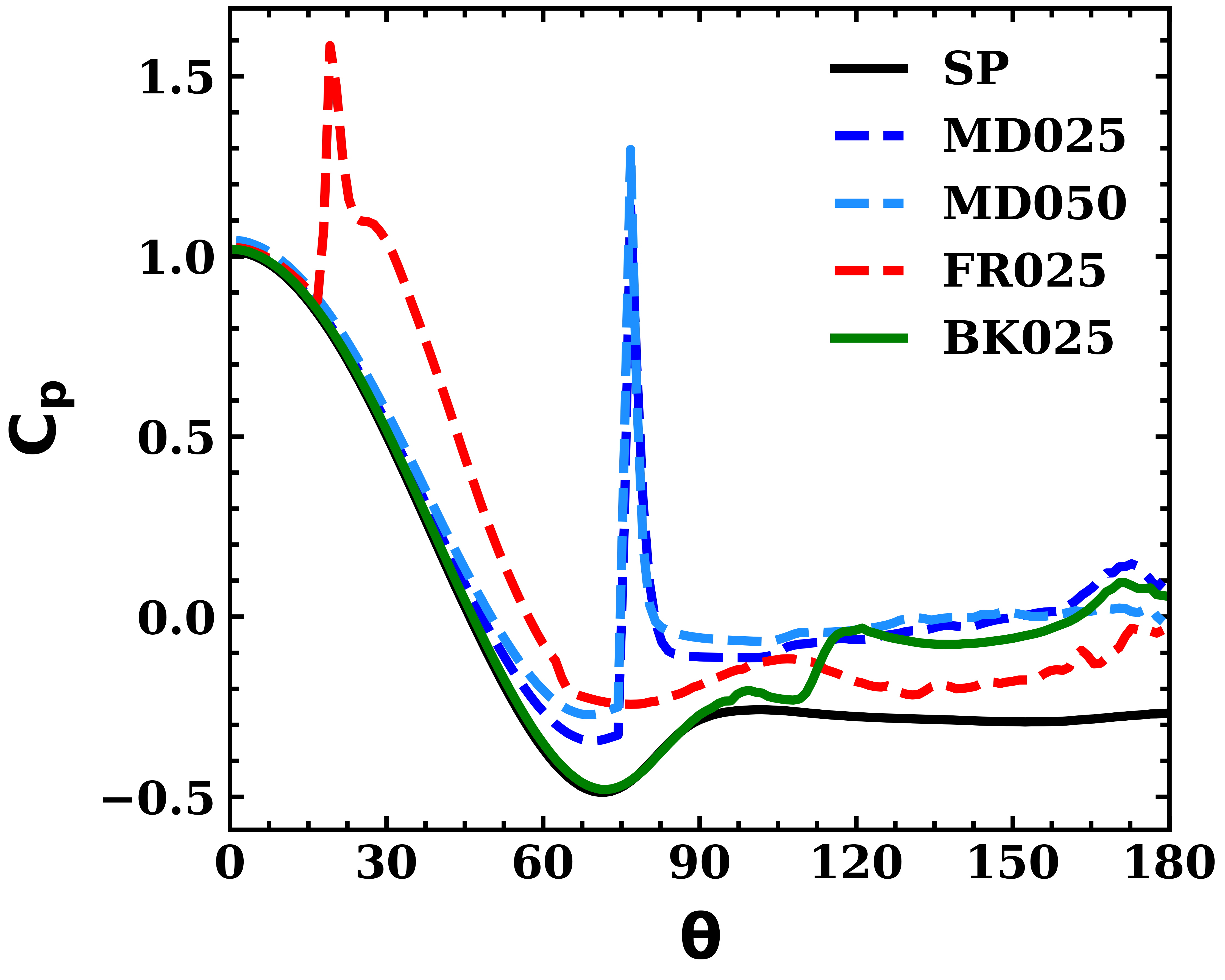}
    \end{subfigure}
    \captionsetup{justification=justified, singlelinecheck=false,width=\textwidth}
    \caption{Time and azimuthally averaged pressure coefficient ($C_p$) distribution for different cases.}
    \label{fig:MP_Cp}
\end{figure}

Figure \ref{fig:MP_Cp} shows the time and azimuthally averaged pressure coefficient ($C_p$) distribution with respect to the azimuthal angle ($\theta$) on the sphere surface for various cases. The pressure coefficient in the mid-injection cases deviates from the SP case at an angle $\sim 30^\circ$. The global minimum pressure around the sphere increases due to ventilation, and this effect is stronger at higher ventilation rates. The formation of a cavity, in an unseparated flow, creates a stagnation region just upstream, thereby increasing the global minimum pressure. \cite{Gnanaskandan_2016} also observed a drop in favorable pressure gradient upstream to the cavity for natural cavitation around the cylinder. A sharp pressure peak is observed in the mid-injection cases near the leading edge of the injection patch (at $\theta = 75^{\circ}$), but within the cavity. The peak occurs due to the puffing phenomenon on the cavity's leading edge. The peak is higher in MD050 than in MD025 due to the higher injection velocity. Away from the pressure peak caused by puffing, the pressure should remain constant over the injection patch and further downstream, due to the cavity formation \citep{Franc_1985, Gnanaskandan_2016}. As seen in figure \ref{fig:MP_Cp}, there is a slight increase in $C_p$ near the base of the sphere. This increase is due to the impingement of the re-entrant jet onto the sphere's base region. The pressure downstream of the air injection, in both MD025 and MD050, is higher than in SP due to cavity formation. \\

In FR025, there is a pressure peak at $\theta= 18^{\circ}$ similar to the mid-injection cases due to puffing.  After this pressure peak, the pressure remains higher than in other cases. This is due to puffing across the entire injection patch, resulting in high pressure in the downstream swept fluid. In contrast to the peaks observed in the front and mid-injection cases, there is no pressure spike in BK025. In this case, the flow is separated from the sphere surface; hence, there is no strong impingement of fluid onto the cavity surface.  
In BK025, a pressure jump is observed at the leading edge of the cavity. The pressure jump is due to higher pressure within the air cavity. The pressure remains higher in the cavity because a less dense fluid than the ambient fluid fills it. The pressure coefficient matches the single-phase flow until the point of cavity inception in the back-injection case. After cavity inception, the pressure exceeds the single-phase flow pressure due to the cavity formation. In all the stable cavity cases (MD025, MD050, and BK025), there is a pressure jump in the neighborhood of cavity detachment, as also demonstrated by \cite{Franc_1985}, and the onset of puffing creates a peak at the cavity leading edge in addition to the pressure jump. \\

\begin{figure}
    \begin{subfigure}[b]{\textwidth}
    \hspace{-2cm}    \includegraphics[width=1.25\textwidth]{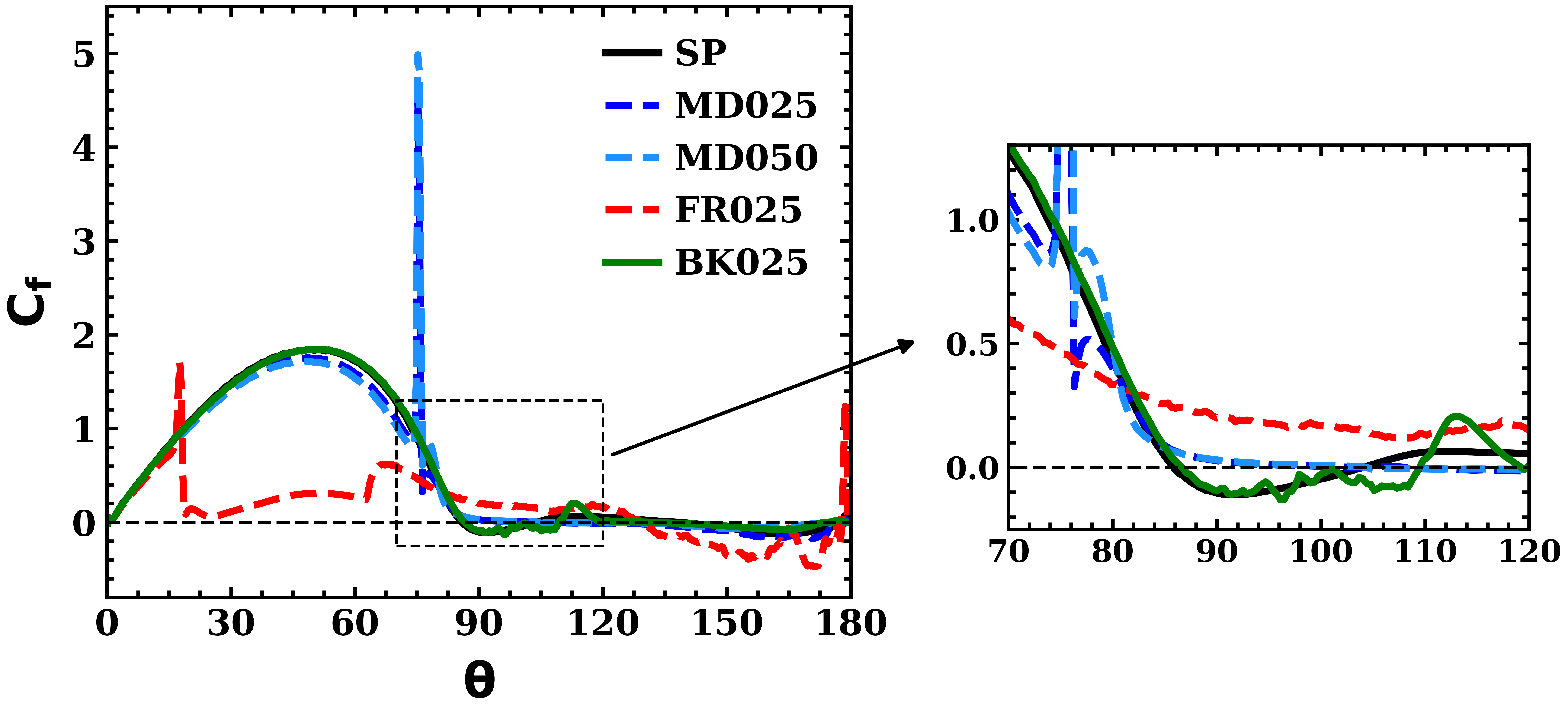}        
    \end{subfigure}
    \captionsetup{justification=justified, singlelinecheck=false,width=\textwidth}
    \caption{Time averaged skin friction coefficient ($C_f$) distribution for different cases.}
    \label{fig:MP_Cf}
\end{figure}

Figure \ref{fig:MP_Cf} shows the time and azimuthally averaged skin friction coefficient ($C_f$) distribution with respect to $\theta$ on the sphere surface for various cases. In the mid-injection cases, the maximum value of $C_f$ drops upstream of the cavity detachment compared with the SP case, indicating a deceleration of the flow \citep{Gnanaskandan_2016}. A slightly larger decrease in $C_f$ in MD050 indicates that the upstream flow deceleration due to cavity formation increases with increased air ventilation. An acute peak in the skin friction coefficient is observed near the leading edge ($\theta=75^{\circ}$) of the injection patch in mid-injection cases. The presence of a stagnation region just upstream of the cavity accelerates the water crossflow. Additionally, since air is injected within the cavity at a specified velocity, a strong velocity gradient across the cavity interface is generated, leading to the formation of the sharp peak.
There is a shorter rise and drop in the friction coefficient adjacent to the injection peak. Here, the secondary rise is due to air pushed downstream by the leading-edge puffing. As seen in the pressure plot (figure \ref{fig:MP_Cp}), there is a sharp adverse pressure gradient near the leading edge(within the cavity) produced by puffing. The air, which is accelerated downstream by the adverse pressure gradient, quickly decelerates further in the absence of an adverse pressure gradient due to nearly constant cavity pressure. 
The injection peak, or primary peak, is responsible for the higher interfacial velocity in mid-injection cases, and the secondary peak is responsible for the strong entrainment of air along the internal cavity boundary. The momentary acceleration of the gas is higher in MD050 than in MD025 (as seen in the inset of figure \ref{fig:MP_Cf}), leading to a higher velocity of air entrainment along the cavity wall. Further, the flow decelerates over the injection patch, hence the friction coefficient decreases. Downstream to the injection patch, the friction coefficient remains nearly zero due to the complete replacement of water by air near the sphere surface. The minor deviations in the friction coefficient are due to the impingement of the re-entrant jet, entrained by the upstream flow internally. \\
 
In FR025, the air injection creates a peak in the friction coefficient, similar to the mid-injection cases. The peak in the front-injection cases is smaller than the peaks obtained in the MD cases, owing to the difference in the direction of velocity injection with respect to the bulk cross flow. We consider the angle between the direction of injection velocity (normal to the $x$ axis) and the local tangent to the sphere's surface along the flow direction. This angle indicates the alignment of the air injection with the cross-flowing bulk fluid. This angle is $18^{\circ}$ for the FR025 case and $75^{\circ}$ for the MD cases at the leading edge of the injection patches, respectively. The lower angle here indicates that the injections are more along the crossflow direction and produce a lower velocity gradient. The higher angle in the mid-injection cases creates a larger velocity gradient, resulting in a taller peak than in the front-injection case, despite the difference in magnitude of the injection velocity. The flow gradually accelerates over the injection patch due to a constant momentum addition to the flow \citep{Gad_el_Hak_2000}. Further downstream, the flow decelerates and is influenced by the wake. \\

The friction coefficient plot for BK025 deviates from the SP case at $ \sim 90^\circ$, but it largely resembles it. A small peak near the cavity detachment location is created by the velocity gradient across the cavity interface. Just upstream of the cavity interface peak, $C_f$ shortly becomes negative owing to the recirculation zone. After cavity inception, $C_f$ decreases to nearly zero due to the presence of less viscous air inside the cavity. \\

With both pressure and skin friction altered with respect to the SP case owing to ventilation, it is intuitive to study the location of flow separation on the sphere's surface. Table \ref{tab:fl_param} lists the location of flow separation ($\varphi_s$) for the stagnation point for various cases. In both front-injection and mid-injection cases, the location of flow separation is delayed. By injecting air, we impart momentum to the flow. Momentum coefficient ($C_{\mu}$) is a non-dimensional parameter used to quantify the momentum imparted or removed from the flow in flow control \citep{Chen_2013, Song_2025} and is defined as 
\begin{equation}
    C_{\mu} = (\rho_{in}Q_{in}V_{in})/(0.5 \rho_{\infty} U_{\infty}^2d_c^2),
\end{equation}
where $\rho_{in}$ is the density of the injected fluid. $C_\mu$ for the present cases is in the order of $10^{-4}$ due to the injection of the less dense air. Since the injected momentum is very low, it is insufficient to disrupt the flow and cause early separation; hence, it augments the flow, delaying separation. \\

In the mid-injection cases, stable attached cavities form despite the delayed flow separation. This is contrary to the natural cavity experiments of \cite{Franc_1985} where it was observed that natural cavities cease to exist in the absence of an upstream flow separation. They attributed it to an increase in the ambient pressure, leading to the disappearance of the cavity. Low ambient pressure is required in natural cavitation for the conversion (evaporation) of liquid into its vapor to form the cavity. With ventilated cavities, there is a constant supply of air with which a cavity can be formed stably. Stable cavities were formed in MD cases, but not in FR025. This suggests that, in the absence of upstream flow separation, the extent of puffing determines the formation of a stable cavity. In BK025, the location of flow separation does not change much because ventilation is performed much downstream. \\ 

\subsection{Loads on the Sphere}\label{sec:load}
\begin{figure}
        (a) \includegraphics[width=0.9\textwidth]{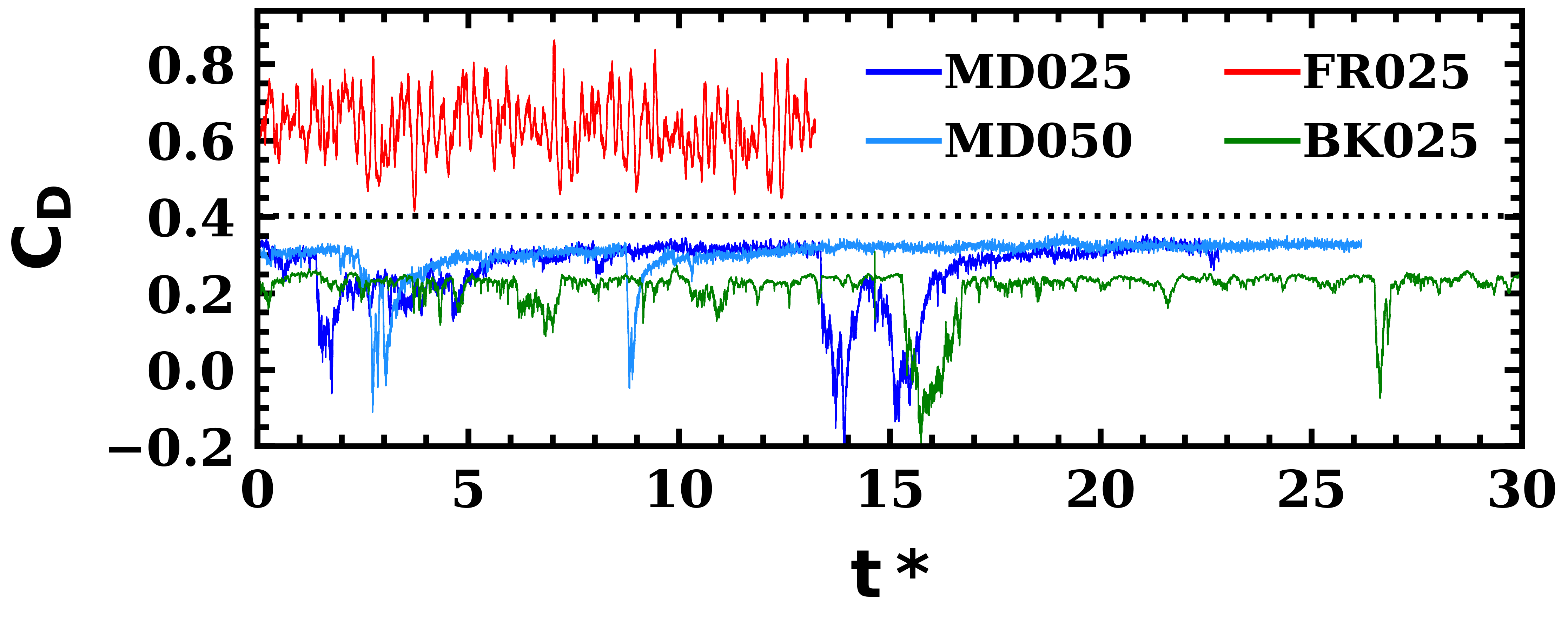}\\
        (b) \includegraphics[width=0.9\textwidth]{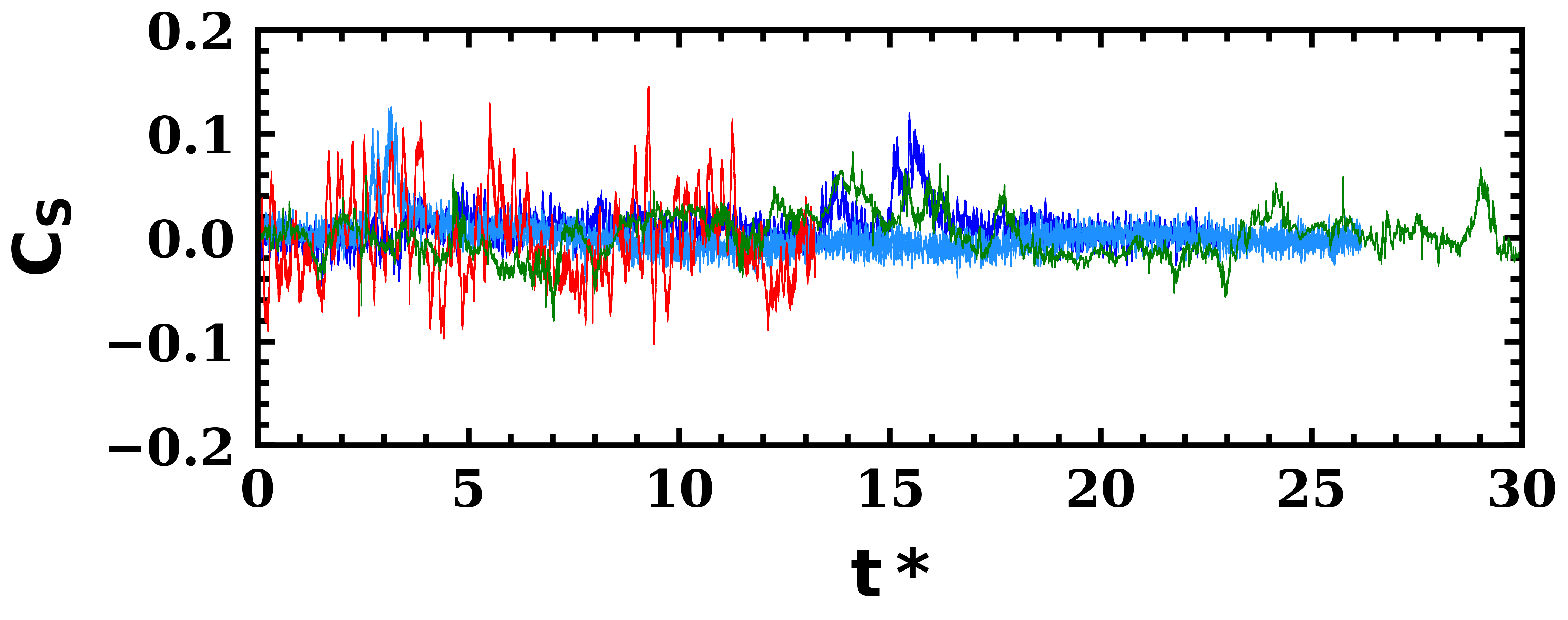}\\
        (c) \includegraphics[width=0.9\textwidth]{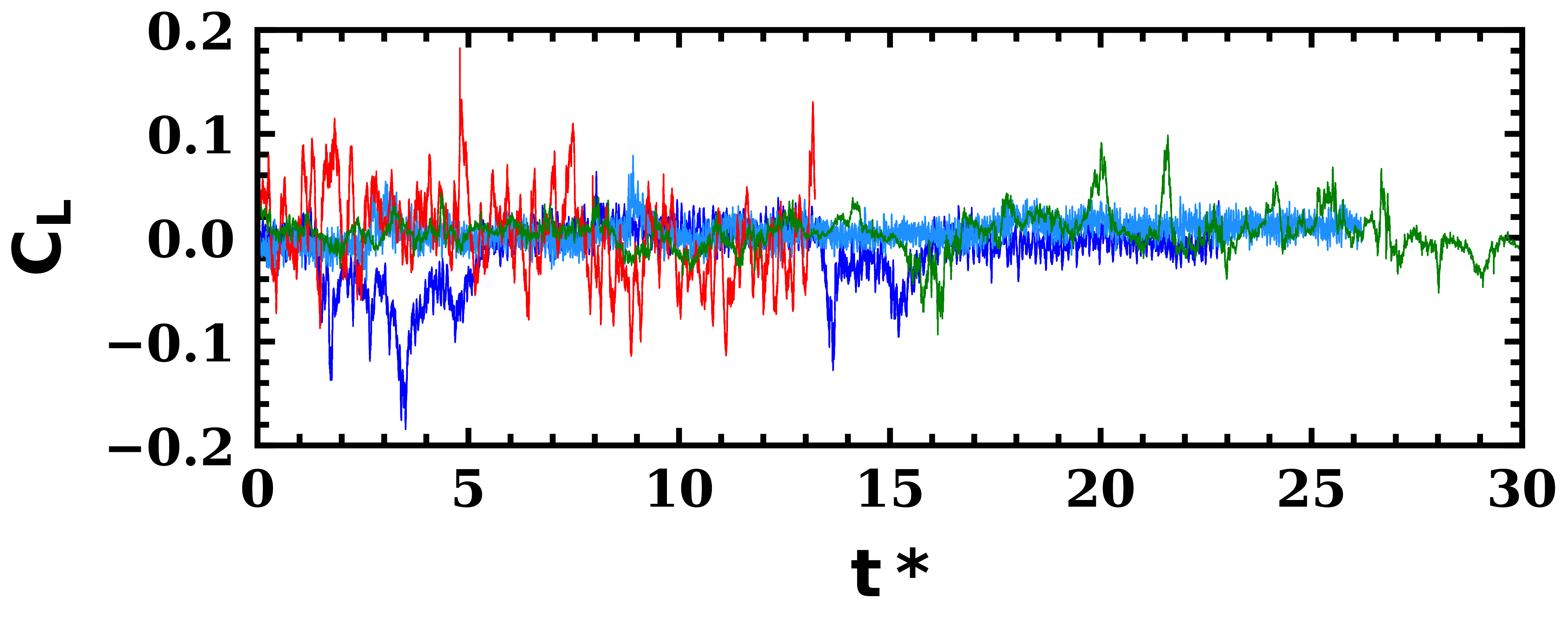}\\
    \captionsetup{justification=justified, singlelinecheck=false,width=\textwidth}
    \caption{Load history on the Sphere. (a) Drag Force ($C_D$), (b) Side Force ($C_S$), and (c) Lift Force ($C_L$). The black line corresponds to the mean drag coefficient in SP.}
    \label{fig:Load_History}
\end{figure}
The pressure distribution, the skin-friction distribution, and the location of flow separation around the sphere are significantly altered owing to the injection of air into the domain. The aggregate effect of the new, altered pressure and skin-friction distributions can be understood by studying the loads acting on the sphere. The force coefficient vector ($\boldsymbol{C_{F}}$), $\boldsymbol{C_{F}} = C_{D}\boldsymbol{i} + C_{L}\boldsymbol{j}+ C_{S}\boldsymbol{k}$; $C_{D}$, $C_{S}$ and $C_{L}$ are the drag, side and lift coefficients.
$\boldsymbol{C_{F}}$ around the sphere is normalized as 
\begin{equation}
    \boldsymbol{C_{F}} = \frac{\boldsymbol{F}}{\frac{1}{8}\rho_{\infty} U_{\infty}^2 \pi d_c^2},
\end{equation}
where $F$ is the force vector. $\boldsymbol{F} = \boldsymbol{F_{p}} + \boldsymbol{F_{f}}$  , $\boldsymbol{F_{p}}$ and $\boldsymbol{F_{f}}$ are the form and the friction drag respectively. The pressure force and the friction force are computed as 
\begin{equation}
    \boldsymbol{F_{p}} = \sum_{i=1}^{N} (p_{i}-p_{\infty})\boldsymbol{S_{f,i}}, 
\end{equation}
\begin{equation}
    \boldsymbol{F_{f}} = \sum_{i=1}^{N} \mathsfbi{T}_{i} \bcdot \boldsymbol{S_{f,i}},
\end{equation}
where $p_{i}$ and $p_{\infty}$ are local and freestream pressure, $\mathsfbi{T}_{i}$ is the shear stress tensor and $\boldsymbol{S_{f,i}}$ are area vector of face $i$ on the sphere surface, and $N$ is total number of faces on the sphere surface. The time history of force coefficients is shown for different injection cases in the figure \ref{fig:Load_History}. In the mid-injection cases, the forces fluctuate at a smaller amplitude but at a higher frequency. The smaller fluctuations are attributed to puffing, which is limited by air injection at lower ambient pressure and by cavity formation. In FR025, all three force coefficients exhibit large-amplitude oscillations about their means, and their frequency is lower compared to the MD cases. Puffing across the cavity injection patch, in the leading hemisphere, is responsible for these spurious oscillations of loads on the sphere in FR025. The drag history in all stable cavity cases (MD025, MD050, and BK025) exhibits random, large-amplitude downward fluctuations from their mean values. These are due to the splashing of the re-entrant jet onto the sphere's base. In all cases, the lateral force coefficients fluctuate primarily around zero due to axisymmetric injection, and their means are nearly zero.\\

The percentage drag reduction owing to ventilation, based on the mean drag coefficients relative to the SP case, is listed in Table \ref{tab:fl_param}. The mid-injection and back-injection cases show drag reduction, whereas the drag increases for FR025. We achieve $35\%$ and $25\%$ drag reduction for MD025 and MD050, respectively. It is interesting to note that the drag reduction is higher for the low ventilation rate. This is attributed to the frequency of the unsteady re-entrant jet splashing onto the sphere's base. The shorter cavity in MD025 frequently generates a re-entrant jet from the closure, which splashes at the base of the sphere and travels upstream, reducing drag. On the other hand, in MD050,  re-entrant jet splashing onto the sphere's base is less frequent, evident from the drops in drag history in figure \ref{fig:Load_History} (a). For BK025, we find that the drag is significantly reduced by $46\%$, indicating that the air injection location dictates the ventilation rate required to achieve a fixed amount of mean drag reduction after stable cavity formation. In contrast, for FR025, the drag is $\sim 57\%$ higher than the SP case owing to the extensive puffing occurring in the sphere's leading hemisphere. \\

\subsection{Cavity Internal Dynamics}
\subsubsection{Near Wake Mean Velocity Profiles}


\begin{figure}
\centering
    (a) \includegraphics[width=0.75\textwidth]{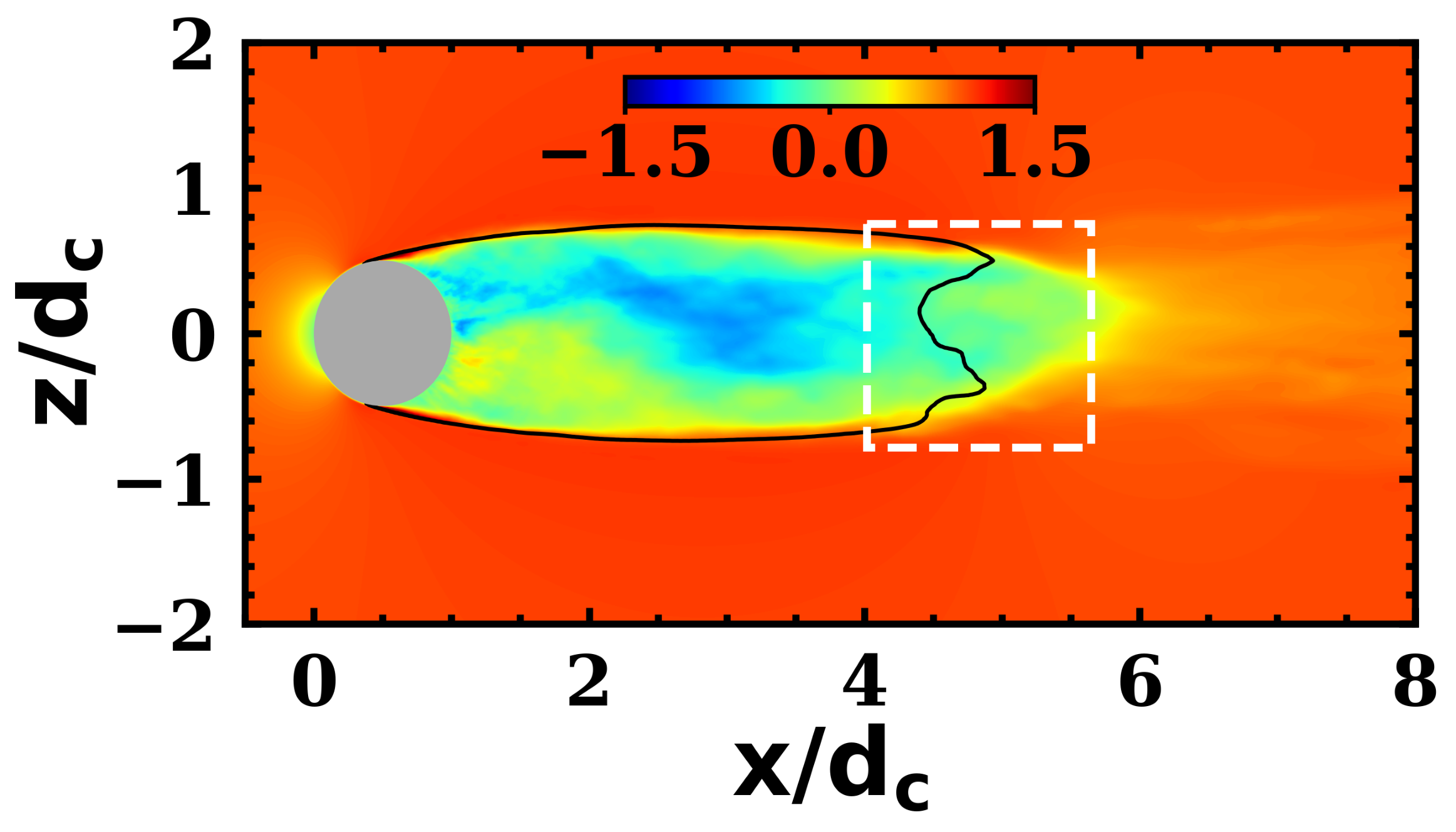} \\
    (b) \includegraphics[width=0.25\linewidth]{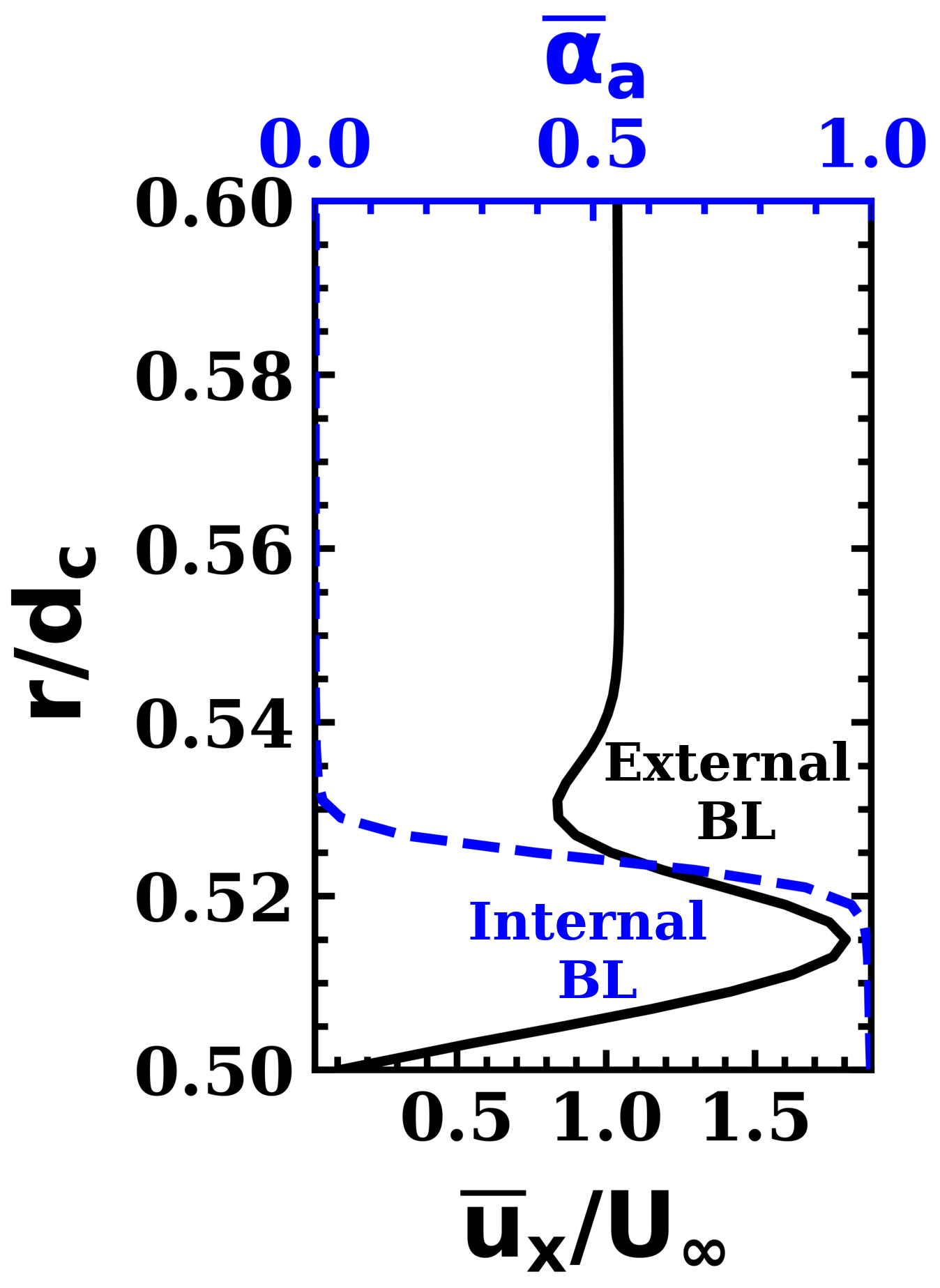}
    (c) \includegraphics[width=0.25\linewidth]{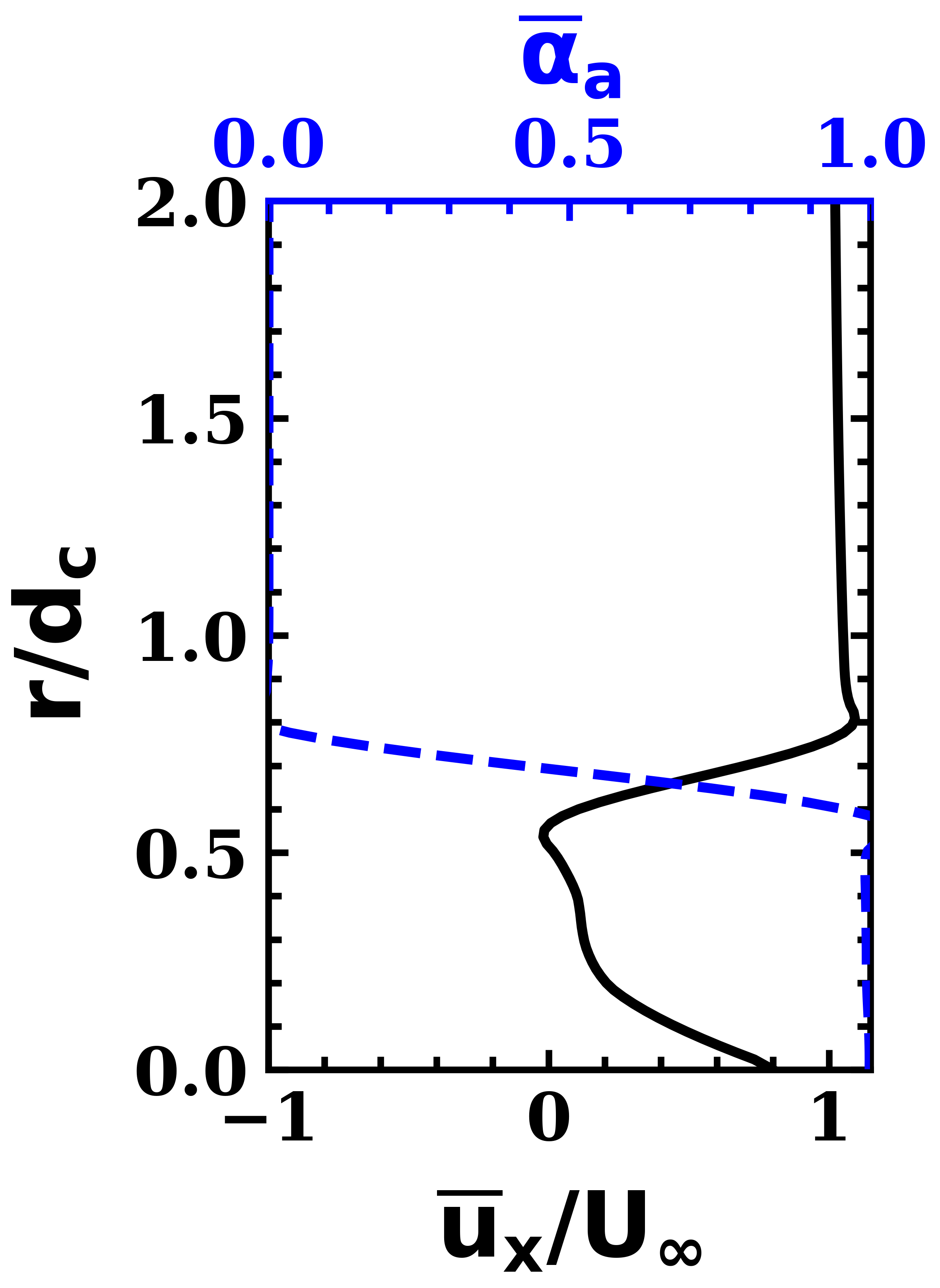}
    (d) \includegraphics[width=0.25\linewidth]{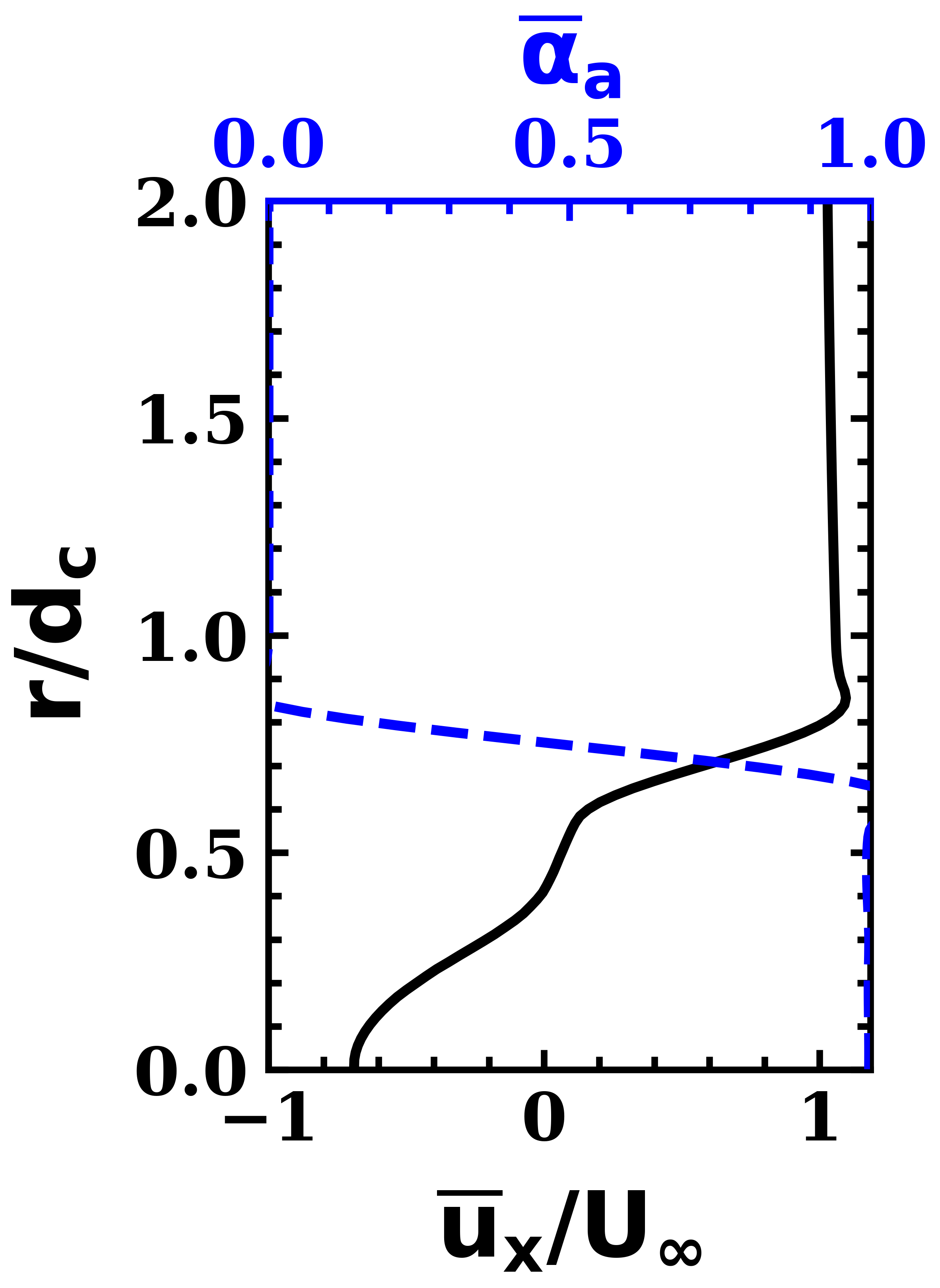} \\
    (e) \includegraphics[width=0.25\linewidth]{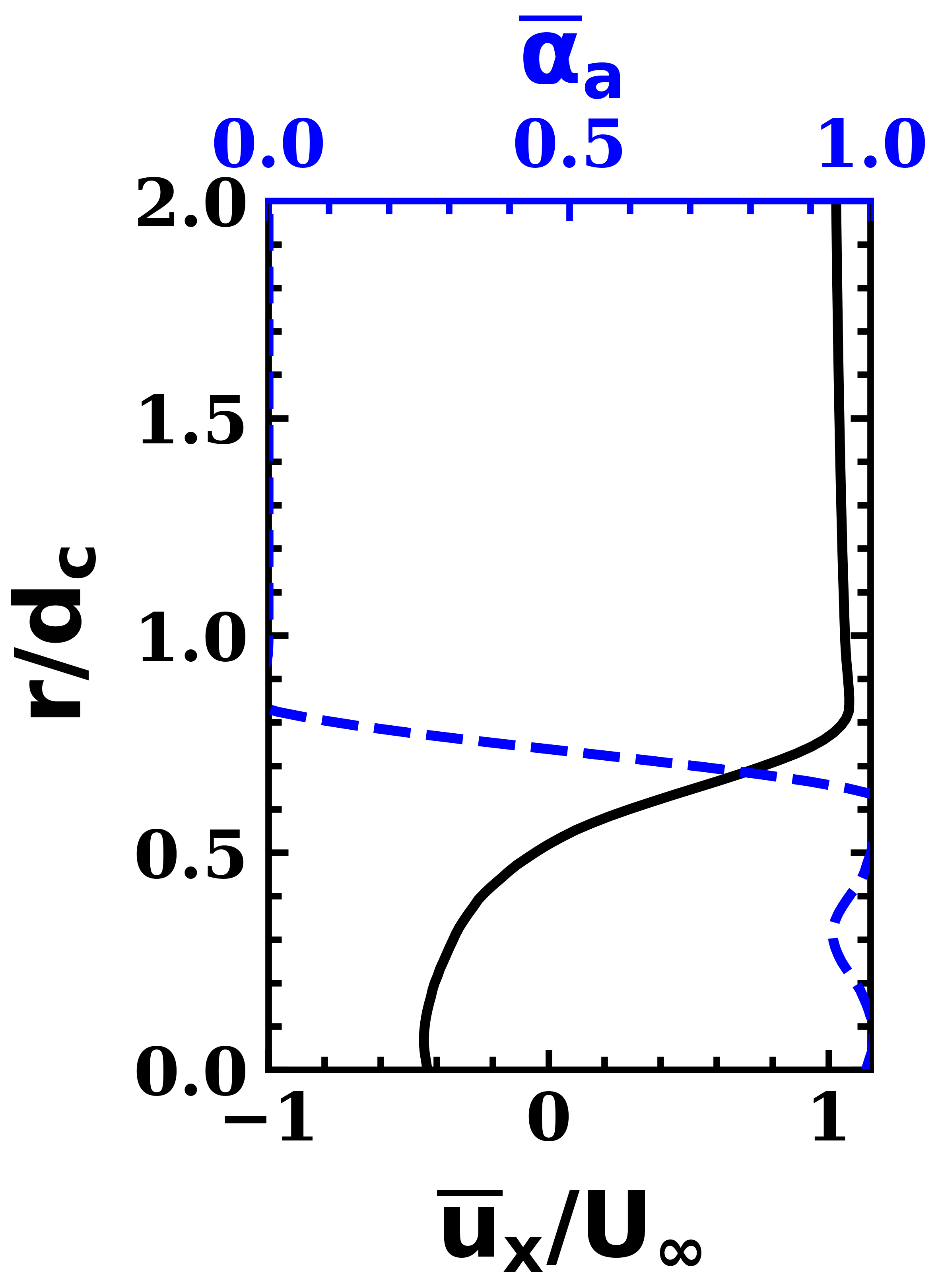}
    (f) \includegraphics[width=0.25\linewidth]{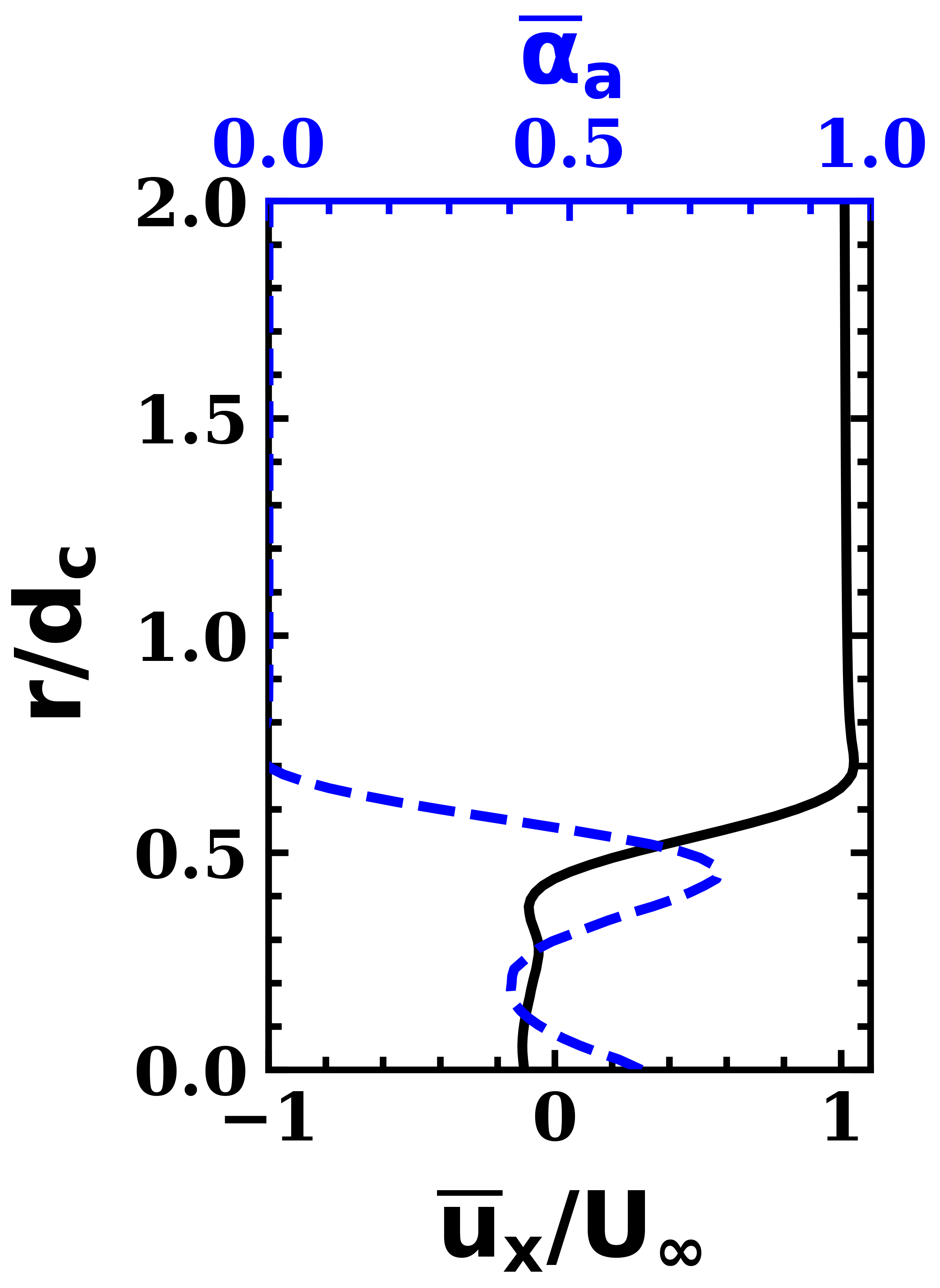}
    \captionsetup{justification=justified, singlelinecheck=false,width=\textwidth}
    \caption{Near wake velocity profiles of MD025. (a) Mean axial velocity contour on the $x-z$ plane plotted with volume fraction isocurve. (b-f) Horizontal velocity profiles in the wake: (b) $x/d_c=0.5$, (c) $x/d_c=1.5$, (d) $x/d_c=2.5$, (e) $x/d_c=3.75$, and (f) $x/d_c=5$.}
    \label{fig:MP_MD025_Wake_Velo}
\end{figure}

\begin{figure}
    \centering
     (a) \includegraphics[width=0.75\linewidth]{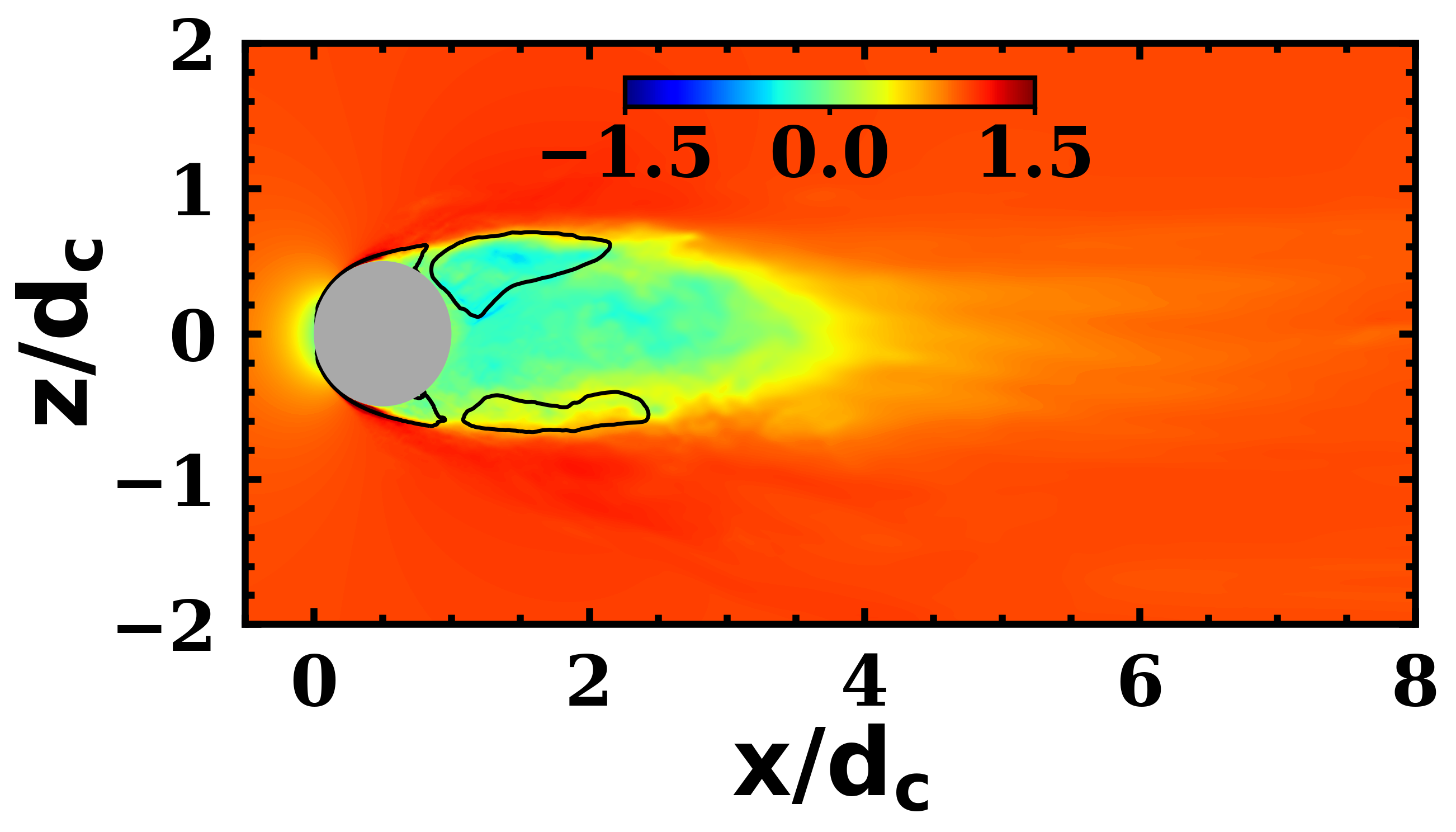}\\
     (b) \includegraphics[width=0.25\linewidth]{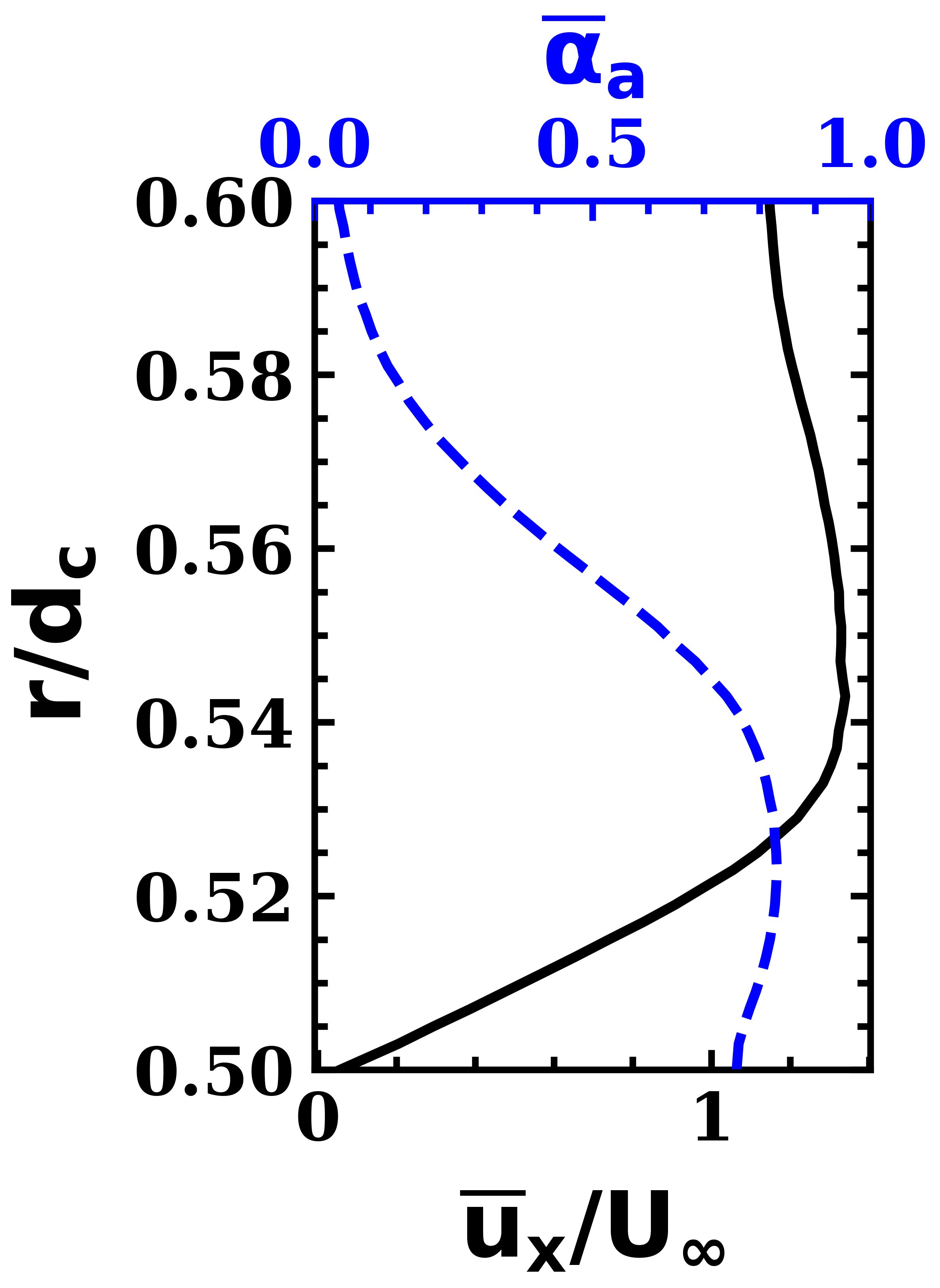}
     (c) \includegraphics[width=0.25\linewidth]{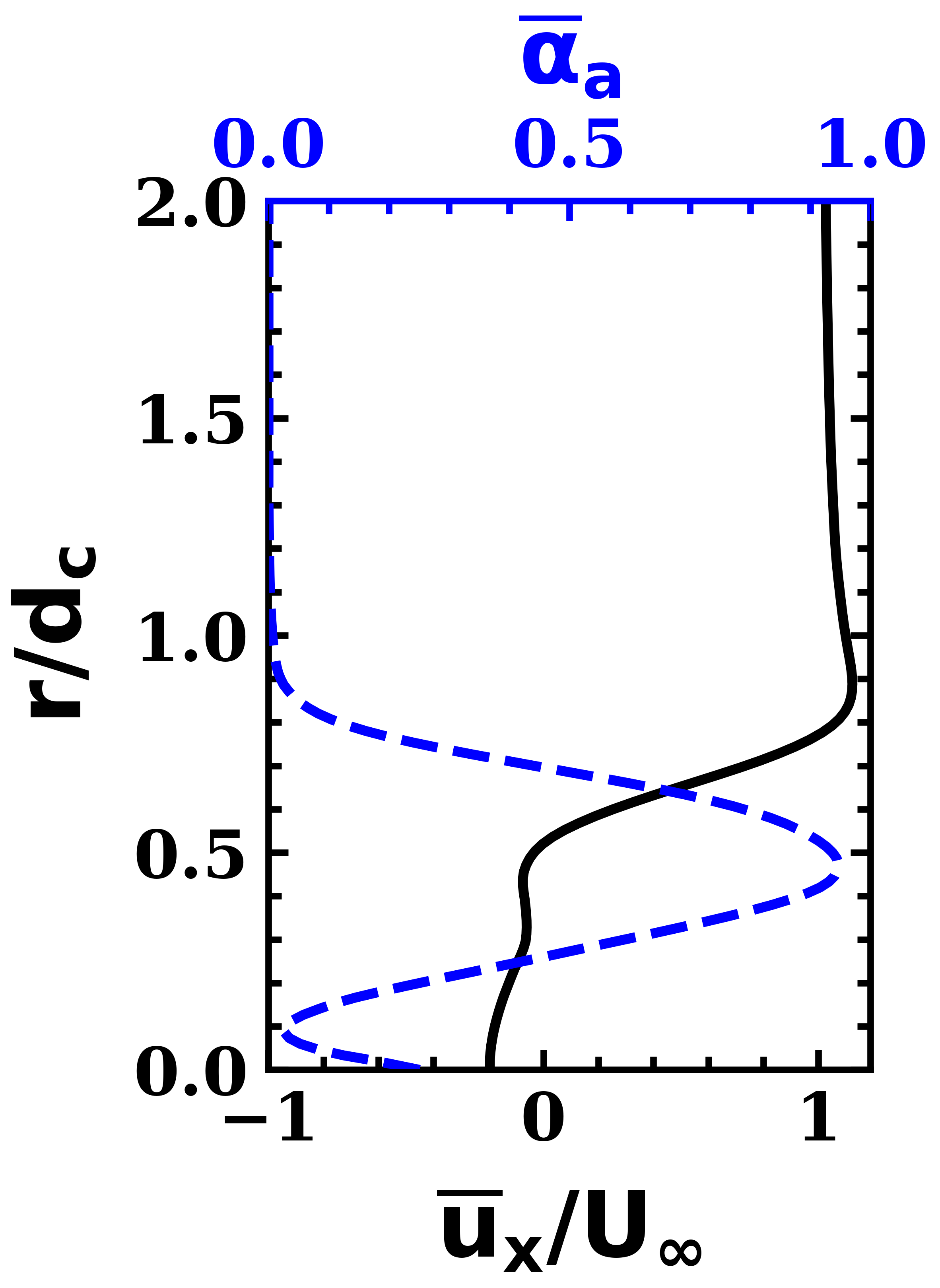}
     (d) \includegraphics[width=0.25\linewidth]{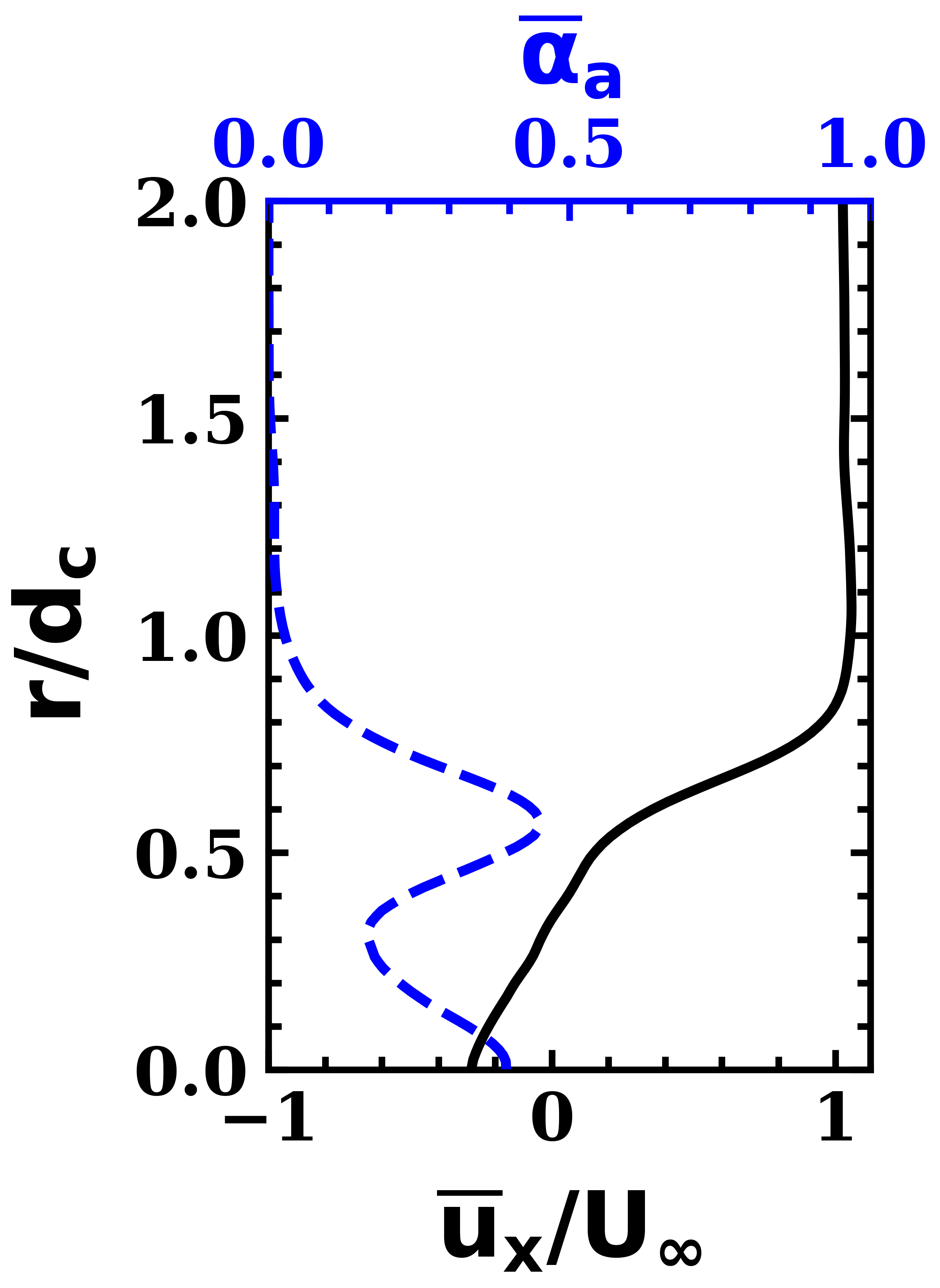}
 
    \captionsetup{justification=justified, singlelinecheck=false,width=\textwidth}
    \caption{Near wake velocity profiles of FR025. (a) Mean axial velocity contour on the $x-z$ plane plotted with volume fraction isocurve. (b-d) Horizontal velocity profiles in the wake: (b) $x/d_c=0.5$, (c) $x/d_c=1.5$, and (d) $x/d_c=2.5$. 
    }
    \label{fig:MP_FR025_Wake_Velo}
\end{figure}

\begin{figure}
 \begin{center}
     (a) \includegraphics[width=0.75\textwidth]{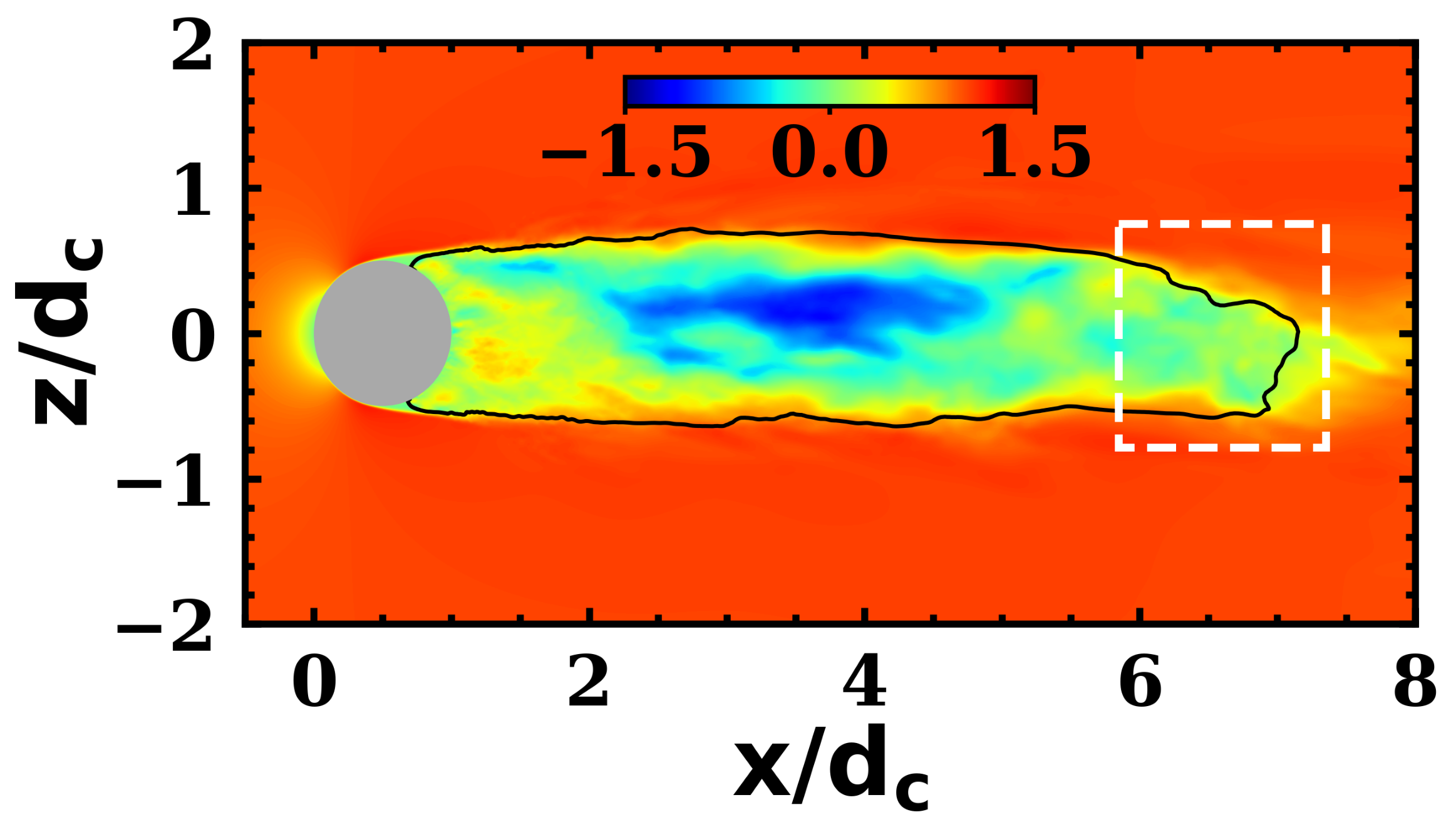}\\
 \end{center}    
    \hspace{-0.75in} (b) \includegraphics[width=0.25\textwidth]{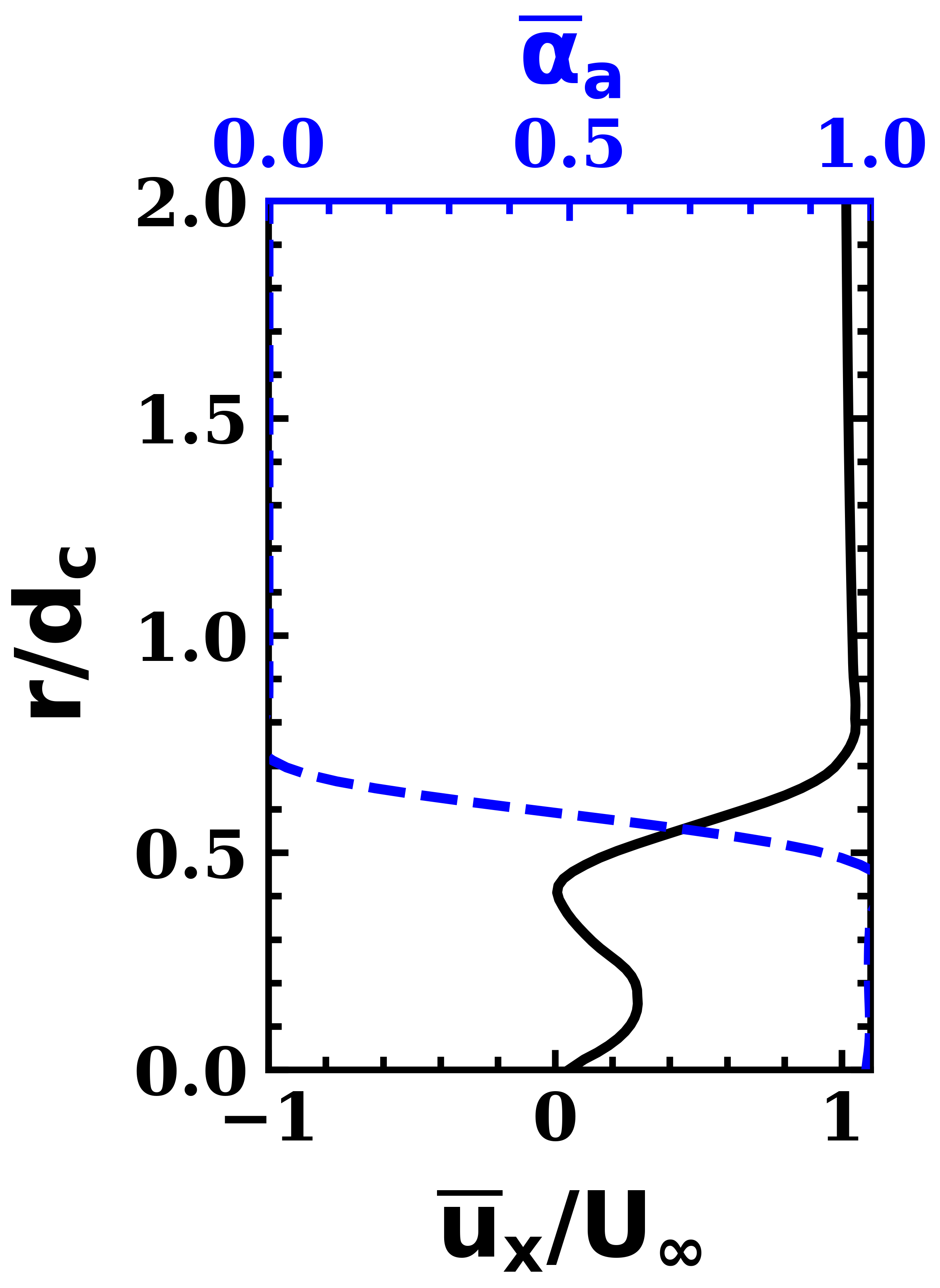}
     (c) \includegraphics[width=0.25\textwidth]{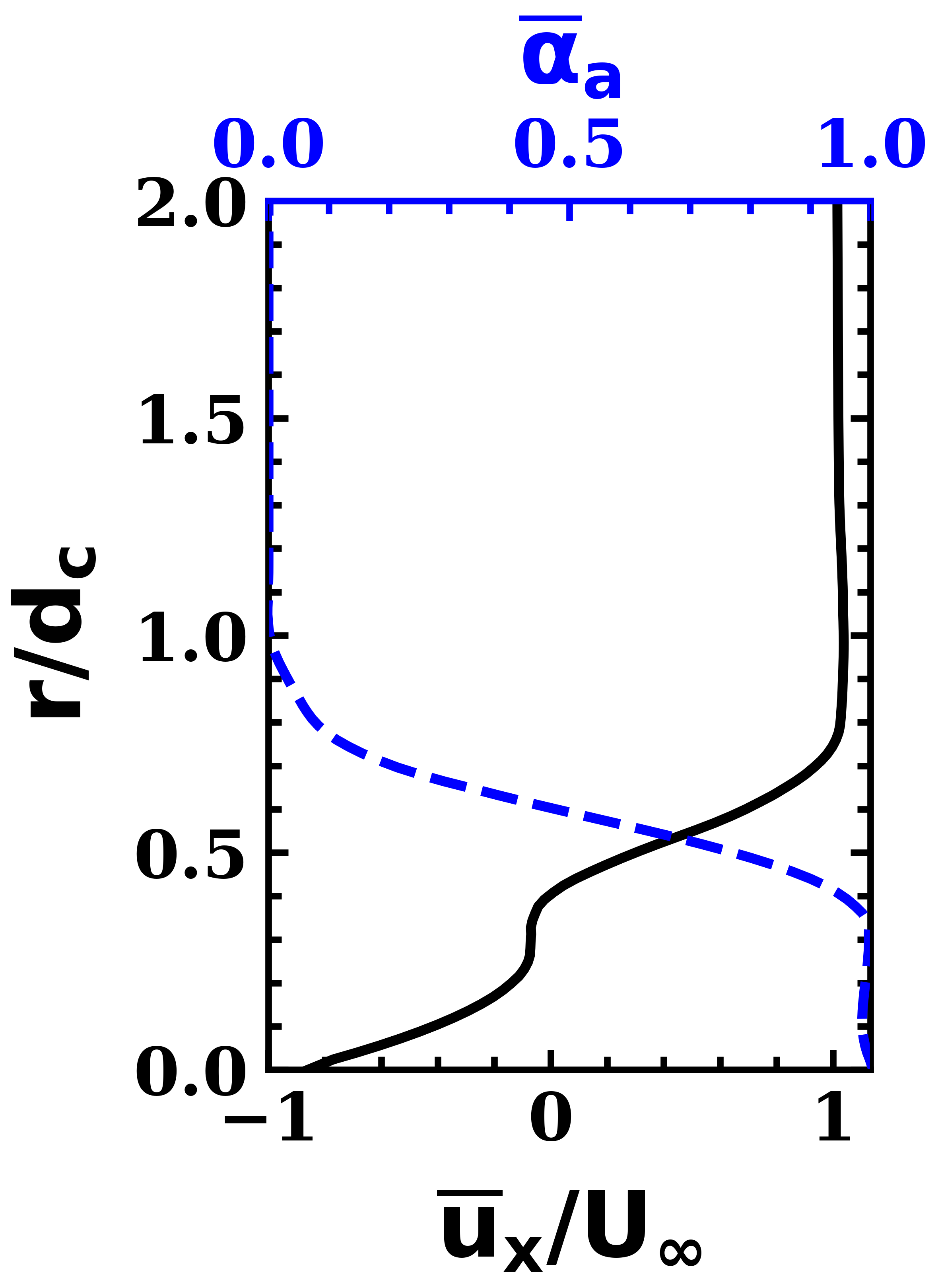}
     (d) \includegraphics[width=0.25\textwidth]{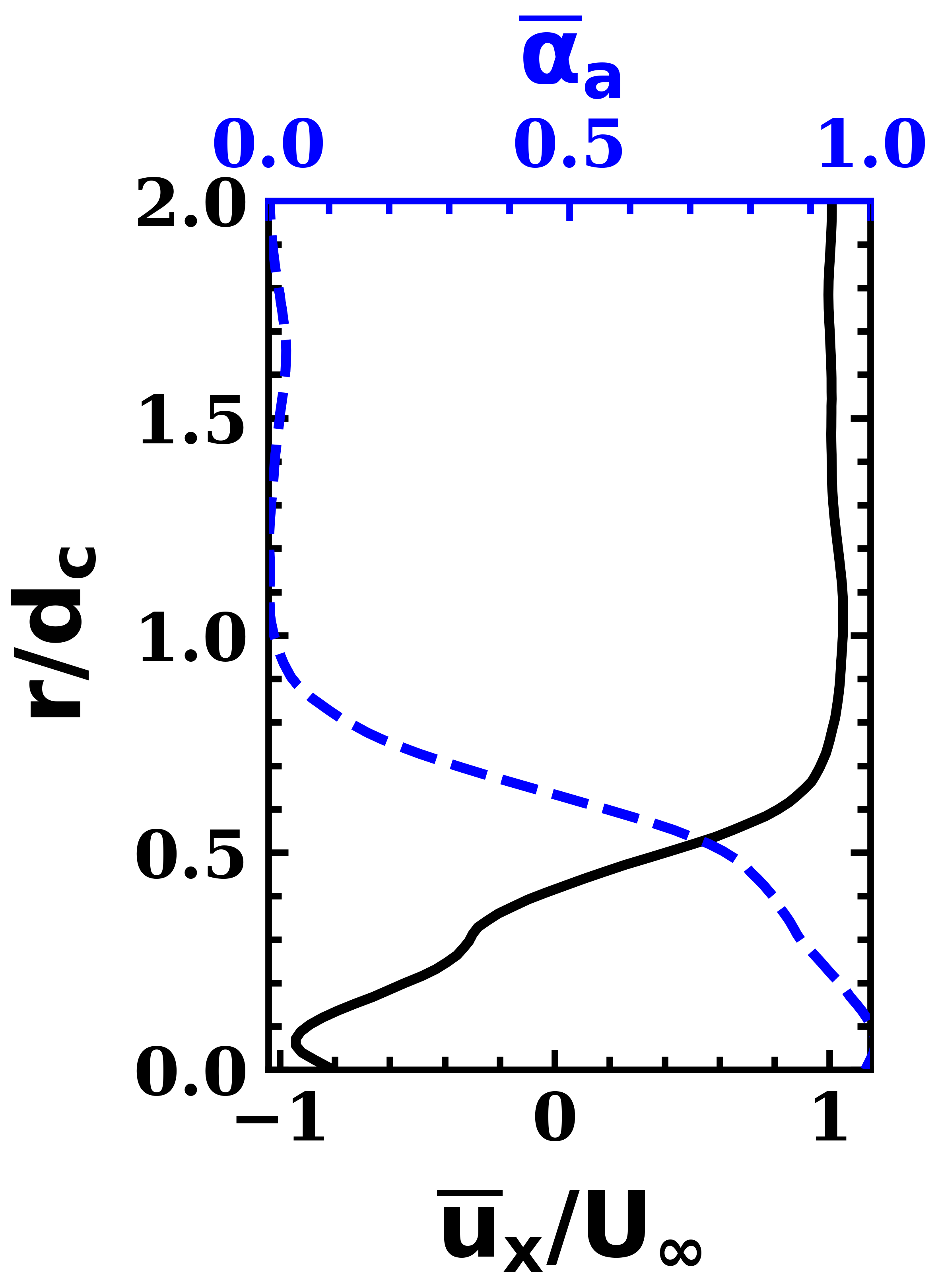}
     (e) \includegraphics[width=0.25\textwidth]{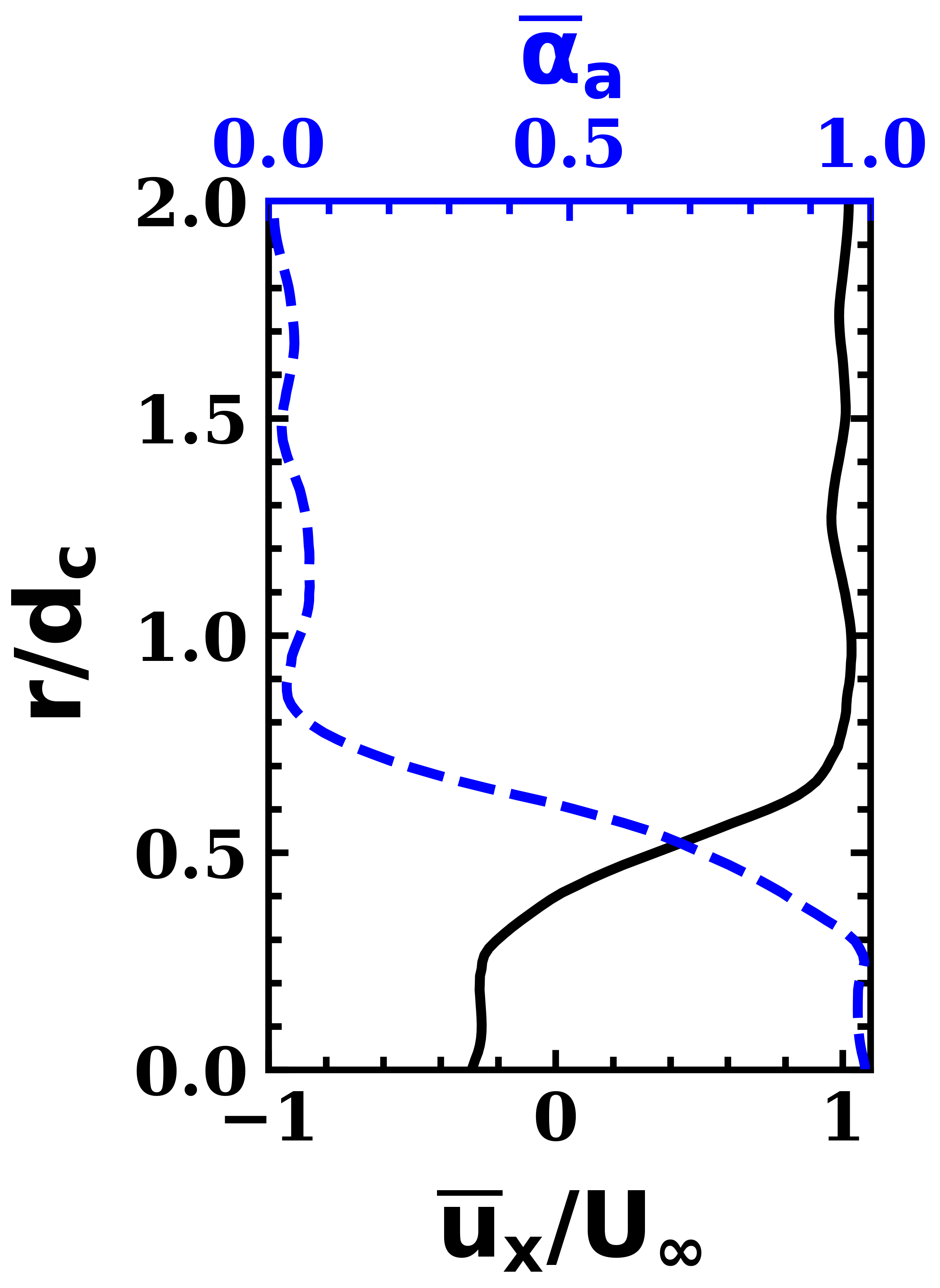}
    \captionsetup{justification=justified, singlelinecheck=false,width=\textwidth}
    \caption{Near wake velocity profiles of BK025. (a) Mean axial velocity contour on the $x-z$ plane plotted with volume fraction isocurve. (b-f) Horizontal velocity profiles in the wake: (b) $x/d_c=1.5$, (c) $x/d_c=2.5$, (d) $x/d_c=3.75$, and (e) $x/d_c=5$.}
    \label{fig:MP_BK025_Wake_Velo}
\end{figure}
Studying the internal dynamics of a cavity is key to understanding the airflow within the cavity and its evacuation to the outside \citep{Spurk_2002, Wu_2019}. In the present simulation, since the gravitational effects are excluded, the closure in all cases is a re-entrant jet closure. Figure \ref{fig:MP_MD025_Wake_Velo} to \ref{fig:MP_BK025_Wake_Velo} plots the time-averaged axial velocity contours with velocity and air's volume fraction profiles at different axial locations for MD050, FR025, and BK025, respectively. We marked the closure region in MD025 (figure \ref{fig:MP_MD025_Wake_Velo} (a)) and BK025 (figure \ref{fig:MP_BK025_Wake_Velo} (a)) with a white box. Figure \ref{fig:MP_MD025_Wake_Velo} (b) confirms the presence of two boundary layers across the cavity interface, an internal boundary formed by gas inside the interface and an external boundary formed by water on the outer surface of the cavity. 
Near the leading edge, the injected air is directly entrained along the cavity's internal surface. This region is referred to as the ventilation influence region by \cite{Wu_2019}. The cavity detaching at the leading edge of the injection patch in mid-injection cases results in strong entrainment of air along the cavity interface, leading to higher air entrainment velocities. Ventilation region is clearly visible in MD025 and MD050 as seen in figure \ref{fig:MP_Cav_Lead_Edge} (a,b). The cavity diameter and length are $1.485d_c$ and $4.055d_c$ in MD025. With increasing ventilation rate, the cavity diameter and length increase to $1.730d_c$ and $7.625d_c$, respectively, in MD050. The diameter of the cavity in MD025 gradually increases along the axial direction from the sphere and reaches a maximum at $x/d_c \sim 2.5$, which is halfway along the cavity. After $x/d_c = 2.5$, the diameter gradually decreases, similar to the observation by \cite{Wu_2019}. There is a strong reverse flow (figure \ref{fig:MP_MD025_Wake_Velo} (c,d,e)) of air along the central portion of the cavity, which is decelerated as it travels near the sphere due to its blockage. \\

In figure \ref{fig:MP_MD025_Wake_Velo} (a), the cavity is concaved around the central axis at the closure; this is due to the inward flow of water from the wake into the cavity, forming the re-entrant jet \citep{Liu_2023}. The volume fraction profile at the closure, figure \ref{fig:MP_MD025_Wake_Velo} (f), shows a peak $y/d_c = 0.495$ due to the concavity of the cavity. The injection position of MD050 is similar to that of MD025, but with a higher ventilation flux. Since the injection location is the same, the cavity dynamics are similar to those of MD025; hence, the results are not presented. \\

Figure \ref{fig:MP_FR025_Wake_Velo} (a) shows a very thin layer iso-contour of air around the leading hemisphere formed due to puffing.  Figure \ref{fig:MP_FR025_Wake_Velo} (b) plots the $\overline{\alpha}_a$ and $\overline{u}_x/U_{\infty}$ within this layer.
The extensive puffing in FR025 results in a continuous sweeping of a bubbly mixture of air and water along the sphere’s surface. Air volume fraction profile, which never reaches unity, in \ref{fig:MP_FR025_Wake_Velo} (b) confirms the presence of the bubbly mixture, and the corresponding velocity profile reaching a maximum of $\sim 1.35U_{\infty}$ corroborates sweeping. Downstream to the sphere, the bubbly mixture is entrained by the shear layer, which is evident from the peaks in $\overline{\alpha}_a$ profile around $r/d_c = 1.5$ (figure \ref{fig:MP_FR025_Wake_Velo} (c)). As the entrained air moves farther downstream, it breaks into bubbles and is advected into the wake. Along the central region near the base of the sphere, up to $x/d_c=2.5$ (figure \ref{fig:MP_FR025_Wake_Velo} (d)), there is a re-entrant jet that majorly entrains water. From the abscissa in the velocity profiles, the re-entrant jet flow is not as strong as in the stable cavity cases.\\

Figure \ref{fig:MP_BK025_Wake_Velo} (a) shows that the ventilation influence region is absent in the back-injection case, as the injection patch is located away from the cavity wall. The cavity diameter ($ 1.362d_c$) is smaller compared to the mid-injection cases (table \ref{tab:fl_param}), and the increase in diameter along the length of the cavity is more gradual owing to the lower interface velocity of the cavity compared to MD025. The cavity length of $5.75d_c$ is higher to the cavity length in MD025. The larger cavity is attributed to the higher magnitude of injection velocity in BK025 compared to MD025 despite the same ventilation rate. The internal dynamics of the cavity in BK025 are similar to the MD025 case, with a reverse flow inside the cavity blocked by the sphere. The reverse flow is evident from the abscissa of the axial velocity plots in figures \ref{fig:MP_BK025_Wake_Velo} (c,d,e). \\

\subsubsection{Re-entrant jet within the cavities} \label{sec:splash}
\begin{figure}    
    \begin{subfigure}[b]{0.5\textwidth}
     (a)\includegraphics[width=1.0\textwidth]{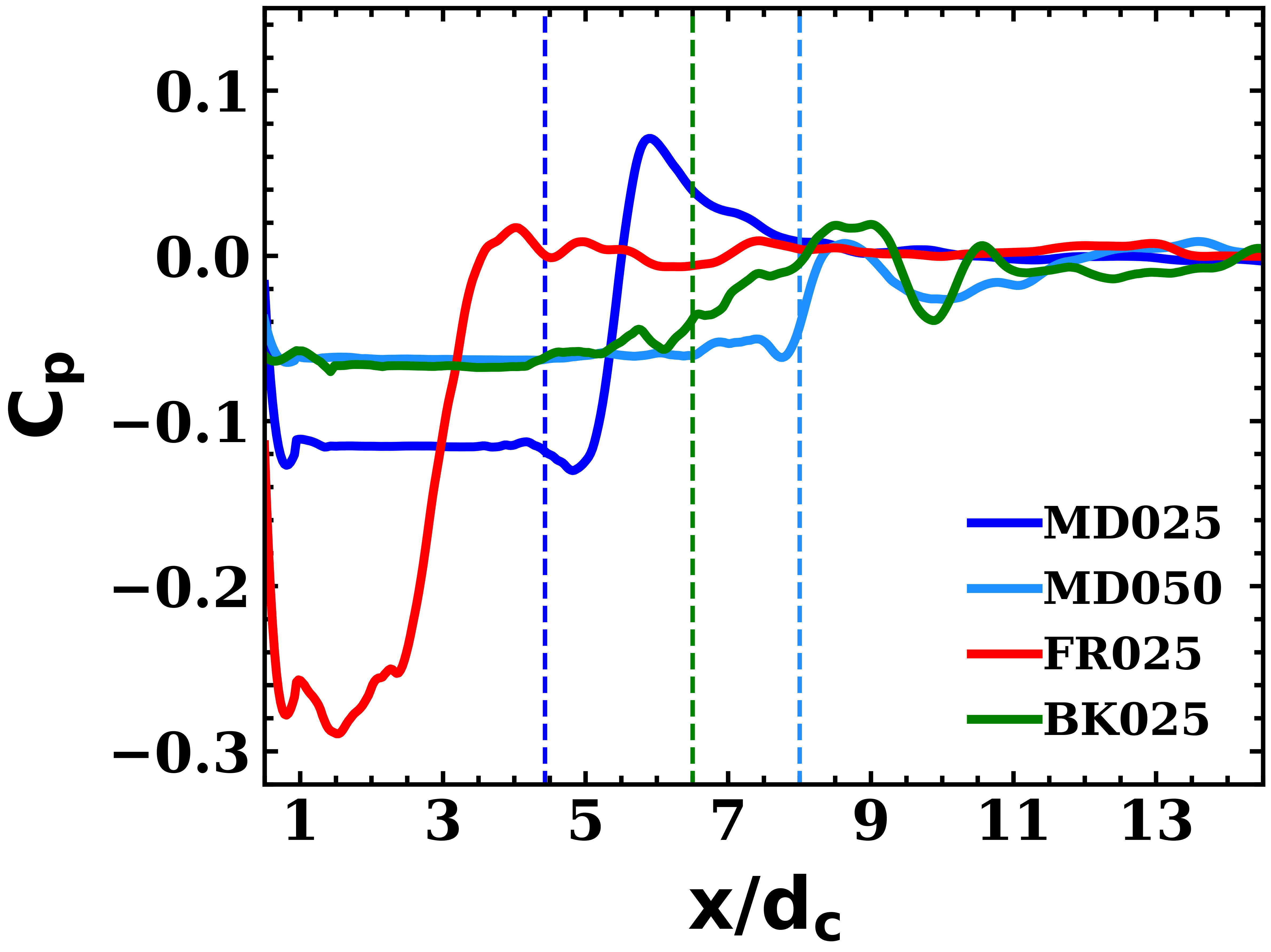}
    \end{subfigure}
    \begin{subfigure}[b]{0.5\textwidth}
     (b)\includegraphics[width=1.0\textwidth]{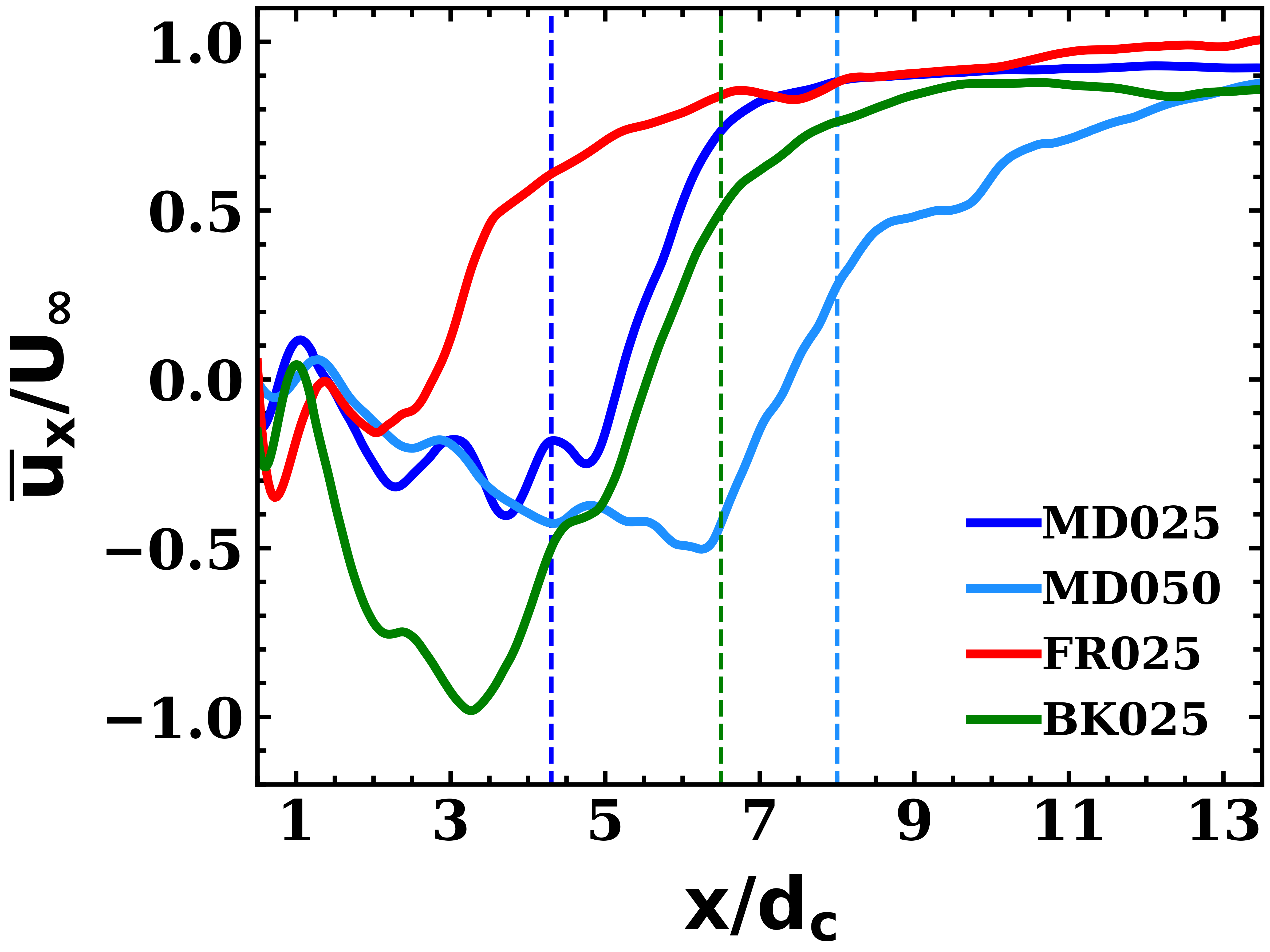}
    \end{subfigure}
    \captionsetup{justification=justified, singlelinecheck=false,width=\textwidth}
    \caption{Time averaged pressure coefficient ($C_p$) and mean horizontal velocity ($\overline{u}_x/U_{\infty}$) distribution along the wake central axis for different injection cases. The colored dashed line marks the respective cavity closure location.}
    \label{fig:MP_Axial_Cp_Ux}
\end{figure}

In all stably attached cavity cases, the closure type is a re-entrant jet. The re-entrant jet entraining water originates at the cavity closure region, travels upstream, and interacts with the sphere, affecting its surface statistics. In figure \ref{fig:MP_Axial_Cp_Ux}, we plot the time-and-azimuthally averaged pressure coefficient and the horizontal velocity along the wake central axis for different cases to understand the dynamics of a re-entrant jet. The dashed lines show the cavity closure locations of the different cases. The pressure coefficient (figure \ref{fig:MP_Axial_Cp_Ux} (a)) remains nearly uniform within the cavity and exhibits a steep adverse pressure gradient near the closure \citep{Spurk_2002, Wu_1972}. There is an increase in the pressure coefficient near the base of the sphere due to the splashing of the re-entrant jet. The adverse pressure gradient at the cavity closure is higher in MD025 than in MD050 and BK025. Similarly, \cite{Cao_2017} found that the closure adverse pressure gradient decreases with an increase in cavity length. The axial velocity reaches a minimum near the cavity closure and increases towards upstream. The decrease in velocity of the re-entrant jet is due to blockage by the sphere (figure \ref{fig:MP_Axial_Cp_Ux} (b)). A re-entrant jet also forms in FR025, as it is evident from the negative axial velocity up to a distance of $2D$ from the sphere's base (figure \ref{fig:MP_Axial_Cp_Ux} (b)).\\

The cavitation number ($\sigma_c$) for all the cases is listed in the table \ref{tab:fl_param}. The nearly constant pressure in the wake axis, from figure \ref{fig:MP_Axial_Cp_Ux} (a), is chosen for calculating $\sigma_c$. This is similar to calculating $\sigma_c$ based on the data obtained by placing a hypodermic tube \citep{Karn_2016a, Wu_2019} or from a pressure transducer \citep{Gawandalkar_2025} placed downstream of the body. Cavitation number is maximum in FR025, where there is a bubbly flow, and there is no stable cavity formation. In stable cavity cases, MD025 has the highest cavitation number. Increasing the ventilation coefficient reduces the cavitation number to 0.06 in MD050. The cavitation numbers are nearly identical in BK025 and MD050, despite the ventilation coefficient being doubled. This shows that the extent of cavitation is greater in an injection position more downstream from the stagnation point. In other words, the ventilation demand to form a stable cavity decreases with the downstream shift of the injection patch. \\

During the growth phase of the cavity, the inward unsteady stream of water-entrant jet exited through the rear walls of the cavity before reaching upstream. However, after the formation of a stable cavity,  leakage of the re-entrant jet through the cavity walls reduces as the walls become stronger. Hence, the re-entrant jet formed at the cavity closure entrains water from the closure and travels upstream along the cavity axis. The base of the sphere obstructs this jet of water. When the velocity of the re-entrant jet is relatively low, the water bubbles get attached to the base of the sphere. \\

\begin{figure}
    \begin{subfigure}[b]{0.49\textwidth}
    (a)\includegraphics[width=1.0\textwidth]{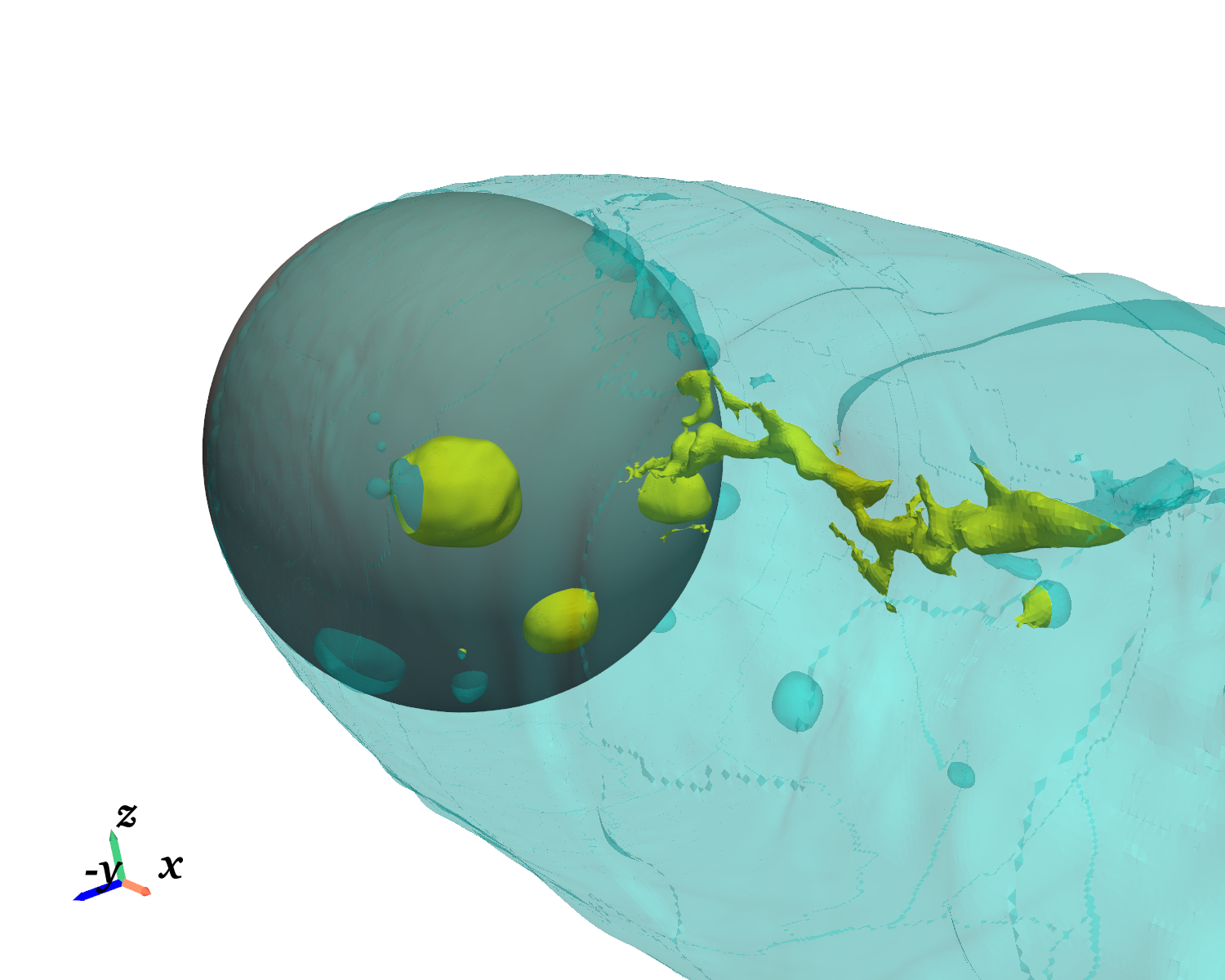}
    \end{subfigure}
    \begin{subfigure}[b]{0.49\textwidth} 
     (b)\includegraphics[width=1.0\textwidth]{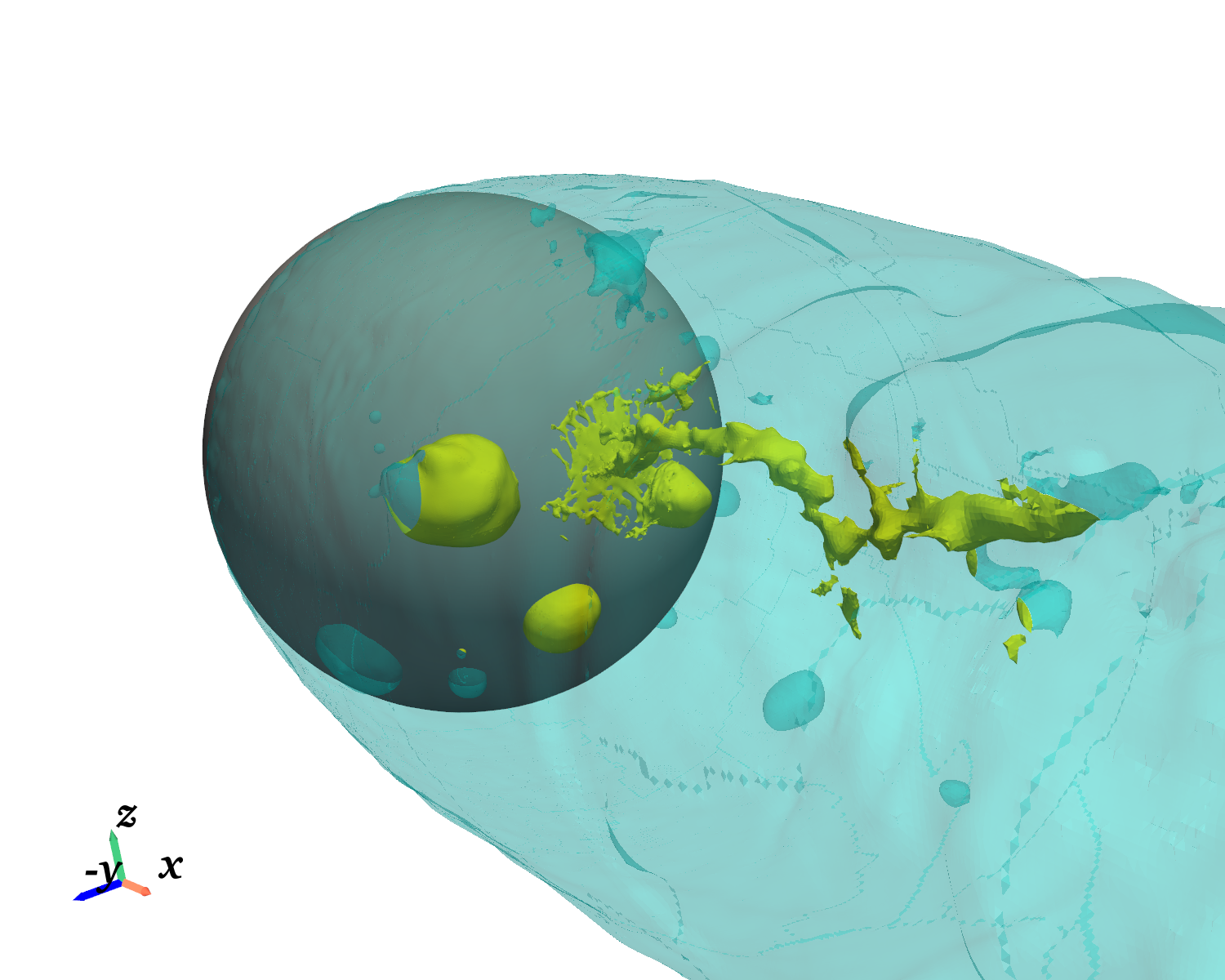}
    \end{subfigure} \\
    \begin{subfigure}[b]{0.49\textwidth}
    (c)\includegraphics[width=1.0\textwidth]{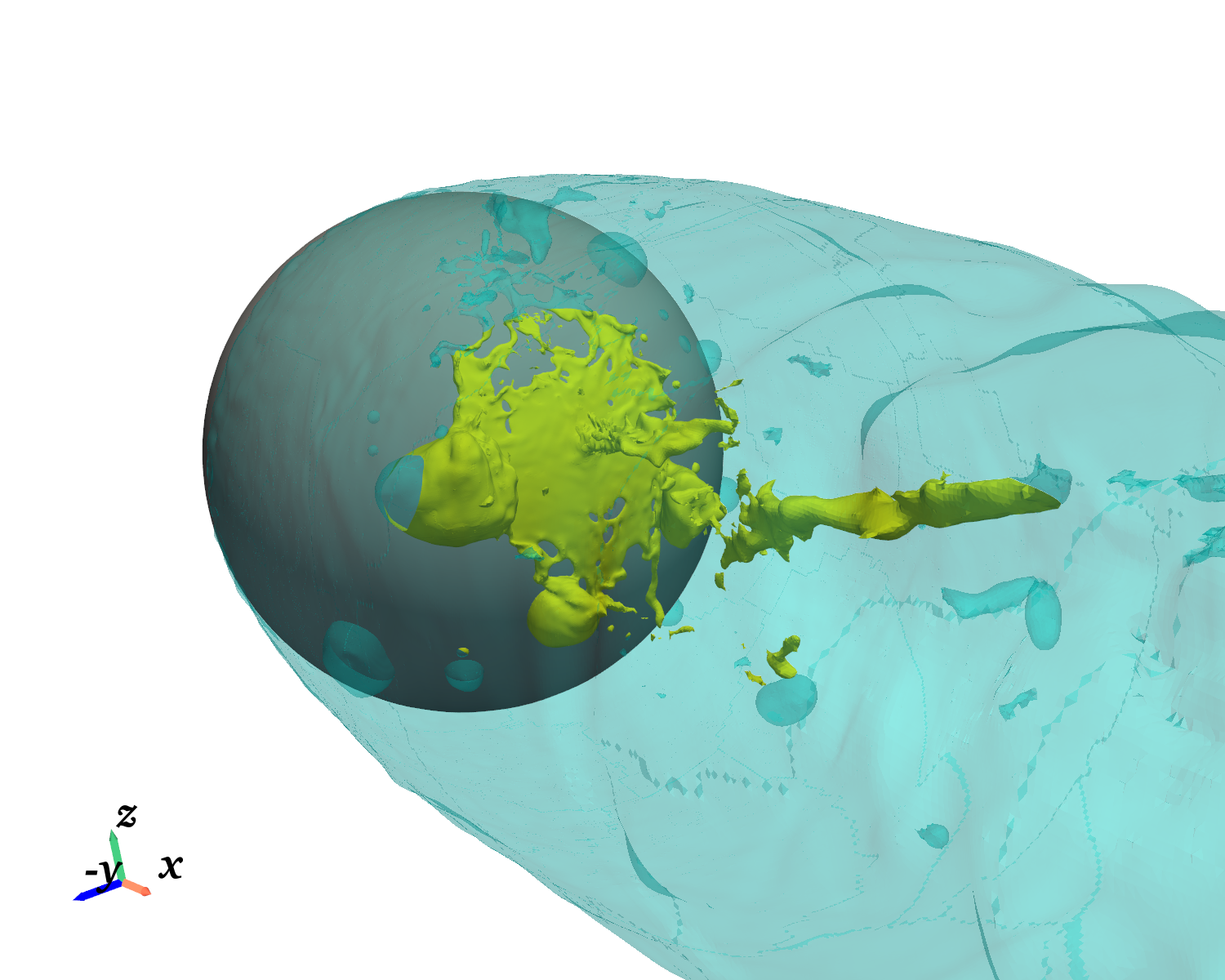}
    \end{subfigure}
    \begin{subfigure}[b]{0.49\textwidth}
    (d)\includegraphics[width=1.0\textwidth]{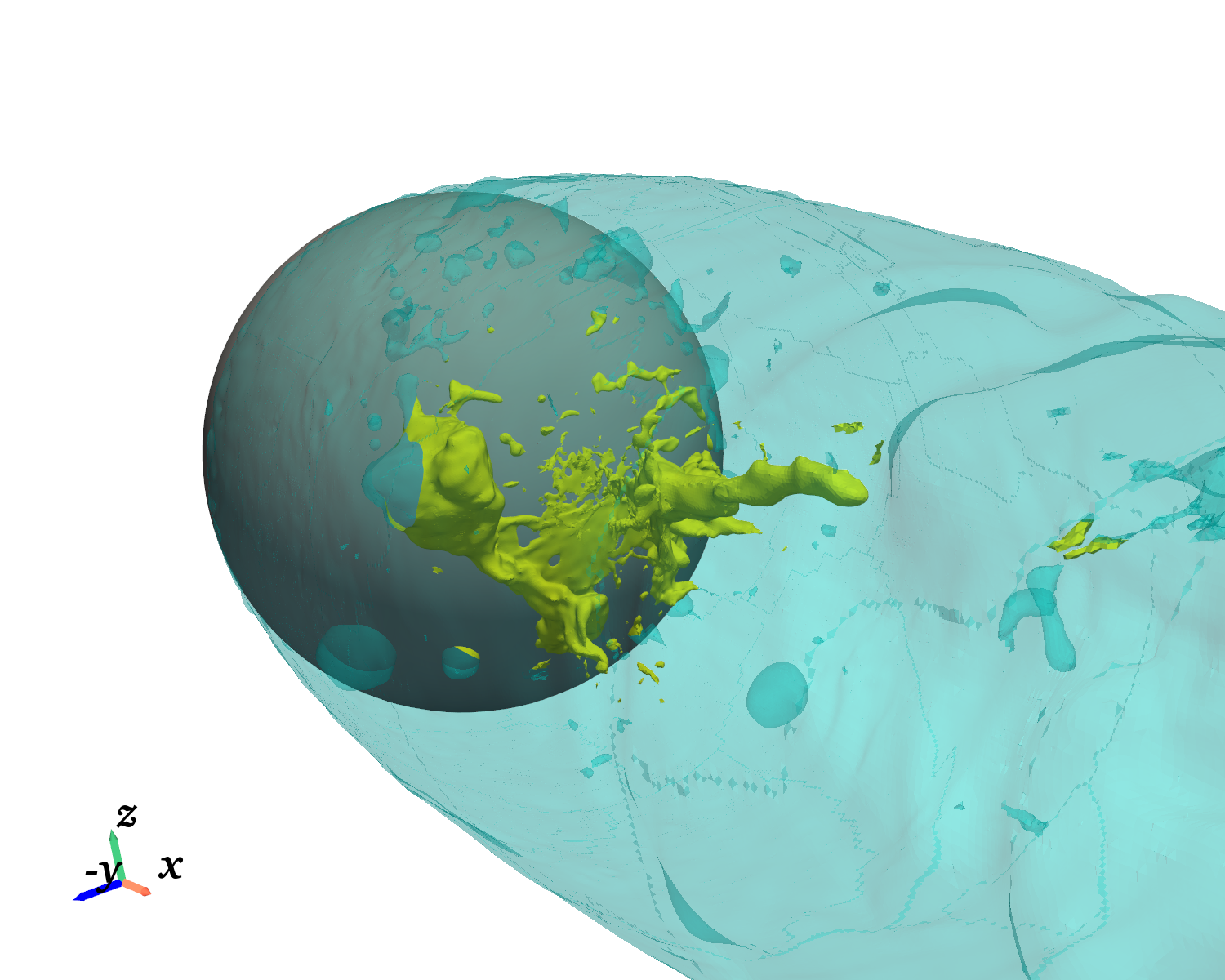}
    \end{subfigure} 
    \captionsetup{justification=justified, singlelinecheck=false,width=\textwidth}
    \caption{Splashing of re-entrant jet onto sphere's base for MD025. The air volume fraction isocontour is split by a cylindrical clip with radius $r=0.3d_c$. The outer clip is colored by a translucent cyan, and the inner clip is colored by opaque yellow to show internal splashing of the re-entrant jet. }
    \label{fig:MP_Splash}
\end{figure}

In contrast, at higher velocities, the impinging jet splashes onto the sphere's trailing surface, breaks into small drops that penetrate the cavity walls, and generates local disturbances. This splashing effect was observed to occur in all stable cavity cases. One such event of a reentrant jet splashing onto the sphere's base is shown in figure \ref{fig:MP_Splash}. The supplementary movie S5 demonstrates this phenomenon. Figure \ref{fig:MP_Splash} plots the air volume fraction isocontour (translucent cyan color). To highlight the flow within the cavity, the isocontour is colored with opaque yellow up to a radial location of $r/d_c \leq 0.3$ from the center axis. Figure \ref{fig:MP_Splash} (a) shows the re-entrant jet containing water moving upstream within the cavity, and figure \ref{fig:MP_Splash} (b) shows the blocking of the upstream-moving jet and the beginning of the splashing event. In the subsequent instant, the span of splashing increases (figure \ref{fig:MP_Splash} (c)), and the water from the splash moves laterally towards the cavity wall. 
The splashed water exits the cavity surface through the wall, as shown in figure \ref{fig:MP_Splash} (d).
\cite{Brennen_1970a} also observed that the roughness in the cavity wall is due to the penetration of the re-entrant jet. The splashing of water depends on the strength of the re-entrant jet at the closure. Splashing also momentarily alters the force coefficients around the sphere. This is evident from the frequent drops in the drag coefficient in stable cavity cases. Similarly, this splashing is responsible for large-scale disturbances in the lateral force coefficients. \\

\section{Summary and conclusions}\label{sec:conc}

We perform the first direct numerical simulation of ventilated cavitation around a sphere at $Re=10,000$, by injecting air from the sphere’s surface into the domain by modifying the boundary conditions for velocity and volume fraction. Unlike the previous numerical investigations, our simulations include the influence of surface tension. The inclusion of surface tension severely restricts the time-step size owing to the requirement of resolving the capillary wave criterion for numerical stability and accurate air-water interface evolution. We have used $\sim 179$ million hexahedral cells within the domain, in accordance with the mesh-resolution study of \cite{Liu_2023}, to resolve all the relevant scales of motion. Our grid resolution near the sphere surface is sufficient to accurately capture pressure and shear stress variations along the sphere's surface for single-phase flow, consistent with the existing literature. The ventilation coefficient ($C_q=0.2,0.4$) and injection position (front, mid, back) are varied, and their effects on the cavity dynamics are studied. The injection locations are chosen based on different flow dynamics around the sphere for the corresponding single-phase flow. In the front-injection case, the water near the injection surface accelerates; in mid-injection cases, it decelerates; and in the back-injection case, the flow has separated from the sphere's surface. This affects the dynamics of cavity separation and the position of cavity detachment. \\

The position of the injection patch strongly affects the leading edge dynamics and the nature of the cavity formed. The front-injection case can only form a bubbly cavity because the injected air is broken into bubbles as it puffs along the leading edge. Stable axisymmetric cavities are formed in the mid-injection and back-injection cases. The cavity surface was rougher in the back-injection case owing to Kelvin-Helmholtz instabilities. Such K-H instabilities were suppressed in the mid-injection case due to a higher interface velocity; instead, the puffing instability was dominant. \\

The cavity in the mid-injection cases detached from the leading edge of the injection patch, irrespective of the injection velocity. The flow separation was delayed due to the momentum imparted by the air injection. This shows that flow separation need not be upstream of a stable attached cavity. In ventilated cavities, the extent of puffing dictates the stability of cavity formation. In the back-injection case, the cavity detached upstream of the injecting patch due to the adverse pressure gradient created by cavity formation. The location of cavity separation was in good agreement with the empirical formula of \cite{Arakeri_1975}. \\

The puffing phenomenon was observed near the cavity leading edge in front-injection and mid-injection cases, where the flow was attached to the sphere. In the front-injection case, puffing spanned the entire injection patch; in the mid-injection cases, it was limited to the leading edge, and in the back-injection case, it was absent. This behavior is attributed to a reduction in the pressure imbalance between the liquid cross-flow and the injected gas's static pressure, as the injection patch was shifted downstream, and to the formation of a stable cavity. The intermittent downstream air supply due to puffing in front-injection prevents cavity formation. In mid-injection cases, puffing generates interfacial instabilities. Similar puffing behavior was observed in vertical gas injection in a horizontal crossflow over a flat plate by \cite{Makiharju_2017, Xiong_2024}. \\

We observed divots near the leading edge of the cavity in the back-injection case. These were observed for higher Reynolds numbers in natural cavitating flows by \cite{Tassin_Leger_1998b, Brandner_2010}. The divots formed at random locations over the circumference, and their formation was less frequent. The divots quickly collapsed due to low liquid jet speeds within them. The collapse of divots results in the shedding of bubbles that remain attached to the sphere due to a recirculation zone just ahead of the cavity. \\

The shear layer in the front-injection case entrains the injected air. Along the central line, a weak re-entrant jet formed. The cavity's internal dynamics were similar in the mid- and back-injection cases, where stable cavities formed. These cavities exhibited three regions, as also observed by \cite{Wu_2019}, namely: i) the ventilation influence region, ii) the internal boundary layer which entrains gas, and iii) the major reverse flow region inside the cavity. There was a strong entrainment of air along the cavity interface in the mid-injection cases, as the cavity detached from the leading edge of the injection patch. Air entrainment increases with increased ventilation.  The sphere blocks the reverse flow moving upstream inside the cavity. The closure mode was a re-entrant jet in all cases due to the absence of gravity. There was no ventilation influence region in the back-injection region, as the injection patch was away from the cavity wall. \\

The load history on the sphere showed that there is no lateral force coefficient, as the air injection is axisymmetric. The mid- and back-injection cases demonstrated a substantial drag reduction.  We obtained $35\%$ and $25\%$ drag reduction in MD025 and MD050, respectively. Interestingly, for the low-ventilation rate case, a higher drag reduction is observed owing to more frequent splashing of the re-entrant jet on the backside of the sphere in MD025 than in MD050. The maximum drag reduction of $46\%$ was obtained in BK025. This indicates that the air-injection location should be beyond the separation point on the sphere's surface to achieve the maximum drag reduction. Such significant drag reduction is consistent with the experimental results of \cite{Chung_2018, Jiang_2019}, where up to $73\%$ reduction in drag is reported owing to the formation of a ventilated cavity. In contrast, the vigorous puffing in the front-injection case yielded a $\sim 56\%$ increase in the drag coefficient. The puffing behavior observed near the leading edge of the cavity had a detrimental effect on the drag force. \\

The present simulation does not include the effect of gravity. Therefore, an obvious extension of this work would be the quantification of the drag reduction and study of the ventilated cavity dynamics and its closure mechanism under the influence of gravity. It will also be interesting to understand the behavior of the ventilated cavity in the presence of density stratification. Moreover, it would be worth exploring whether similar drag reduction could be achieved for other bluff bodies relevant to marine engineering applications. \\

\appendix

\section{Grid independence Study}\label{appA}
\begin{figure}    
    \begin{subfigure}[b]{0.5\textwidth}
     (a)\includegraphics[width=1.0\textwidth]{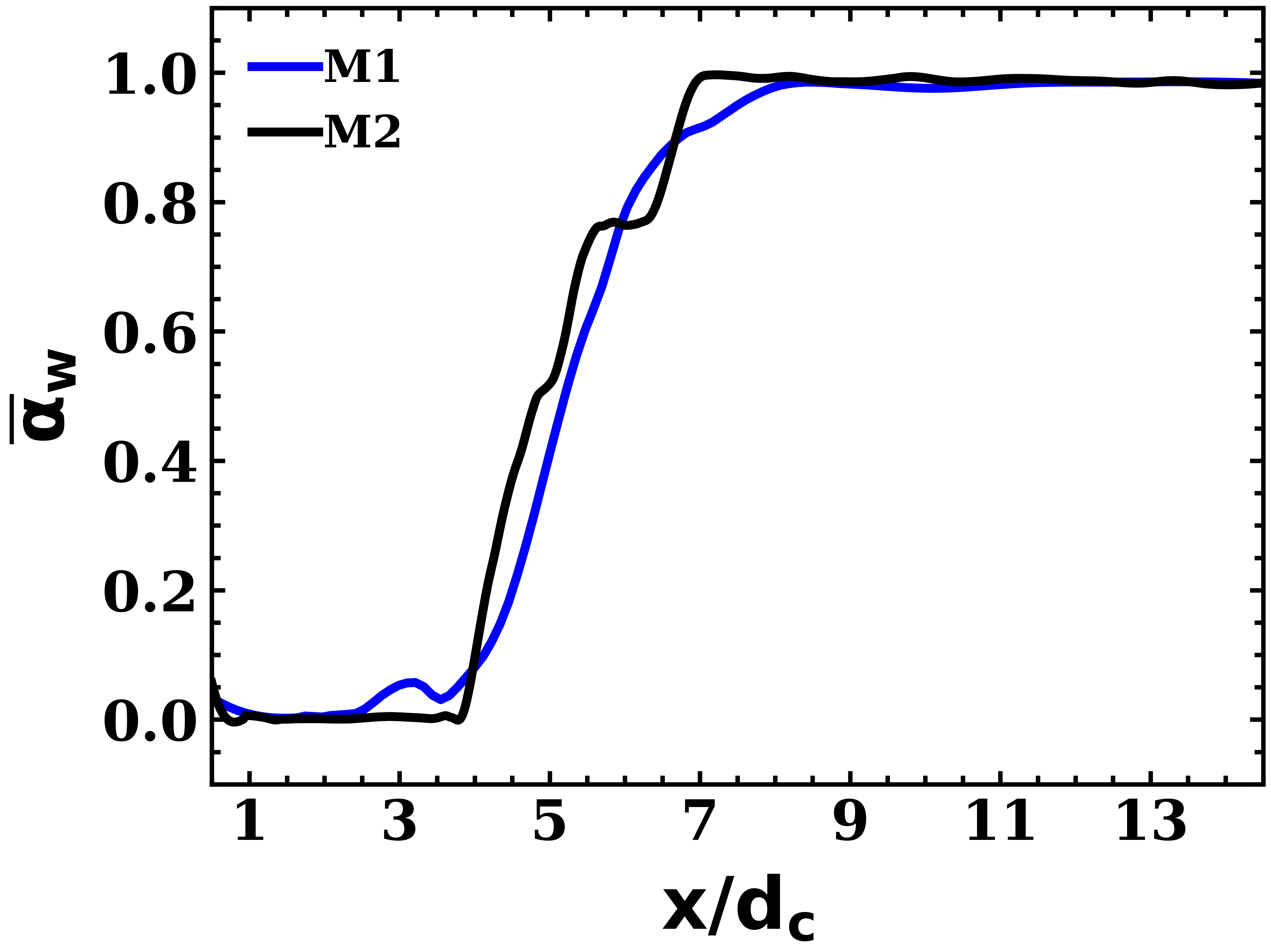}
    \end{subfigure}
    \begin{subfigure}[b]{0.5\textwidth}
     (b)\includegraphics[width=1.0\textwidth]{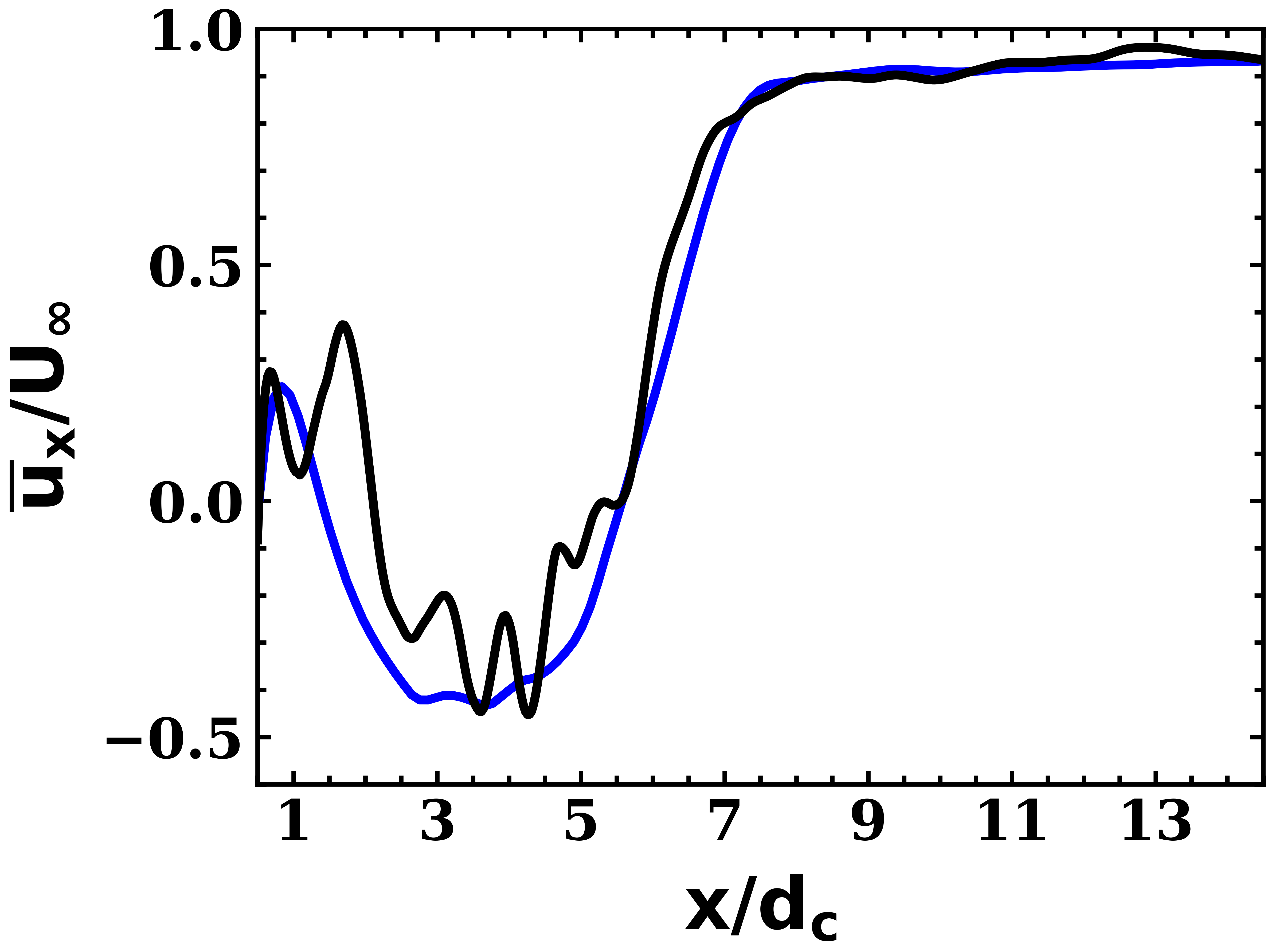}
    \end{subfigure} 
    \captionsetup{justification=justified, singlelinecheck=false,width=\textwidth}
    \caption{Time averaged water volume fraction ($\overline{\alpha}_w$) and mean horizontal velocity ($\overline{u}_x/U_{\infty}$) distribution along the wake central axis. }
    \label{fig:MP_Axial_Alp_Ux}
\end{figure}

To ensure the present solution is independent of the grid resolution, we perform a grid-independence study for case MD025, using two different grids, M1 and M2. M1 is a coarser grid comprised of $\sim 31$M cells, and M2 is a finer grid containing $\sim 179$M cells. We choose both the grids in such a way that they lie between the intermediate grid (M2) and finer grid (M3) of \cite{Liu_2023}. We plot the time averaged water volume fraction ($\overline{\alpha}_w$) and axial velocity ($\overline{u}_x/U_{\infty}$) along the wake centerline in figure \ref{fig:MP_Axial_Alp_Ux}. Both $\overline{\alpha}_w$ and $\overline{u}_x/U_{\infty}$ are independent of the grid resolution. Hence, we conducted our study using the M2 grid.

\bibliographystyle{jfm}
\bibliography{jfm}

\end{document}